%% file: paper_I.tex
\documentclass{emulateapj}
\usepackage{savesym}
\savesymbol{tablenum}
\usepackage{siunitx}
\restoresymbol{SIX}{tablenum}
\let\oldenddeluxetable\enddeluxetable
\let\olddeluxetable\deluxetable
\makeatletter
\renewenvironment{deluxetable}[1]{
\olddeluxetable{[#1]}
\def\pt@format{#1}
}{\oldenddeluxetable}
\makeatother
\usepackage{color}
\definecolor{white}{rgb}{1,1,1}

\usepackage{colortbl}
\definecolor{lmugreen}{rgb}{0.01,0.58,0.25}

\usepackage{amsmath}
\usepackage{natbib}
\usepackage{float}
\usepackage{dcolumn}
\usepackage{booktabs}
\usepackage[
	colorlinks=true,
	urlcolor=lmugreen,
	linkcolor=lmugreen,
	citecolor=lmugreen
	]{hyperref}
\usepackage{rotating}
\usepackage{graphicx}
\usepackage{tablefootnote}
\usepackage[nolist,nohyperlinks]{acronym}

\usepackage{listings}
\newcommand{\citeorder}[1]{}

\defcitealias{KDR1:inprep}{Paper I}
\defcitealias{KDR2a:inprep}{Paper II}
\defcitealias{KDR2b:inprep}{Paper III}
\defcitealias{KDR3:inprep}{Paper IV}
\defcitealias{KoepferlRobitaille:inprep}{\textsc{FluxCompensator} Paper}

\DeclareSIUnit\yr{yr}
\DeclareSIUnit\ergs{ergs}
\DeclareSIUnit\Jy{Jy}
\DeclareSIUnit\Hz{Hz}
\DeclareSIUnit\sr{sr}
\DeclareSIUnit\pc{pc}
\DeclareSIUnit\ppm{ppm}
\DeclareSIUnit\kpc{\kilo\pc}
\DeclareSIUnit\AU{AU}
\DeclareSIUnit\eqsign{\text{\ensuremath{=}}}
\DeclareSIUnit\simeqsign{\text{\ensuremath{\simeq}}}
\DeclareSIUnit\mag{mag}
\DeclareSIUnit\gramms{g}
\DeclareSIUnit\gccm{\gramms\per\cubic\centi\metre}
\DeclareSIUnit\microns{\micro\metre}
\DeclareSIUnit\Myr{\mega\yr}
\DeclareSIUnit\Msun{\text{\ensuremath{M_\odot}}}
\DeclareSIUnit\Lsun{\text{\ensuremath{L_\odot}}}
\DeclareSIUnit\Rsun{\text{\ensuremath{T_\odot}}}
\DeclareSIUnit\Minfall{\Msun\per\yr}

\slugcomment{The Astrophysical Journal Supplement}

\shorttitle{Synthetic Star-forming Regions - I. Reliable Mock Observations from SPH Simulations}
\shortauthors{Koepferl, Robitaille, Dale, Biscani}

\begin{document}

\title{Insights from Synthetic Star-forming Regions: \\I. Reliable Mock Observations from SPH Simulations}
\author{Christine M. Koepferl$^{1,2}$, Thomas P. Robitaille$^{1,3}$, James E. Dale$^4$, and Francesco Biscani$^1$}
\affil{$^1$ Max Planck Institute for Astronomy, K\"onigstuhl 17, D-69117 Heidelberg, Germany\\
$^2$ Scottish Universities Physics Alliance (SUPA), School of Physics and Astronomy, University of St Andrews\\North Haugh, St Andrews, KY16 9SS, UK\\
$^3$ Freelance Consultant, Headingley Enterprise and Arts Centre, Bennett Road Headingley, Leeds LS6 3HN\\
$^4$ University Observatory Munich, Scheinerstr. 1, D-81679 Munich, Germany}
\email{cmk8@st-andrews.ac.uk}
\received{29 January 2016}
\accepted{1 March 2016}

\begin{abstract}
Through synthetic observations of a hydrodynamical simulation of an evolving star-forming region, we assess how the choice of observational techniques affects the measurements of properties which trace star formation. Testing and calibrating observational measurements requires synthetic observations which are as realistic as possible. In this part of the paper series \citepalias{KDR1:inprep}, we explore different techniques for how to map the distributions of densities and temperatures from the particle-based simulations onto a Voronoi mesh suitable for radiative transfer and consequently explore their accuracy. We further test different ways to set up the radiative transfer in order to produce realistic synthetic observations. We give a detailed description of all methods and ultimately recommend techniques. We have found that the flux around \SI{20}{\microns} is strongly overestimated when blindly coupling the dust radiative transfer temperature with the hydrodynamical gas temperature. We find that when instead assuming a constant background dust temperature in addition to the radiative transfer heating, the recovered flux is consistent with actual observations. We present around \num{5800} realistic synthetic observations for \emph{Spitzer} and \emph{Herschel} bands, at different evolutionary time-steps, distances and orientations. In the upcoming papers of this series (\citetalias{KDR2a:inprep}, \citetalias{KDR2b:inprep} and \citetalias{KDR3:inprep}), we will test and calibrate measurements of the \ac{SFR}, gas mass and the \ac{SFE} using our realistic synthetic observations.
\end{abstract}

%%%%%%%%%%%%%%%%%%%%%%%%%%%%%%%%%%%%%%%%%%%%%%%%%%%%%%%
%%%%%%%%%%%%%%%%%%%%%%%%%%%%%%%%%%%%%%%%%%%%%%%%%%%%%%%
\section{Introduction}
\label{C4:Sec:intro}
%%%%%%%%%%%%%%%%%%%%%%%%%%%%%%%%%%%%%%%%%%%%%%%%%%%%%%%
%%%%%%%%%%%%%%%%%%%%%%%%%%%%%%%%%%%%%%%%%%%%%%%%%%%%%%%
Over the last decade, progress in the field of star formation has been fueled by an increasing number of large infrared/optical surveys of our Galaxy (e.\,g.~from space missions, such as \emph{Hubble}, \emph{Spitzer}, \emph{Herschel}, \emph{Planck}, and ground missions, such as \ac{2MASS}, \ac{UKIDSS} and \acs{ATLASGAL} and a growing number of state-of-the-art simulations of entire star-forming regions with increasing quality and diversity (e.\,g.~\textsc{Orion}, \textsc{SPH-NG}, \textsc{Arepo}; for more details and recent application, see e.\,g.~\citealt{Offner:2012}, \citealt{Bate1:2014}, \citealt{SpringelVORO:2010}). Thanks to recent advances in radiative transfer techniques (in particular 3-d Monte-Carlo radiative transfer; for more details, see the review of \citealt{Steinacker:2013}) we are now able to produce synthetic observations of star-forming regions. Synthetic observations have the potential, when constructed with great care and effort, to close the loop between simulations and observations. Through \textbf{synthetic observations}: 
\begin{enumerate}
    \item On the one hand, simulations can be tested to see whether they reproduce features seen in observations. Whether certain physical processes are dominant over other processes can be tested with synthetic observations by comparing the mock observations of simulated parameter studies to real observations.\\[-0.5cm] 
    \item On the other hand, synthetic observations allow us to test and calibrate techniques that are used by the community to infer properties related to star formation. Since the intrinsic properties of the simulations (e.\,g.~\ac{SFR}s, gas masses, filament widths) are known, the accuracy of the observational determination can be explored. Such an approach enables us also to improve existing techniques and to develop new techniques that can produce better measurements. 
\end{enumerate}

Lately, there have been many attempts to use synthetic observations with respect to their potential listed above. For example, \cite{Kurosawa:2004} used one small-scale simulation of a star-forming region from \cite{Bate:2002a,Bate:2002b,Bate:2003}. They mapped their initial particle-based simulation output onto an \ac{AMR} grid and included the density profile of a protoplanetary disk close to their stellar accretion particles (represented by sink particles, for more details on sink particles see \citealt{BateSink:1995}). With their parameter study that included set-ups with and without accretion disks, they developed classification criteria for accretion disks in the \ac{MIR}.

\cite{Offner:2012} also used a simulation of a small-scale star-forming region to estimates the quality and reliability of properties inferred for protostars from \acs{SED}-fitting techniques. In contrast to \cite{Kurosawa:2004}, they did not refine the input density of the simulation further, e.\,g.~with a protoplanetary disk, since the original resolution was sufficient to recover the flux of contributing regions, while unresolved regions were too optically thick to produce flux in the "observed" wavelength bands.
\begin{figure*}[t]
\includegraphics[trim=0cm 22cm 16cm 0cm, width=\textwidth]{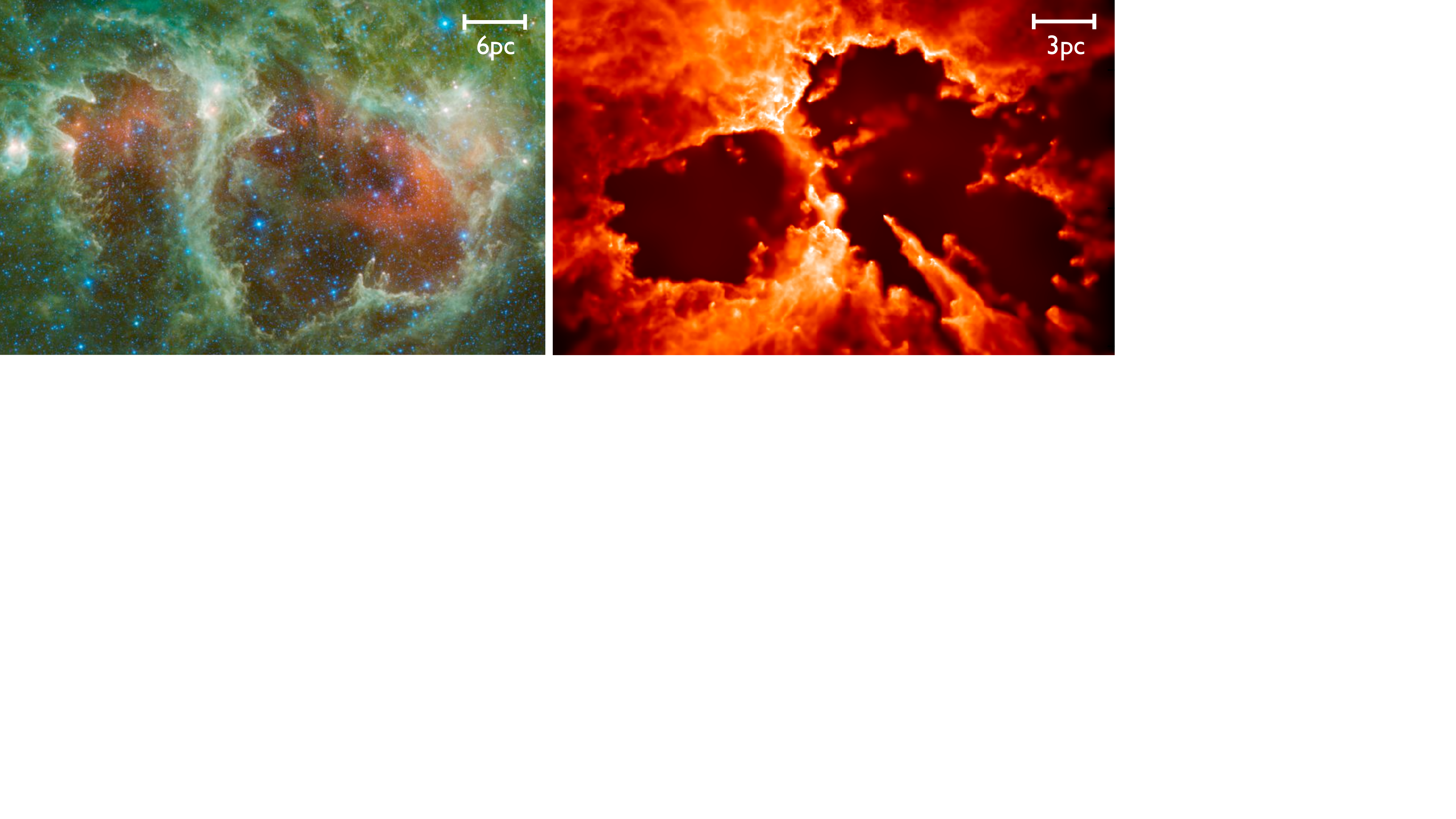}
\caption[Observation and simulation of a star-forming region]{\label{C4:Fig:apples_light}(Left) \ac{MIR} \ac{WISE} observation\footnote{Online references accessed: 30.08.2015;
\url{http://www.jpl.nasa.gov/spaceimages/details.php?id=PIA13112} \\Courtesy NASA/JPL-Caltech/UCLA} of the Soul Nebula star-forming region (blue: \SI{3.4}{\microns}, cyan: \SI{4.6}{\microns}, green: \SI{12}{\microns}, red: \SI{22}{\microns}), (right) column density structure of a simulated star-forming region from the \acs{D14} \ac{SPH} simulations\footnote{Surface density maps provided by Jim Dale.} with high-mass stellar feedback (ionization-only).}
\end{figure*}

Up until now, most studies with synthetic observations have focused on simulations of star-forming regions below the \si{\pc}-scale. Moreover, usually only one snapshot from a simulation is tested, rather than several evolutionary time-steps of the same simulation for different orientations and distances, which would improve the reliability of the conclusions. 

We believe that the sequence of simulations initiated by \cite{DaleI:2011} is a good choice for producing and exploring the applications of synthetic observations, since they cover large-scale single star-forming regions with sizes on the order of several tens of \si{\pc}. The initial simulation in this sequence of many published \acs{SPH} simulated star-forming regions covered only one simulated star-forming cluster which satisfies the modified\footnote{The \ac{EOS} has been adjusted to recover a Jeans mass $M_{\textup{Jeans}}$ of \SI{0.5}{\Msun}.} \cite{Larson:2005} \ac{EOS} with an isothermal dust cooling part at \SI{7.5}{\kelvin} and a line cooling process, but lacks stellar feedback mechanisms such as heating through radiation, ionization\footnote{ \cite{DaleI:2011} also had an ionization description implemented but the impact was rather small because their simulated region had a large escape velocity.} \& winds. This initial simulation was extended by different high-mass stellar feedback mechanisms, such as ionization (see \citealt{DaleCodeIoni:2007} for the formalism), introduced by \cite{DaleIoni:2012} for bound and \cite{DaleIoni:2013} for unbound clouds. Further, \cite{DaleWind:2013} introduced winds as feedback of bound and unbound clouds (see \citealt{DaleCodeWind:2008} for the formalism). A combination of the initial high-mass stellar feedback parameter study (ionization \& wind) for bound and unbound clouds was presented by \cite{DaleBoth:2014}. With this large set of simulations, these authors are studying the impact of feedback on the star-forming regions. For example, in \cite{DaleBoth:2014}, they compare the effects of different feedback mechanisms on the \ac{SFR} and \ac{SFE}. For simplicity, we will hereafter refer to this set of simulations as the \acs{D14} \ac{SPH} simulations. 

In this paper, hereafter referred to \citetalias{KDR1:inprep}, we will produce realistic synthetic observations of a simulated star-forming region at different stages of evolution, orientations and distances from the \acs{D14} \ac{SPH} simulations. We use the 3-d dust continuum Monte-Carlo radiative transfer code \textsc{Hyperion}, developed by \cite{Robitaille:2011}, to produce ideal synthetic observations. \textsc{Hyperion} uses a gridded density distribution. In the radiative transfer code, photon packets are emitted from the stellar sources and travel through the density distribution and interact with the dust. They can get scattered, absorbed or re-emitted by the dust grains. \textsc{Hyperion} calculates the specific energy absorbed by the dust resulting from this stellar heating and estimates the dust temperature, and ideal synthetic images can be produced. For more details about \textsc{Hyperion} and radiative transfer in general see \cite{Robitaille:2011} and \cite{Steinacker:2013}. 

\subsection{Motivation}
In this paper \citepalias{KDR1:inprep}, we will describe the set-up and the post-processing of radiative transfer models using particle-based simulations as input. Note that we will not focus on the calculations with the radiative transfer code itself. Rather we will explore different methods for how to set up the radiative transfer using the data from the \ac{SPH} simulations and in order to produce realistic synthetic observations from the radiative transfer output. We go through this effort, so that we can find reliable techniques, to recover the infrared flux as accurately as possible, which is commonly used by observers to interpret star-forming regions.

\subsection{Outline}
In Section~\ref{C4:Sec:methods_HD}, we will introduce the \acs{D14} \ac{SPH} simulations in more detail, show the evolution of properties over the relevant time-steps and justify the pre-processing steps to make the simulation sample suitable for the radiative transfer set-up. In Section~\ref{C4:Sec:methods_sph2voro}, we will present different approaches for discretizing an \ac{SPH} simulation onto a radiative transfer grid such as a Voronoi mesh. Aspects of setting up radiative transfer photon sources are discussed in Section~\ref{C4:Sec:methods_RT}, before the careful set-up of circumstellar material is described in Section~\ref{C4:Sec:methods_YSO}. In Section~\ref{C4:Sec:methods_dust_temp}, different dust properties and temperature combinations and their effects on the radiative transfer output are explored. We give suggestions in the respective sections on which of the methods (explored in Section~\ref{C4:Sec:methods_HD} to Section~\ref{C4:Sec:methods_dust_temp}) to favor. The post-processing of the radiative transfer output is described in Section~\ref{C4:Sec:methods_synobs} and the resulting realistic synthetic observations are presented there. In Section~\ref{C4:Sec:discuss}, we summarize caveats when transferring simulations to "observations". 
\begin{table}[t]
\caption[Summary of Properties of the \emph{run I} Simulated Star-forming Region]{Summary of Properties\\ of the \emph{run I} Simulated Star-forming Region.}
	\label{C4:Tab:runI}
\begin{center}
\begin{tabular*}{0.45\textwidth}{crrrcc}
\hline\\[-13pt]
\hline\\[-5pt]
&&&&&\\
\multicolumn{1}{c}{time-step} &
\multicolumn{1}{c}{time} &
\multicolumn{1}{c}{$r(M_{1/2})$} &
\multicolumn{1}{c}{sink} & 
\multicolumn{1}{c}{$M_{\textup{gas}}$} &
\multicolumn{1}{c}{feedback}\\
\multicolumn{1}{c}{(ID)} &
\multicolumn{1}{c}{(\si{\mega\yr})} &
\multicolumn{1}{c}{(\si{\pc})} &
particles& 
\multicolumn{1}{c}{$(M_\odot)$} &
on?\\[6pt]
\hline\\[-5pt]
024 &	3.576 	&	 8.45 	&	3 	&	9873 	&	no	\\
025	&	3.725	&	 8.52 	&	4	&	9825	&	no	\\
026	&	3.874	&	 8.60 	&	12	&	9750	&	no	\\
027	&	4.023	&	 8.70 	&	15	&	9664	&	no	\\
028	&	4.172	&	 8.77 	&	17	&	9593	&	no	\\
029	&	4.321	&	 8.83 	&	21	&	9536	&	no	\\
030	&	4.470 	&	 8.89 	&	30	&	9476	&	no	\\
031	&	4.619	&	 8.94 	&	32	&	9414	&	no	\\
032	&	4.768	&	 9.01 	&	44	&	9341	&	no	\\
042	&	4.917	&	 9.08 	&	48	&	8833	&	yes	\\
052	&	5.066	&	 9.27 	&	48	&	8643	&	yes	\\
062	&	5.215	&	 9.58 	&	53	&	8598	&	yes	\\
072	&	5.364	&	 9.91 	&	61	&	8495	&	yes	\\
082	&	5.513	&	 10.30 	&	73	&	8377	&	yes	\\
092	&	5.662	&	 10.79 	&	85	&	8187	&	yes	\\
102	&	5.811	&	 11.30 	&	91	&	7987	&	yes	\\
112	&	5.960 	&	 11.84 	&	98	&	7722	&	yes	\\
122	&	6.109	&	 12.37 	&	100	&	7500	&	yes	\\
132	&	6.258	&	 12.94 	&	108	&	7254	&	yes	\\
142	&	6.407	&	 13.50 	&	115	&	6996	&	yes	\\
152	&	6.556	&	 14.09 	&	122	&	6687	&	yes	\\
162	&	6.705	&	 14.69 	&	125	&	6409	&	yes	\\
172	&	6.854	&	 15.33	&	128	&	6105	&	yes	 \\[6pt]
\hline\\[-5pt]
 \end{tabular*}
\end{center}
\vspace*{-0.3cm}
\end{table}

In \cite[submitted]{KDR2a:inprep} and \cite[submitted]{KDR2b:inprep}, hereafter referred to \citetalias{KDR2a:inprep} and \citetalias{KDR2b:inprep}, we use the resulting realistic synthetic observations at different evolutionary stages to test and calibrate techniques which measure properties, such as the \ac{SFR}, the gas mass and the \ac{SFE}.
%%%%%%%%%%%%%%%%%%%%%%%%%%%%%%%%%%%%%%%%%%%%%%%%%%%%%%%
%%%%%%%%%%%%%%%%%%%%%%%%%%%%%%%%%%%%%%%%%%%%%%%%%%%%%%%
\section{Hydrodynamical Simulations}
\label{C4:Sec:methods_HD}
%%%%%%%%%%%%%%%%%%%%%%%%%%%%%%%%%%%%%%%%%%%%%%%%%%%%%%%
%%%%%%%%%%%%%%%%%%%%%%%%%%%%%%%%%%%%%%%%%%%%%%%%%%%%%%%
The \acs{D14} \ac{SPH} simulations are a set of over 20 different realizations with different initial conditions for initially bound and unbound star-forming regions. Furthermore, for every realization, different combinations of high-mass stellar feedback mechanisms, such as ionization alone, winds alone, winds \& ionization coupled, are calculated. The morphological features and physical scale of the simulations (half-mass radii $r(M_{1/2})$ from \SIrange{1}{100}{\pc}) are of the same order as those of observed regions with high-mass star formation in the Galactic plane. This can be seen in Figure~\ref{C4:Fig:apples_light}, where we show a three-color \ac{MIR} \ac{WISE} observation of the Soul Nebula (left, \citealt{Koenig:2012,Wright:2010}) and the column density structure of one snapshot from the \acs{D14} \ac{SPH} simulations including high-mass stellar feedback in form of ionization (right). As mentioned in Section~\ref{C4:Sec:intro}, the \acs{D14} \ac{SPH} simulations are in some sense unique in the community since they cover single star-forming regions on large scales of the order of several tens of parsec with different feedback scenarios. 
%%%%%%%%%%%%%%%%%%%%%%%%%%%%%%%%%%%%%%%%%%%%%%%%%%%%%%%
\subsection{Simulation Properties}
\label{C4:Sec:properties}
%%%%%%%%%%%%%%%%%%%%%%%%%%%%%%%%%%%%%%%%%%%%%%%%%%%%%%%
The density distribution in particle-based simulations, such as the \acs{D14} \ac{SPH} simulations, is approximated by the sum of many individual \ac{SPH} particles with an analytical description.
%%%%%%%%%%%%%%%%%%%%%%%%%%%%%%%%%%%%%%%%%%%%%%%%%%%%%%%
\subsubsection{SPH Particles}
\label{C4:Sec:properties_sph}
%%%%%%%%%%%%%%%%%%%%%%%%%%%%%%%%%%%%%%%%%%%%%%%%%%%%%%%
From the simulation output, we know the position of an \ac{SPH} particle $i$, its peak density $\rho_{i,\textup{SPH}}$, its hydrodynamical gas temperature $T_{i,\textup{SPH}}$, its constant particle mass of $m_{i,\textup{SPH}}=\SI{e-2}{\Msun}$\footnote{ This is only true for the simulated clouds with a total mass of  $M_{\textup{cloud}}=\SI{e4}{\Msun}$  for instance run I, since $m_{i,\textup{SPH}}=M_{\textup{cloud}}/N_{\textup{SPH}}$, with the total \ac{SPH} particle number $N_{\textup{SPH}}$ being \si{1e6}.} and its smoothing length $h_i$. A \ac{SPH} particle $i$ has a 3-d Gaussian-like profile. The profile strength at position $j$ with distance $r_{ij}$ from the \ac{SPH} particle $i$ is defined by the kernel function \citep[for a detailed review, see][]{SpringelSPH:2010}:
\begin{eqnarray}
    \label{C4:Eq:kernel}
    W_{ij}(r_{ij},h_i)=\frac{1}{h_{i}^3 \pi}\times\ \ \ \ \ \ \ \ \ \ \ \ \ \ \ \ \ \ \ \ \ \ \ \ \ \ \ \ \ \ \ \ \ \ \ \ \ \nonumber\\
    \Bigg\{\begin{matrix}
  1 - 1.5(r_{ij}/h_i)^2 + 0.75 (r_{ij}/h_i)^3 && \forall\ r_{ij} < h_i\\
  0.25(2 -(r_{ij}/h_i))^3 && \forall \ h_i \leq r_{ij} < 2h_i. \\
  0 && \forall\ r_{ij} > 2 h_i
 \end{matrix}
\end{eqnarray}

%%%%%%%%%%%%%%%%%%%%%%%%%%%%%%%%%%%%%%%%%%%%%%%%%%%%%%%
\subsubsection{Sink Particles}
\label{C4:Sec:properties_sinks}
%%%%%%%%%%%%%%%%%%%%%%%%%%%%%%%%%%%%%%%%%%%%%%%%%%%%%%%
The stellar particles in an \ac{SPH} simulations are represented by sink particles \citep{BateSink:1995}. They form from \ac{SPH} particles which are bound by the gravitational potential of a forming stellar object and follow the following conditions: 

\begin{itemize}
    \item \textbf{Density Threshold}\\
    The density threshold is evaluated from the minimal resolvable Jeans mass $M_{\textup{Jeans}}$ and the isothermal temperature in the simulations. \ac{SPH} particles, which have higher densities than the threshold density, are tracked and the nearest neighbors are tested by the following conditions.
    \item \textbf{Jeans Limit}\\
    The average density and temperature of the group of \ac{SPH} particles. The resulting mean total mass of the group should be higher than the Jeans mass $M_{\textup{Jeans}}$.
    \item \textbf{Potential}\\
    The group should be gravitationally bound.
    \item \textbf{Velocity Field}\\
    The group should have a negative divergence of the velocity field at the location of the densest particle of the group which indicates a collapse.
    \item \textbf{Energy Balance}\\
    The group of collapsing \ac{SPH} should also have a higher gravitational energy than its rotational kinetic energy to ensure that the collapse cannot be stabilized by angular momentum conservation.
\end{itemize}
In the \acs{D14} \ac{SPH} simulations, a sink particle forms once more than 50 \ac{SPH} particles satisfy the above conditions at a given position. In Table~\ref{C4:Tab:runI} we give the number of sink particles at different time-steps in one of the \acs{D14} \ac{SPH} simulations. 

From the minimum resolved Jeans mass $M_{\textup{Jeans}}$, a respective length scale, called the accretion radius $r_{\textup{acc}}$, is calculated, which is comparable to the Jeans length $\lambda_{\textup{Jeans}}$. \ac{SPH} particles are accreted by a sink particle when they are within the accretion radius $r_{\textup{acc}}$, are gravitationally bound by the sink particle's potential and have low enough rotational energies. The sink particles\footnote{For the intermediate mass clouds of the \acs{D14} \ac{SPH} simulations sink particles as low as \SI{0.5}{\Msun} can be formed and the sink particles and represent stars. For clouds above \SI{e5}{\Msun}, which are not subject of this paper, sink particles represent subclusters.} grow over time through accretion of many \ac{SPH} particles.
%
%%%%%%%%%%%%%%%%%%%%%%%%%%%%%%%%%%%%%%%%%%%%%%%%%%%%%%%
\subsubsection{Feedback}
\label{C4:Sec:properties_feedback}
%%%%%%%%%%%%%%%%%%%%%%%%%%%%%%%%%%%%%%%%%%%%%%%%%%%%%%%
High-mass stellar feedback is switched on in the \acs{D14} \ac{SPH} simulations as soon as three high-mass sink particles above \SI{20}{\Msun} have formed\footnote{This is only true for intermediate mass clouds below \SI{e5}{\Msun} in the \acs{D14} \ac{SPH} simulations.}. The simulation ends before the first supernova of a high-mass sink particle would presumably go off. 

%%%%%%%%%%%%%%%%%%%%%%%%%%%%%%%%%%%%%%%%%%%%%%%%%%%%%%%
\subsubsection{Run I}
\label{C4:Sec:properties_runI}
%%%%%%%%%%%%%%%%%%%%%%%%%%%%%%%%%%%%%%%%%%%%%%%%%%%%%%%
In this work, we explore the \emph{run I} of the \acs{D14} \ac{SPH} simulations an intermediate mass cloud with a total gas mass of \SI{e4}{\Msun} and initially \num{e6} \ac{SPH} particles. We choose this run because it is of relatively high resolution with a sink accretion radius of $r_{\textup{acc}}=\SI{0.005}{\pc}$. Individual stellar objects in the form of sink particles with masses between \SI{0.5}{\Msun} and \SI{70}{\Msun} are produced and can be resolved, and therefore can be interpreted as single stars. The first stars formed after about \SI{3.6}{\Myr} after the start of the simulation (see also Section~\ref{C4:Sec:properties_age}). The stars in the simulation grow through accretion (see also Section~\ref{C4:Sec:properties_accretion}) and ensures that the stellar population aquires its own \ac{IMF} without imposing an initial \ac{IMF} to the simulation (see \citealt{DaleIoni:2012}).

We choose \emph{run I} because we aim to produce synthetic observations which are as close to real star-forming regions as possible. Therefore, for our studies, we limit ourselves to the coupled feedback scenario: ionization \& winds. The ionizing sources in the simulations emit a flux of ionized photons ranging from \SI{1.3e48}{\per\second} to \SI{1.0e49}{\per\second}. They also emit winds\footnote{only modeled as momentum fluxes} with a speed ranging from \SI{1780}{\kilo\meter\per\second} to \SI{3125}{\kilo\meter\per\second} and a respective mass-loss rate between \SI{3.1e-7}{\Msun\per\yr} to \SI{2.8e-6}{\Msun\per\yr}. For a detailed description of the high-mass feedback formalism, see \cite{Dale:Review:2015,DaleWind:2013,DaleIoni:2012,DaleCodeWind:2008,DaleCodeIoni:2007}. 

In \emph{run I} the first stars formed after about \SI{3.6}{\Myr}, high-mass stellar feedback is switched on about \SI{4.8}{\Myr} and the simulation is stopped at about \SI{7}{\Myr} after the start of the simulation. The simulation is stopped before the first supernova\footnote{Note that the supernova explosion is not part of the \acs{D14} \ac{SPH} simulations. For further information about the supernova (e.\,g.~timing) see the \acs{D14} \ac{SPH} simulations.} would presumably go off. We select \num{23} time-steps with a constant step size $\Delta t = \SI{149000}{\yr}$, starting once the first star has formed. At every time-step $t_{\textup{step}}$ and for every sink particle, we keep track of the physical properties, such as the sink particle mass $M_{\textup{sink}}(t_{\textup{step}})$, the age of the sink particle $t_{\textup{age}}(t_{\textup{step}})$ and the accretion rate $\dot{M}_{\textup{acc}}(t_{\textup{step}})$. For more details see Table~\ref{C4:Tab:runI}.
%%%%%%%%%%%%%%%%%%%%%%%%%%%%%%%%%%%%%%%%%%%%%%%%%%%%%%%
\subsubsection{Sink Particle Mass}
\label{C4:Sec:properties_mass}
%%%%%%%%%%%%%%%%%%%%%%%%%%%%%%%%%%%%%%%%%%%%%%%%%%%%%%%
The sink particle mass is defined as the sum over all $N_{\textup{acc}}$ accreted \ac{SPH} particles $i$:
\begin{eqnarray}
    \label{C4:Eq:M_sink}
    M_{\textup{sink}}(t_{\textup{step}})&=&\sum\limits_{i=1}^{N_{\textup{acc}}}m_{i,\textup{SPH}}.
\end{eqnarray}

%%%%%%%%%%%%%%%%%%%%%%%%%%%%%%%%%%%%%%%%%%%%%%%%%%%%%%%
\subsubsection{Sink Particle Age}
\label{C4:Sec:properties_age}
%%%%%%%%%%%%%%%%%%%%%%%%%%%%%%%%%%%%%%%%%%%%%%%%%%%%%%%
Hereafter, we define the age of a sink particle as 
\begin{eqnarray}
    \label{C4:Eq:t_age}
    t_{\textup{age}}(t_{\textup{step}})&=&t_{\textup{step}}-t_{\textup{appear}} + \Delta t/2,
\end{eqnarray}
where $t_{\textup{appear}}$ is the time of the time-step when a sink particle first forms (binds more then 50 \ac{SPH} particles). We add the time $\Delta t/2$ to the age because the sink particle may have formed at any time between the time-step when it appeared and the previous one, and therefore on average we assume that it formed half way in between the two.
%%%%%%%%%%%%%%%%%%%%%%%%%%%%%%%%%%%%%%%%%%%%%%%%%%%%%%%
\subsubsection{Sink Accretion Rate}
\label{C4:Sec:properties_accretion}
%%%%%%%%%%%%%%%%%%%%%%%%%%%%%%%%%%%%%%%%%%%%%%%%%%%%%%%
We consider a sink particle to be accreting between two different time-steps, when 
\begin{eqnarray}
    \mbox{accretion}|_{t_{\textup{step}}\rightarrow t_{\textup{step}+1}}&=&M_{\textup{sink}}(t_{\textup{step}+1}) > M_{\textup{sink}}(t_{\textup{step}}).\ \ \ \ 
\end{eqnarray}
The accretion definition becomes important when mapping the \ac{SPH} simulation to the Voronoi mesh as described in Section~\ref{C4:Sec:methods_sph2voro}. We define the accretion rate at a time $t_{\textup{step}}$ as the  mass-gain rate between two time-steps:
\begin{eqnarray}
    \label{C4:Eq:infall}
    \dot{M}_{\textup{acc}}(t_{\textup{step}})&=&\frac{M_{\textup{sink}}(t_{\textup{step}+1}) - M_{\textup{sink}}(t_{\textup{step}-1})}{2\Delta t}.
\end{eqnarray}
%

%%%%%%%%%%%%%%%%%%%%%%%%%%%%%%%%%%%%%%%%%%%%%%%%%%%%%%%
\subsection{Pre-processing the SPH Simulation Output}
\label{C4:Sec:clipping}
%%%%%%%%%%%%%%%%%%%%%%%%%%%%%%%%%%%%%%%%%%%%%%%%%%%%%%%
In subsequent sections, we want to extract high-resolution radiative transfer images (see Section~\ref{C4:Sec:methods_synobs}) with constant pixel number and pixel-size scale over all time-steps. Since we are only interested in the inner region of the simulation, we spatially clip the simulations in order to save computational resources. The clear trade-off is that we are then not always tracing the same amount of mass in each time-step.

We clip the \acs{D14} \ac{SPH} simulation output in physical size to a cube with length \SI{30}{\pc}. For the first time-steps, the star-forming region is completely contained inside the box; once the high-mass stellar feedback has been switched on, the region expands beyond the boundaries of the cube. In the final time-steps, the boundaries of the cube are approximately the half-mass radius $r(M_{1/2})$ of the simulations. More detailed values are provided in Table~\ref{C4:Tab:runI}, where we summarize some properties of \emph{run I} at every time-step.

In the following, we investigate two possible selection methods to single out the appropriate sample of \ac{SPH} particles for the respective radiative transfer set-up:
\begin{figure*}[t]
\includegraphics[width=1\textwidth]{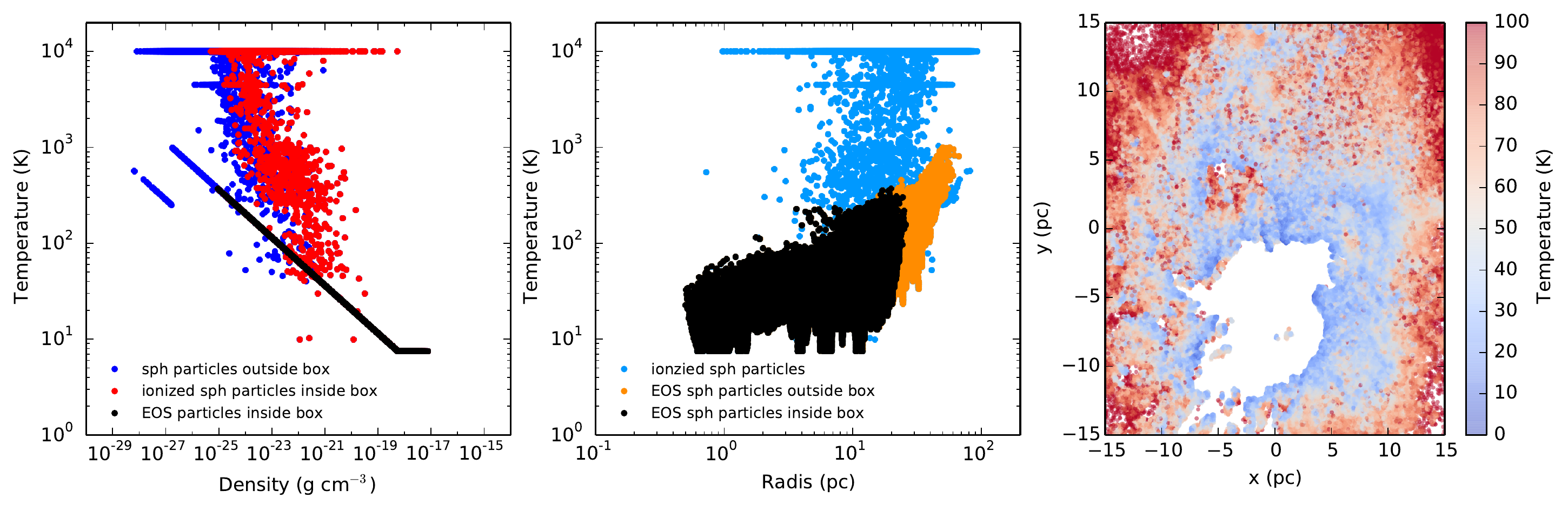}
\caption[Temperature relation in density and radius of SPH particles]{\label{C4:Fig:EOS}Temperature relation in density and radius of \ac{SPH} particles and a 2-d projection showing the temperature increase at the edges of the simulation at time-step 122 (\SI{6.109}{\Myr}).}
\end{figure*}

%%%%%%%%%%%%%%%%%%%%%%%%%%%%%%%%%%%%%%%%%%%%%%%%%%%%%%%
\subsubsection{Method c1 | Neutral SPH Particles in Box}
\label{C4:Sec:methods_c1}
%%%%%%%%%%%%%%%%%%%%%%%%%%%%%%%%%%%%%%%%%%%%%%%%%%%%%%%
While the \acs{D14} \ac{SPH} simulations have high-mass stellar feedback mechanisms implemented, the simulations do not take into account the dust heating from the stars. In \acs{D14} \ac{SPH} simulations, the external heating through the diffuse Galactic radiation field is approximated by the \cite{Larson:2005} \ac{EOS}\footnote{ The \ac{EOS} has been adjusted by the \acs{D14} \ac{SPH} simulations of intermediate clouds to recover a Jeans mass $M_{\textup{Jeans}}$ of \SI{0.5}{\Msun}.} rather than just an isothermal approximation. The \ac{EOS} contains the following regimes (following Equation~1 and Figure~1 of \citealt{DaleIoni:2012}):
\begin{itemize}
    \item line cooling processes
\begin{eqnarray}
    \label{C4:Eq:line}
 &P& \sim \rho^{3/4}\\
 &\forall& \ \rho < \SI{5.5e-19}{\gccm}\nonumber
\end{eqnarray}
    \item dust cooling processes
\begin{eqnarray}
    \label{C4:Eq:dust}
&P& \sim \rho^{1.0}\\
&\forall& \ \SI{5.5e-19}{\gccm} < \rho < \SI{5.5e-15}{\gccm},\nonumber
\end{eqnarray}
\end{itemize}  
where $P$ is the gas pressure and $\rho$ the gas density.
Since in the ideal gas approximation \citep[see][]{Feynman1}, the pressure $P$ always follows 
\begin{eqnarray}
\label{C4:Eq:line2}
P=N k_B T / V = \rho k_B T / m_p &\sim&\rho T,
\end{eqnarray}
with gas particle number $N$, the Boltzmann constant $k_B$, the gas temperature $T$, the gas particle mass $m_p$ and the gas Volume $V$, we can also express the line cooling term from Eq.~\ref{C4:Eq:line} as a power-law of density $\rho$ and temperature $T$:
\begin{eqnarray}
\label{C4:Eq:line3}
P \sim \rho T \sim \rho^{3/4} &\rightarrow& \rho \sim T^{-4}.
\end{eqnarray}
At high densities, dust cooling is assumed to dominate and the \ac{EOS} becomes isothermal with $T_{\textup{iso}}=\SI{7.5}{\kelvin}$. The densities in large parts of the cloud are low enough that the gas and dust should not be well coupled thermally.

In Figure~\ref{C4:Fig:EOS}\textcolor{lmugreen}{(a)}, we have plotted temperature versus density for time-step 122 (at \SI{6.109}{\Myr}) of \emph{run I} \SI{1.341}{\Myr} after switching on high-mass stellar feedback (c.\,f.~Table~\ref{C4:Tab:runI}). Points regardless of color (blue, red and black points together) represent all \num{e6} \ac{SPH} particles of \emph{run I}. The red and black points represent \ac{SPH} particles which lie within our selected box of \SI{30}{\pc} width. The red points are (partly or fully) ionized particles (c.\,f.~with \citealt{DaleIoni:2012,DaleCodeIoni:2007}) within the box. The horizontal line at \SI{e4}{\kelvin} shows all the fully ionized \ac{SPH} particles. The other red points are partly ionized and lie within the selected box. The black points are neutral \ac{SPH} particles in the selected box. The state of these neutral particles is described by the \ac{EOS} which is describing the dependency of density and temperature (diagonal line: Eq.~\ref{C4:Eq:line}, Eq.~\ref{C4:Eq:line2}, Eq.~\ref{C4:Eq:line3}; horizontal: Eq.~\ref{C4:Eq:dust}).

In this work, we will use a dust radiative transfer code, therefore we will only use \ac{SPH} particles in the box which are completely neutral since dust gets destroyed during the ionization process \citep{DiazMiller:1998}. Hereafter, we call the clipping for size and only neutral \ac{SPH} particles, resulting in the black particles in Figure~\ref{C4:Fig:EOS}\textcolor{lmugreen}{(a)}, method \acs{c1}.
%%%%%%%%%%%%%%%%%%%%%%%%%%%%%%%%%%%%%%%%%%%%%%%%%%%%%%%
\subsubsection{Method c2 | Neutral SPH Particles in Box \& Temperature Limit}
\label{C4:Sec:methods_c2}
%%%%%%%%%%%%%%%%%%%%%%%%%%%%%%%%%%%%%%%%%%%%%%%%%%%%%%%
In Figure~\ref{C4:Fig:EOS}\textcolor{lmugreen}{(b)}, we can see that the \ac{SPH} temperature increases with radius from the center of the simulation. Hot ionized particles (cyan) move to larger radii, but \ac{SPH} particles that still follow the \ac{EOS} also have higher temperatures at larger radii. This is a result of the \ac{EOS}, since the density decrease with radius in the centrally-condensed clouds. These can also be seen in Figure~\ref{C4:Fig:EOS}\textcolor{lmugreen}{(b)} and the 2-d temperature plot of Figure~\ref{C4:Fig:EOS}\textcolor{lmugreen}{(c)}, where we only plot the black sample of Figure~\ref{C4:Fig:EOS}\textcolor{lmugreen}{(a,b)}. The maximum temperature at the boundary of the box at this time-step lies around \SI{400}{\kelvin}. 

The \ac{SPH} temperature of neutral particles in the box is lower than the typical dust sublimation temperature of \SI{1600}{\kelvin} \citep[e.\,g.][]{Whitney:2003}, but much higher than the expected ambient dust temperature around a star-forming region of about \SI{18}{\kelvin} (see \citetalias{KDR2a:inprep}). We explore further temperature cuts of the line cooling part of the \ac{EOS}. Hereafter, we call the clipping in size using only neutral particles, and removing particles above a certain threshold temperature clipping, method \acs{c2}. Note that method \acs{c2} is the same as \acs{c1}, but additionally particles above a certain threshold temperature are removed.
%%%%%%%%%%%%%%%%%%%%%%%%%%%%%%%%%%%%%%%%%%%%%%%%%%%%%%%
\subsubsection{Results | Pre-processing}
\label{C4:Sec:results_clipping}
\begin{figure*}[t]
\includegraphics[width=\textwidth]{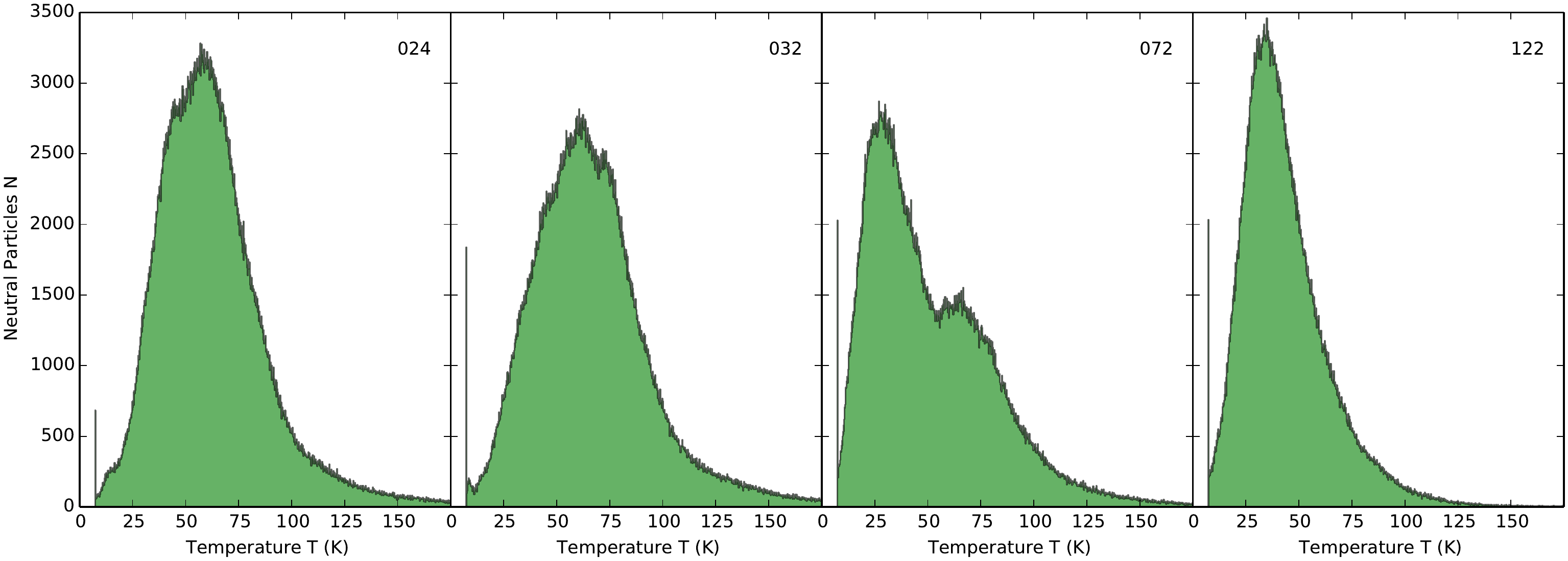}
\caption[Temperature histogram of neutral SPH particles]{\label{C4:Fig:temp_hist} Temperature histogram of neutral \ac{SPH} particles for time-step 024 (\SI{3.576}{\Myr}), 032 (\SI{4.768}{\Myr}), 072 (\SI{5.364}{\Myr}) and 122 (\SI{6.109}{\Myr}). Feedback gets switched on after time-step 32. The sharp spikes at \SI{7.5}{\kelvin} are \ac{SPH} particles in the isothermal phase.}
\end{figure*}

%%%%%%%%%%%%%%%%%%%%%%%%%%%%%%%%%%%%%%%%%%%%%%%%%%%%%%%
In Figure~\ref{C4:Fig:temp_hist}, we show the temperature histogram of neutral particles with a box of \SI{30}{\pc} of the four time-steps 024 (\SI{3.576}{\Myr}), 032 (\SI{4.768}{\Myr}), 072 (\SI{5.364}{\Myr}) and 122 (\SI{6.109}{\Myr}). We can see that the temperature peak shifts from $T=\SI{60}{\kelvin}$ to about $T=\SI{35}{\kelvin}$ once the high-mass stellar feedback is turned on after time-step 32 (\SI{4.768}{\Myr}). This is due to the sweeping up of the low-density neutral gas into dense, cool clumps.

Therefore, a clipping, and hence removing of the \ac{SPH} particles above a certain temperature (e.\,g.~\SI{60}{\kelvin}) in clipping method \acs{c2}, is not helpful, since we would lose too many particles for certain time-steps. Also, a threshold temperature below the sublimation is hard to pick, since it remains unclear when exactly the temperature of the dust decouples from the temperature of the gas. Hereafter, we will only remove non-neutral \ac{SPH} particles from the simulation, as in clipping method \acs{c1}, and keep \ac{SPH} particles at all temperatures within the box, which is always below the sublimation temperature.

Note that such a steep increase in temperature with radius is not observed in the dust species of real star-forming regions. Hence, implementing the hydrodynamical temperature which is a result of the line cooling part (Eq.~\ref{C4:Eq:line}) of the \ac{EOS}, and therefore gas phase of the region, should happen with caution. We will address this in detail in Section~\ref{C4:Sec:methods_dust_temp}, where we present a better treatment of the dust temperature.
%%%%%%%%%%%%%%%%%%%%%%%%%%%%%%%%%%%%%%%%%%%%%%%%%%%%%%%
%%%%%%%%%%%%%%%%%%%%%%%%%%%%%%%%%%%%%%%%%%%%%%%%%%%%%%%
\section{Grid}
\label{C4:Sec:methods_sph2voro}
%%%%%%%%%%%%%%%%%%%%%%%%%%%%%%%%%%%%%%%%%%%%%%%%%%%%%%%
%%%%%%%%%%%%%%%%%%%%%%%%%%%%%%%%%%%%%%%%%%%%%%%%%%%%%%%
To create realistic synthetic observations we use \textsc{Hyperion}, a 3-d dust continuum Monte-Carlo radiative transfer code \citep{Robitaille:2011}.
%%%%%%%%%%%%%%%%%%%%%%%%%%%%%%%%%%%%%%%%%%%%%%%%%%%%%%%
\subsection{Different Grids}
\label{C4:Sec:methods_grid}
%%%%%%%%%%%%%%%%%%%%%%%%%%%%%%%%%%%%%%%%%%%%%%%%%%%%%%%
\textsc{Hyperion} and other Monte-Carlo radiative transfer codes use density and temperature structures that are discretized onto grids. The right choice of grid is very important in terms of recovering the dynamical range of the input structure and in terms of computational efficiency. \textsc{Hyperion} has a variety of grid types implemented. 

Regular Cartesian grids become inefficient when mapping a particle distribution onto a high-resolution grid to recover the dynamic range of a large-scale simulation box. Cartesian related grids, such as Octree grids or Cartesian \ac{AMR} grids, can improve the calculation by providing higher resolution in certain regions and can recover the dynamic range of the density structure in the simulation. But these grids are orientation dependent which can cause observational artifacts when observing along the axes. For example, "observations" along the Cartesian axes will make the edges of a cell visible.

The Voronoi tessellation \citep[for more details, see][]{Camps:2013,SpringelVORO:2010} is a type of irregular grid where each cell is described by an irregular polyhedron. Every polyhedron is set up by a point in 3-d space, which we refer to as a \emph{site} from now onwards. The boundaries of the polyhedra are defined by the set of points that are closest (hence, nearest neighbors see \citealt{NumericalRecipes}) to the site. The polyhedra have no preferred direction and resolution can be increased by adding more Voronoi sites, which enables us to represent large dynamic ranges. Therefore, mapping particle-based simulations onto a mesh using a Voronoi mesh\footnote{ The Voronoi tessellation has been recently implemented into \textsc{Hyperion} by Thomas Robitaille and Francesco Biscani using the \textsc{Voro++} library \citep{2009:Rycroft}.} is beneficial for our proposes. Note, Mapping \ac{SPH} simulations onto a Voronoi mesh has also been carried out by \cite{Hubber:2016}, \cite{Barcarolo:2014}, \cite{Starinshak:2014} and \cite{Zhou:2007}.

Since the primary goal of this paper is to produce reliable synthetic observations of star-forming regions we choose an approach which follows the following condition set by the radiative transfer calculation:
\begin{itemize}
\item satisfactory resolution around the forming stars\\[-0.5cm]
\item mapped simulation properties follow the gradient of simulation properties\\[-0.5cm]
\item number of cells manageable by the radiative transfer calculation
\end{itemize}
In order to fullfil the above requirements we can not refine the output of the \ac{SPH} simulation before assigning the Voronoi sites. Otherwise we would produce too many cells. Therefore we need to construct the Voronoi sites first (see Section~\ref{C4:Sec:methods_sites}) before calculating the resulting properties (see Section~\ref{C4:Sec:methods_parameter}) of the found sites. 
%%%%%%%%%%%%%%%%%%%%%%%%%%%%%%%%%%%%%%%%%%%%%%%%%%%%%%%
\subsection{Voronoi Sites}
\label{C4:Sec:methods_sites}
%%%%%%%%%%%%%%%%%%%%%%%%%%%%%%%%%%%%%%%%%%%%%%%%%%%%%%%
\begin{figure*}[t]
\includegraphics[width=1\textwidth]{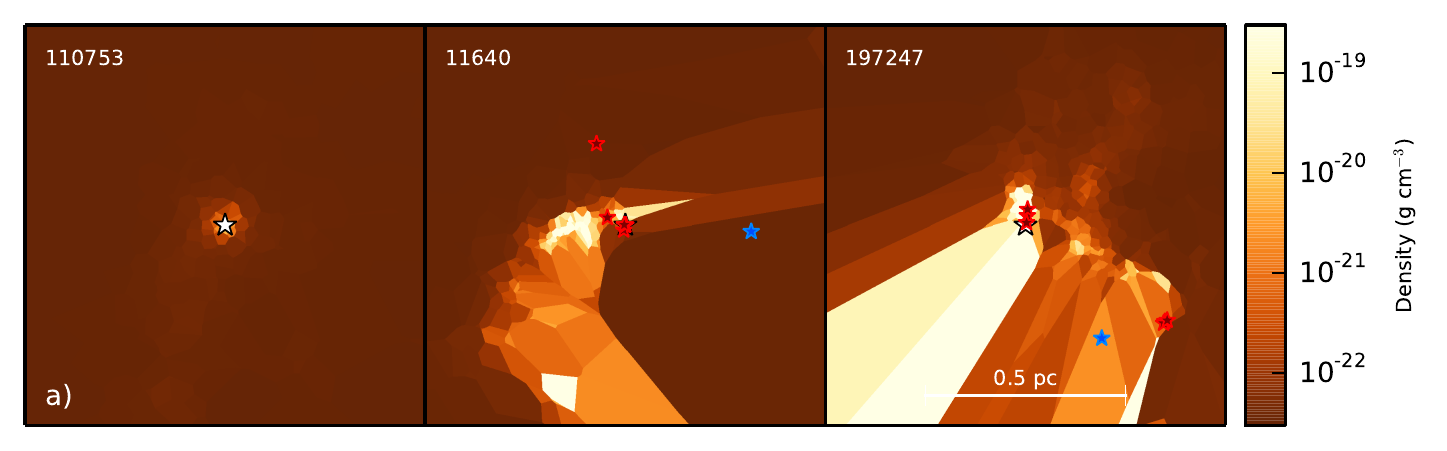}
\includegraphics[width=1\textwidth]{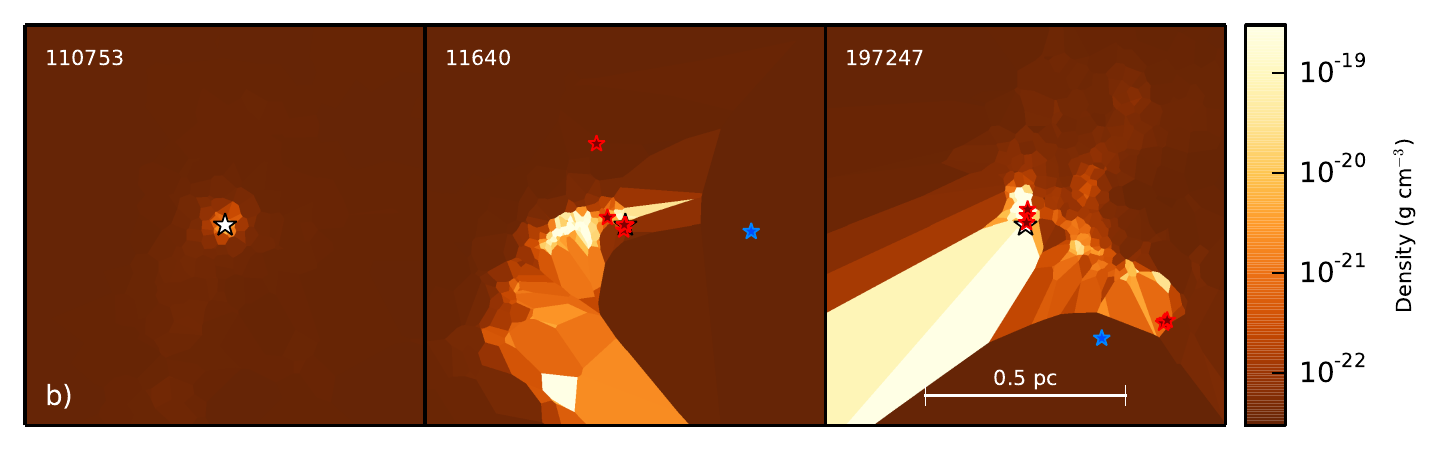}
\includegraphics[width=1\textwidth]{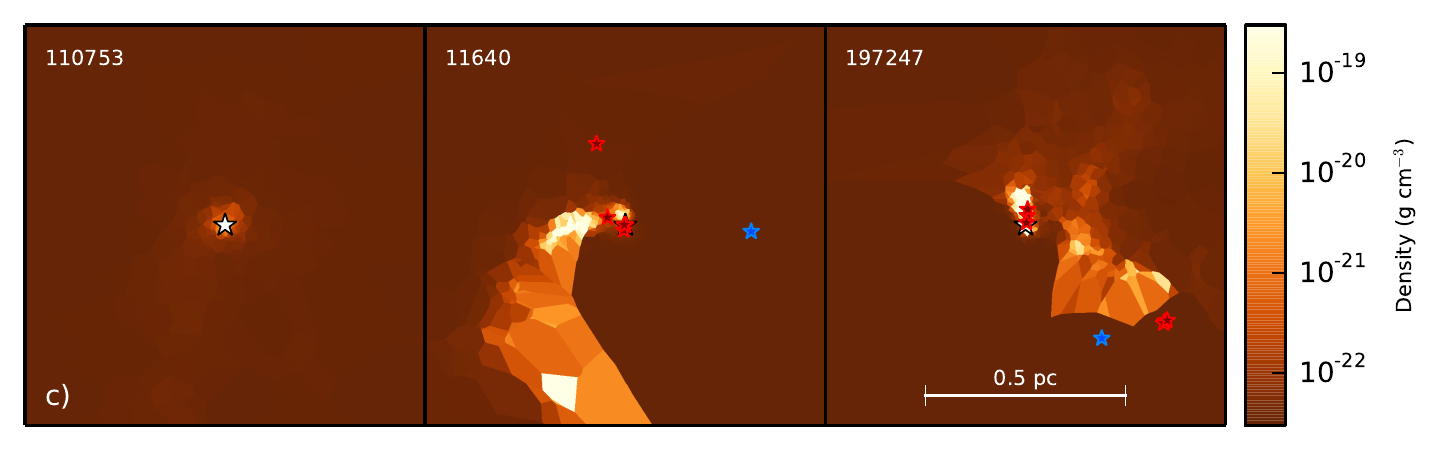}
\caption[2-d density slices of the Voronoi tessellation with different site methods]{\label{C4:Fig:Sites} 2-d density slices of the Voronoi tessellation with different site methods (\acs{s1}: top, \acs{s2}: middle, \acs{s3}: bottom); (white stars) sink particles with ID; (red stars) accreting sink particles and (blue stars) sink particles in the ionizing bubble within a box of \SI{1}{\pc} length. }
\end{figure*}
Choosing the Voronoi sites of the tessellation wisely is one of the critical elements when setting up a Voronoi mesh efficiently. While a rule of thumb is that regions with higher densities should have higher resolution and hence more sites, one has to keep in mind that very optically thick regions will trap photons and many such cells are not beneficial to the radiative transfer calculation \citep[for a discussion of this, see][]{Camps:2013}. In Figure~\ref{C4:Fig:Sites}, we plot density slices of different site distributions within a box of \SI{1}{\pc} and highlight accreting sink particles in red and sink particles in the ionizing bubble in blue. The actual evaluation of the density in the cells will be described in Section~\ref{C4:Sec:methods_parameter}. In this section, we only focus on the changes of the cell sizes when using three different methods of distributing the sites in the Voronoi mesh:
%%%%%%%%%%%%%%%%%%%%%%%%%%%%%%%%%%%%%%%%%%%%%%%%%%%%%%%
\subsubsection{Method s1 | Sites at SPH Particle Position}
\label{C4:Sec:method_s1}
%%%%%%%%%%%%%%%%%%%%%%%%%%%%%%%%%%%%%%%%%%%%%%%%%%%%%%%
A first order approach is, to put sites at the positions of the \ac{SPH} particles of the clipped sample \acs{c1} (for details about the sample selection, see Section~\ref{C4:Sec:clipping}). Hereafter, the site method \acs{s1} is defined by sites at \ac{SPH} positions alone. While this works rather well for isolated sink particles (see sink particle 110753 in Figure~\ref{C4:Fig:Sites}\textcolor{lmugreen}{(a)}), very large high-density Voronoi cells are created close to the ionizing bubble (see 11640 and 197247 in Figure~\ref{C4:Fig:Sites}\textcolor{lmugreen}{(a)}). These large high-density cells are an artifact of the \ac{SPH} to Voronoi mesh mapping and were not part of the initial simulations. Artificial high-density regions create overestimated fluxes in the radiative transfer because now very large high-density regions lie close to very luminous objects. 
%%%%%%%%%%%%%%%%%%%%%%%%%%%%%%%%%%%%%%%%%%%%%%%%%%%%%%%
\subsubsection{Method s2 | Sites at SPH \& Sink Particle Position}
\label{C4:Sec:method_s2}
%%%%%%%%%%%%%%%%%%%%%%%%%%%%%%%%%%%%%%%%%%%%%%%%%%%%%%%
\begin{figure*}[t]
\includegraphics[width=1\textwidth]{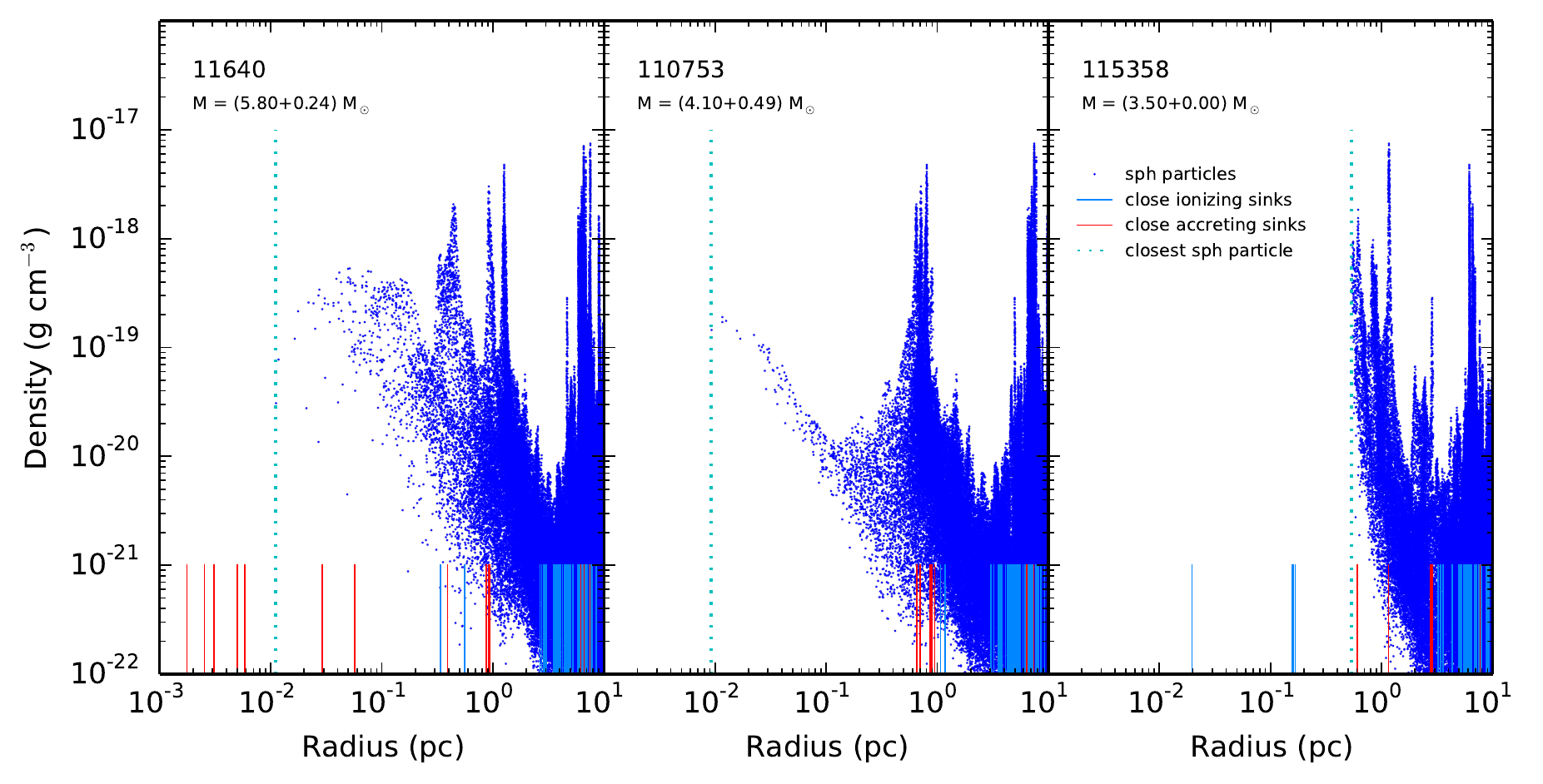}
\caption[Radial density profile centered on three sink particles]{\label{C4:Fig:SPHprofile} 
Radial density profile in time-step 122 (\SI{6.109}{\Myr}): 
    (a) centered on three sink particles: 11640 is an accreting sink particle in a small tight cluster of sink particles; 110753 is an isolated accreting sink particle; 115358 is sink particles in the ionizing bubble which lost all its circumstellar material due to other ionizing sink particles. 
    (b - online material) centered on all sink particles. 
Red lines highlight the distance of other accreting and blue of sink particles in the ionizing bubble. The dashed cyan line shows the radius of the closest \ac{SPH} particle to the sink particle.
}
\end{figure*}
In Figure~\ref{C4:Fig:SPHprofile}, we show the radial peak density distribution of \ac{SPH} particles close to sink particles for an accreting source in a small cluster (11640), an isolated accreting source (110753) and a source in the ionizing bubble (115358), which lost all its material due to a close-by ionizing source; all profiles of all sources are displayed in the online extension of Figure~\ref{C4:Fig:SPHprofile}. Red vertical lines highlight the distance of nearby accreting sink particles, while blue lines highlight nearby sink particles in the ionizing bubble. We can see that accreting sources can cause a radial trend in the distribution of the \ac{SPH} particles, which can be seen very clearly for sink particle 110753 directly in Figure~\ref{C4:Fig:SPHprofile} (middle), but also for the density peaks in the density profile at \SI{1}{\pc} distance of sink particle 11640 in Figure~\ref{C4:Fig:SPHprofile} (middle) and of sink particle 110753 in Figure~\ref{C4:Fig:SPHprofile} (left). 

Ionizing sources create (large) voids in the profile (for 115358 about \SI{0.6}{\pc}). Physically, this void represents ionized low-density bubbles around the ionizing stars. Since there is a low-density region around the ionizing sink particles, very few \ac{SPH} particles are located close to the other sink particles in the ionizing bubble. For example, the closest \ac{SPH} particle to sink particle 115358 is about \SI{0.6}{\pc} away. 

The Voronoi cells around the sink particles are created by distant \ac{SPH} particles beyond the bubble which have a higher density than the bubble. When ionizing sources are present in the particle-based simulation, artificial large density cells can be avoided by putting additional sites at the positions of the sink particles. Hereafter, we will refer to this method as site method \acs{s2}, which combines sites at the position of the \ac{SPH} particles and the sink particles. 

When adding sites at the position of the sink particles, we can reproduce the low-density bubbles around the ionizing sink particles. And hence, the size of high-density cells reaching inside the bubbles is truncated. In Figure~\ref{C4:Fig:Sites}\textcolor{lmugreen}{(b)}, we show this effect for sink particle 11640 and sink particle 197247, which have ionizing bubbles close-by. 

%%%%%%%%%%%%%%%%%%%%%%%%%%%%%%%%%%%%%%%%%%%%%%%%%%%%%%%
\subsubsection{Method s3 | Sites at SPH Particle \& Sink Particle Position \& Circumstellar Sites}
\label{C4:Sec:method_s3}
%%%%%%%%%%%%%%%%%%%%%%%%%%%%%%%%%%%%%%%%%%%%%%%%%%%%%%%
When inspecting Figure~\ref{C4:Fig:Sites}\textcolor{lmugreen}{(b)}, we see that there are still some large high-density cells relatively close to the accreting sink particles, which is partly due to lack of resolution of the \ac{SPH} simulation. Adding additional circumstellar sites around the accreting sink particles increases the resolution around the accreting sink particles and minimizes the artificial high-density cells as can be seen in Figure~\ref{C4:Fig:Sites}\textcolor{lmugreen}{(c)}. Hereafter, site method \acs{s3} combines sites at the position of \ac{SPH} particles, the sink particle position and the circumstellar sites around accreting sink particles.

On average, the distance between an accreting sink particle and the closest \ac{SPH} particle is \SI{0.01}{\pc} which is about \SI{2000}{\AU} as can be seen for the accreting sink particles 11640 and 110753 in Figure~\ref{C4:Fig:SPHprofile} (left, middle). Concretely, in the method \acs{s3} set-up, the position of the sites is constructed using a constant \ac{PDF} \citep[for more details, see Appendix~\ref{C4:Appendix_PDF} and][]{NumericalRecipes} in log-space of radius: 
\begin{eqnarray}
r &=& 10^{\xi\left[\log_{10}{r_{max}} - \log_{10}{r_{min}}\right]+ \log_{10}{r_{min}}},
\end{eqnarray}
with $r_{min}=\SI{e-4}{\pc}$ and $r_{max}=\SI{e-1}{\pc}$. The angles $\theta$ and $\phi$ are spaced isotropically within the sphere:
\begin{eqnarray}
    \label{C4:Eq:angles}
    \theta &=& \arccos{(2 \xi -1)}\\
    \label{C4:Eq:angles2}
    \phi&=& 2 \pi \xi.
\end{eqnarray}
Once we set the circumstellar sites of method \acs{s3} in this way, we can approximate the density gradient of the envelope more smoothly. This will be discussed in more detail in Section~\ref{C4:Sec:methods_envelope}.
%%%%%%%%%%%%%%%%%%%%%%%%%%%%%%%%%%%%%%%%%%%%%%%%%%%%%%%
\subsubsection{Results | Voronoi Sites}
\label{C4:Sec:results_sites}
%%%%%%%%%%%%%%%%%%%%%%%%%%%%%%%%%%%%%%%%%%%%%%%%%%%%%%%
In this section, we already explained in detail different methods of how to distribute the sites in a Voronoi mesh with a particle-based code as input. Site distribution method \acs{s3}, with sites at the neutral \ac{SPH} positions, at the sink particle positions and circumstellar sites around the accreting sink particles, can increase the resolution and remove artifacts from the particle to mesh transition. This is why we highly recommend distributing sites in this manner. We found that the number of circumstellar sites $N_{\textup{circ}}\approx \num{1000}$ is a good estimate to increase resolution around the sink particles but not create too many cells, which might slow down computation for the steps described in the next sections. In the remainder of this paper, we explicitly use the site distribution method \acs{s3}.
%%%%%%%%%%%%%%%%%%%%%%%%%%%%%%%%%%%%%%%%%%%%%%%%%%%%%%%
\newpage
\subsection{Density and Temperature Mapping onto new Sites}
\label{C4:Sec:methods_parameter}
%%%%%%%%%%%%%%%%%%%%%%%%%%%%%%%%%%%%%%%%%%%%%%%%%%%%%%%
In Figure~\ref{C4:Fig:Sites}, we showed slices of the density for different site distribution methods. To make the plot we used for the sites coinciding with \ac{SPH} particle positions the peak density $\rho_{i,\textup{SPH}}$ at that position of the \ac{SPH} particle. When adding sites to a Voronoi mesh, where no \ac{SPH} particles were located before, we need to evaluate the properties such as density or temperature from the \ac{SPH} distribution. We need to preserve the density and temperature structure as precisely as possible because later those property meshes will be passed to the radiative transfer set-up. In this section, we will investigate three techniques to map the density distribution onto a Voronoi mesh:
\begin{figure*}[t]
\includegraphics[width=1\textwidth]{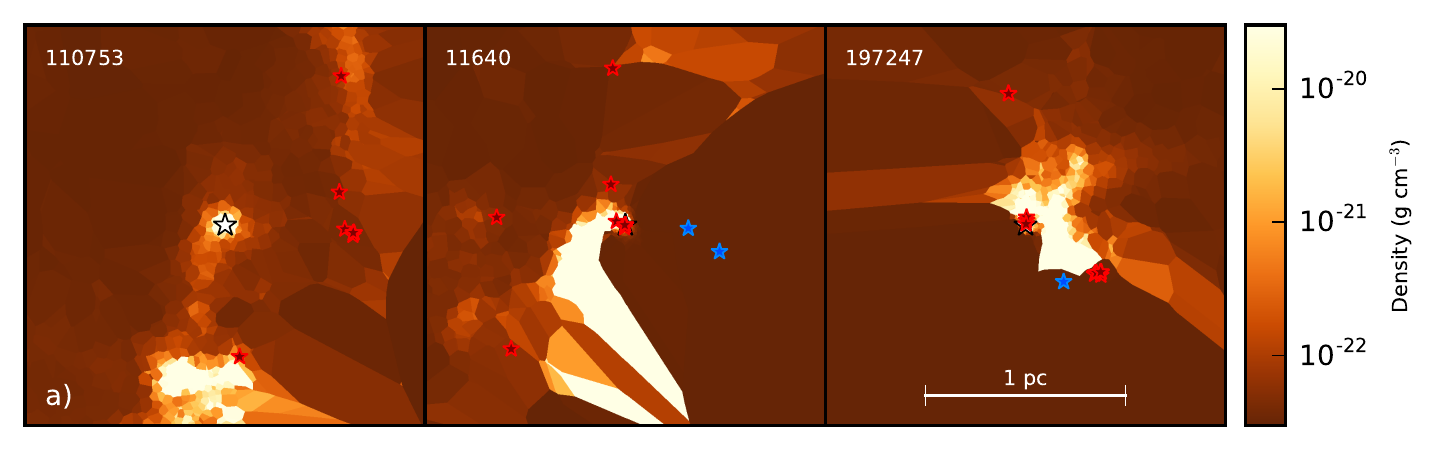}
\includegraphics[width=1\textwidth]{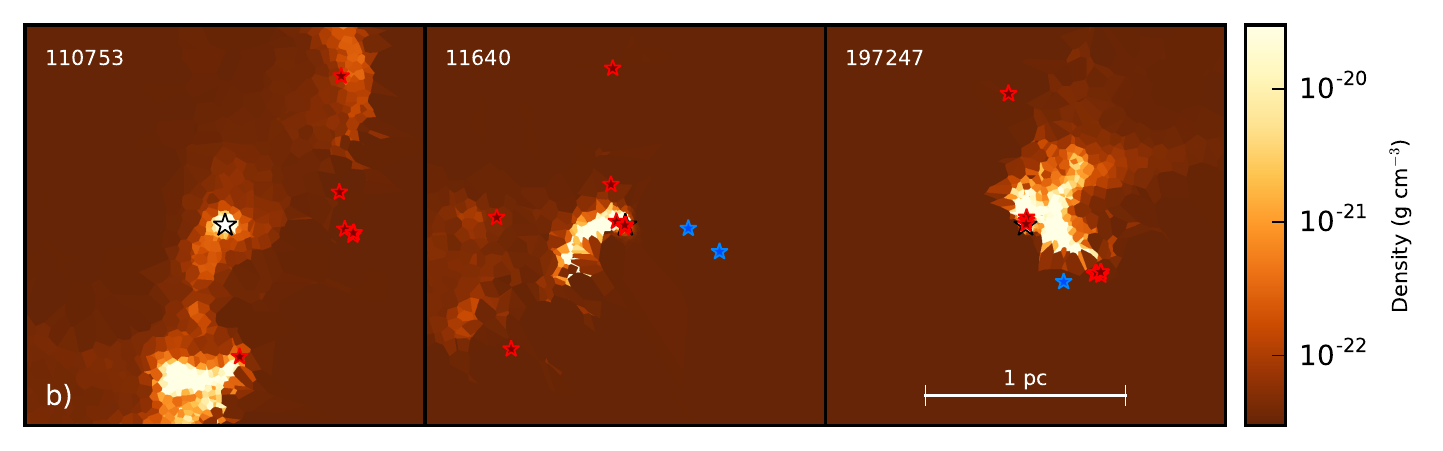}
\includegraphics[width=1\textwidth]{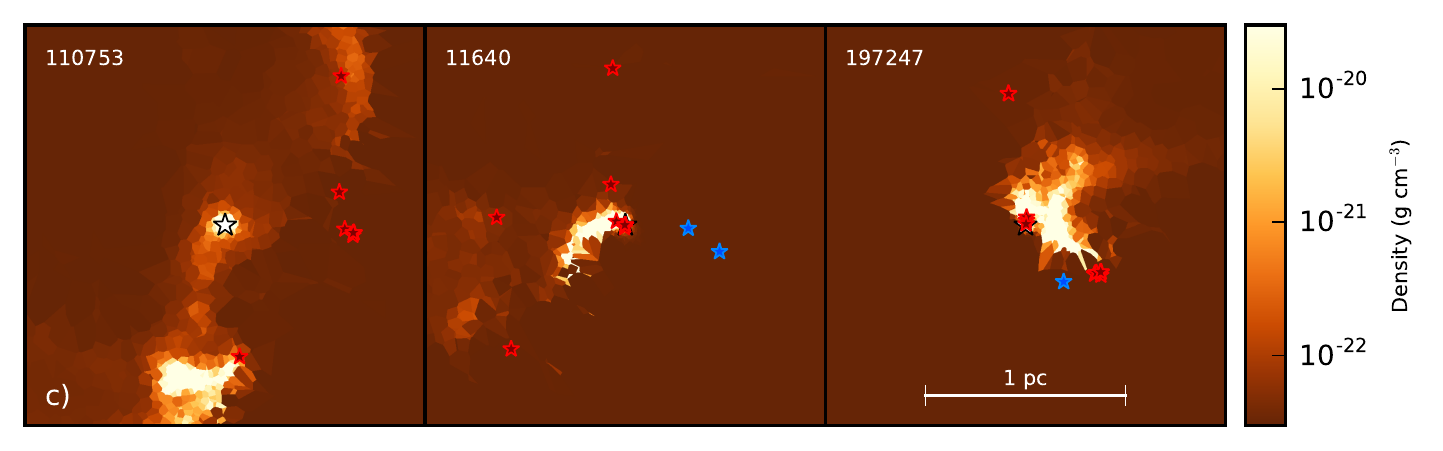}
\caption[Different methods of density translation onto the Voronoi grid]{\label{C4:Fig:Parameters} Different methods of density translation onto the Voronoi grid (\acs{p1}: top, \acs{p2}: middle, \acs{p3}: bottom); (white stars) sink particles with ID; (red stars) accreting sink particles and (blue stars) sink particles in the ionizing bubble within a box of \SI{2}{\pc} length.}
\end{figure*}
%%%%%%%%%%%%%%%%%%%%%%%%%%%%%%%%%%%%%%%%%%%%%%%%%%%%%%%
\subsubsection{Method p1 | Evaluation of SPH Kernel Function}
\label{C4:Sec:method_p1}
%%%%%%%%%%%%%%%%%%%%%%%%%%%%%%%%%%%%%%%%%%%%%%%%%%%%%%%
To evaluate the density at a new site $j$ at the position of the sink particle or at a circumstellar site around a sink particle, we compute the \ac{SPH} density function \citep[for a detailed review, see][]{SpringelSPH:2010} for the $N$ closest \ac{SPH} particles:
\begin{eqnarray}
    \label{C4:Eq:kernel_density}
    \rho_{j,\textup{p1}}&=&\sum\limits_{i=1}^{N} m_{i,\textup{SPH}}W_{ij}(r_{ij},h_i).
\end{eqnarray}
Where $m_{i,\textup{SPH}}$ is the constant \ac{SPH} particle mass with the value $m_{i,\textup{SPH}}=\SI{e-2}{\Msun}$ in \emph{run I} of the \acs{D14} \ac{SPH} simulations and the kernel function $W_{ij}(r_{ij},h_i)$ from Eq.~\ref{C4:Eq:kernel}.

We estimate the temperature at a new site $j$ by weighting \citep[][]{NumericalRecipes} the \ac{SPH} particle temperature $T_{i,\textup{SPH}}$ of the closest $N$ particles with distance\footnote{ The temperature can also be evaluated by weighting with the density function. We will explore this in the next section.} $r_{ij}$
\begin{eqnarray}
    T_{j,\textup{p1}}&=&\frac{\sum\limits_{i=1}^{N} T_{i,\textup{SPH}}\,m_{i,\textup{SPH}}W_{ij}(r_{ij},h_i)}{\rho_{j,\textup{p1}}}
\end{eqnarray}
with the help of Eq.~\ref{C4:Eq:kernel_density}. We explored the effect of setting $N$ to different values and found that setting $N>N_{\textup{neighbor}}$ produces a good approximation\footnote{In \textit{run I} $N_{\textup{neighbor}}=65$. We found that with $N=\num{100}$, which means that $N>N_{\textup{neighbor}}$, we can recover \SI{99.9}{\percent} of the exact solution.} of the exact solution of the density. Hereafter, we call this property evaluation method \acs{p1}. 

In Figure~\ref{C4:Fig:Parameters}\textcolor{lmugreen}{(a)}, we plot the density evaluation with property method \acs{p1} and site method \acs{s3} (note the different scale than Figure~\ref{C4:Fig:Sites}\textcolor{lmugreen}{(c)}). We can see that even with the site distribution method \acs{s3}, we still have artifacts of large high-density cells as already mentioned in Section~\ref{C4:Sec:methods_sites}. These large high-density cells are due to the density overestimation when using the peak density $\rho_{i,\textup{SPH}}$ at the \ac{SPH} particles which lie in large cells. 

When mapping the density with method \acs{p1}, we recover a total mass of \SI{139945}{\Msun} which is several orders of magnitude to high then the actual mass of about \SI{7500}{\Msun}. Therefore, we explore other techniques.
%%%%%%%%%%%%%%%%%%%%%%%%%%%%%%%%%%%%%%%%%%%%%%%%%%%%%%%
\subsubsection{Method p2 | Evaluation of Splitted SPH Particles}
\label{C4:Sec:method_p2}
%%%%%%%%%%%%%%%%%%%%%%%%%%%%%%%%%%%%%%%%%%%%%%%%%%%%%%%
In the community of \ac{SPH} modelers, the density is sometimes approximated in Voronoi cells using an average density $\bar{\rho}=m_{\textup{SPH}}/V_{\textup{cell}}$ (David Hubber, private communication). We do not use this technique, since it can only be used to evaluate the density in cells containing \ac{SPH} particles. When adding new sites, this approximation becomes untrue because the volume $V_{\textup{cell}}$ shrinks, while $m_{\textup{SPH}}$ remains constant by definition. Also, the density evaluation of cells which contain the new sites is not covered by that technique. 

Thus, we virtually split the mass of an \ac{SPH} particle $i$ according to the 3-d Gaussian-shaped \ac{SPH} kernel function (see Eq.~\ref{C4:Eq:kernel}). We repeat this for all \ac{SPH} particles. Through simple counting, we reevaluate the mass in the cells of the sites $j$ of the existing \ac{SPH} particle sites plus the newly added sites of method \acs{s3}. Hereafter, we call this property evaluation method \acs{p2}. 

To space the split \ac{SPH} particle $k$ with respect to the \ac{SPH} particle $i$ again, we use a linear spaced \ac{PDF} (as described in detail in Appendix~\ref{C4:Appendix_PDF}) and solve for the positions of the split particles $r_{ik}$. We distribute the split \ac{SPH} particles isotropically (see Eq.~\ref{C4:Eq:angles} and Eq.~\ref{C4:Eq:angles2}). We convert all the $N^{\textup{p2}}_{\textup{split}}$ split \ac{SPH} particles to Cartesian coordinates and scale the distance by the respective smoothing length $h_i$ of the \ac{SPH} particle which got split, since we set $h_i=\num{1}$ for the calculation above. The split \ac{SPH} particles do not necessarily fall in the cell of the \ac{SPH} particles they split from. By searching for the nearest neighbor sites \citep{NumericalRecipes} to each split \ac{SPH} particle, we can now reassess the mass of the Voronoi cells, the split \ac{SPH} particles fall into. We then calculate the new density $\rho_{j,\textup{p2}}$ at a cell site $j$ as
\begin{eqnarray}
    \label{C4:Eq:density_p2}
    \rho_{j,\textup{p2}}&=&\sum\limits_{k=1}^{n_j} m_{\textup{SPH},k} / V_j,
\end{eqnarray}
with the split \ac{SPH} particle mass $m_{\textup{SPH},k}=m_{\textup{SPH},i}/N^{\textup{p2}}_{\textup{split}}$ and the number $n_j$ set to the number of split \ac{SPH} particles found in a cell with the volume $V_j$.  We found that splitting by $N^{\textup{p2}}_{\textup{split}}=\num{100}$ is a good trade-off between finding a smooth density distribution and computational efficiency. 

Since the probability of accumulating many split \ac{SPH} particles decreases with decreasing cell size, we introduce a threshold of $N^{\textup{p2}}_{\textup{threshold}}=\num{10}$ split \ac{SPH} particles. For cells which collect fewer than $N^{\textup{p2}}_{\textup{threshold}}$, we evaluate $\rho_{j,\textup{p1}}$ from Eq.~\ref{C4:Eq:kernel_density}. 

In Figure~\ref{C4:Fig:Parameters}\textcolor{lmugreen}{(b)}, we plot the density evaluated with property method \acs{p2}. When comparing to Figure~\ref{C4:Fig:Parameters}\textcolor{lmugreen}{(a)}, one can see that the transition between cells is smoother and artificial high-density Voronoi cells are substantially fewer. 

To evaluate the temperature, we simply keep track from which original \ac{SPH} particle of mass $m_{\textup{SPH},i}$ a split \ac{SPH} particles of mass $m_{\textup{SPH},k}$ comes from and what temperature $T_i$ it had, we can weight the temperature accordingly by the density:
\begin{eqnarray}
    \label{C4:Eq:temp_p2}
    T_{j,\textup{p2}}&=&\sum\limits_{k=1}^{n_j} T_{i}(m_{\textup{SPH},k}) / n_j.
\end{eqnarray}
For the cells with $n_j<N^{\textup{p2}}_{\textup{threshold}}$, we calculate the temperatures as in method \acs{p1}.

When mapping the density with method \acs{p2}, we recover a total mass of \SI{7530}{\Msun} relatively close to the real mass of \SI{7500}{\Msun}.

%%%%%%%%%%%%%%%%%%%%%%%%%%%%%%%%%%%%%%%%%%%%%%%%%%%%%%%
\newpage
\subsubsection{Method p3 | Evaluation Using Random Sampling}
\label{C4:Sec:method_p3}
%%%%%%%%%%%%%%%%%%%%%%%%%%%%%%%%%%%%%%%%%%%%%%%%%%%%%%%
Another possibility is to evaluate the density by sampling $N_{\textup{random}}$ random positions $l$ in the grid and calculating the density from the kernel distribution (as in method \acs{p1}) of the $N$ closest \ac{SPH} particles to the positions $l$. 

By averaging\footnote{ This method was developed by Francesco Biscani and Thomas Robitaille.} the density of the random points $l$ at $n_j$ positions in cell $j$, we get the total density $\rho_{j,\textup{p3}}$ of method \acs{p3}
\begin{eqnarray}
    \label{C4:Eq:rho_v3}
    \rho_{j,\textup{p3}}&=&\sum\limits_{l=1}^{n_j} \rho_{l,\textup{p1}}/n_j.
\end{eqnarray}
Typically, we set $N_{\textup{random}}\approx\num{2e7}$ for $N_{\textup{SPH},i}\approx\num{7e5}$ neutral \ac{SPH} particles in the simulation time-step and $N^{\textup{p3}}_{\textup{split}}=\num{30}$ as above. To ensure that every cell $j$ has at least $N^{\textup{p3}}_{\textup{threshold}}=\num{20}$, we add points to under-filled cells until $n_j=N^{\textup{p3}}_{\textup{threshold}}$. 

Figure~\ref{C4:Fig:Parameters}\textcolor{lmugreen}{(c)} shows the property evaluation method \acs{p3}. Density gradients are smoother and no artificial high-density cells remain.

The temperature of property method \acs{p3} is a weighted version of Eq.~\ref{C4:Eq:rho_v3} using Eq.~\ref{C4:Eq:kernel_density}
\begin{eqnarray}
    \label{C4:Eq:temp_v3}
    T_{j,\textup{p3}}=&\frac{1}{n_j}\sum\limits_{l=1}^{n_j}
    \frac{\sum\limits_{i=1}^{N}T_{i,\textup{SPH}} m_{i,\textup{SPH}}W_{il}(r_{il},h_i)}
    {\rho_{l,\textup{p1}}},
\end{eqnarray}
again $N>N_{\textup{neighbor}}$ as in method \acs{p1}.

When mapping the density with method \acs{p1}, we recover a total mass of \SI{7495}{\Msun} extremely close to the actual mass of \SI{7500}{\Msun}.

%%%%%%%%%%%%%%%%%%%%%%%%%%%%%%%%%%%%%%%%%%%%%%%%%%%%%%%
\subsubsection{Results | Mapping}
\label{C4:Sec:results_parameter}
%%%%%%%%%%%%%%%%%%%%%%%%%%%%%%%%%%%%%%%%%%%%%%%%%%%%%%%
In Figure~\ref{C4:Fig:Parameters}, we show different methods of mapping the \ac{SPH} density and temperature onto the Voronoi mesh. 

We found that kernel evaluation method \acs{p1} is efficient and reliable in small cells. The solving of the kernel function using the $N>N_{\textup{neighbor}}$ nearest neighbors is sufficient to approximate the properties at the new position. But method \acs{p1} creates artifacts for large cells. 

The splitting kernel method \acs{p2} is less computationally efficient but produces a smooth density or temperature gradient between large cells. For small cells this method breaks down, since the probability that a small cell gets "hit" by split particles is small. Therefore, the introduced threshold $N^{\textup{p2}}_{\textup{threshold}}=\num{10}$ is important. For such small cells, we use the method \acs{p1}. We found that a virtual splitting the \ac{SPH} particle by $N^{\textup{p2}}_{\textup{split}}=\num{100}$ produces a solid result while being tolerably efficient in time. 

Method \acs{p3} removes the statistical sampling issues of the small cells (compared to method \acs{p2}) by adding random sampling points within the cell as long as $N^{\textup{p3}}_{\textup{threshold}}$ is reached. We found that a smooth gradient in temperature or density can be reached when setting the number of points to sample to $N_{\textup{random}}\approx\num{2e7}$ and the threshold to $N^{\textup{p3}}_{\textup{threshold}}=\num{20}$. The trade-off is that this method is less efficient computationally when comparing with method \acs{p2}, but it is fast enough for the purposes of our work.

Below, we list the total mass within the simulation box of time-step 122 (\SI{6.109}{\Myr}), recovered from the density mapping for every method present in this section:
\begin{eqnarray}
    \label{C4:Eq:mass_px}
    m^{tot}_{\textup{p1}}&\approx&\SI{139945}{\Msun}\\
    m^{tot}_{\textup{p2}}&\approx&\SI{7530}{\Msun}\\
    m^{tot}_{\textup{p3}}&\approx&\SI{7495}{\Msun}    
\end{eqnarray}
The actual mass within the box of the \ac{SPH} simulation $m^{tot}_{\textup{SPH}}=\SI{7500}{\Msun}$ (c.\,f.~Table~\ref{C4:Tab:runI}). In respect of mass conservation, we note that method \acs{p1} is completely unreliable. The small overestimate of method \acs{p2} results from the threshold mechanism\footnote{ The threshold for the methods \acs{p2} and \acs{p3} mean different things. A high threshold in \acs{p2} improves the gradient but also overestimates the mass, while in the \acs{p3}, a higher threshold improves the mass conservation.}. In method \acs{p2}, when there are less than $N^{\textup{p2}}_{\textup{threshold}}$ split \ac{SPH} particles in the cell, only the density at the position of the site is estimated. This leads to an over-prediction of the cell's density. For this reason, method \acs{p1} and method \acs{p2} are faulty by definition, since they introduce too high densities to certain cells. Therefore, we suggest method \acs{p3} when mapping \ac{SPH} distributions onto a Voronoi tessellation, since this overestimation is reduced (error below \SI{1}{\percent}) by sampling more positions in small cells. In Figure~\ref{C4:Fig:SPH_env_profile} we show how well method \acs{p3} works in recovering a density gradient from the initial simulation. Blue points represent the simulated peak density of the \ac{SPH} particles and the black points represent the \acs{p3} circumstellar sites. Figure~\ref{C4:Fig:SPH_env_profile} shows that the technique succeeds to recover the density structure over 2 orders of magnitude over less than \SI{0.1}{\pc}.
%%%%%%%%%%%%%%%%%%%%%%%%%%%%%%%%%%%%%%%%%%%%%%%%%%%%%%%
%%%%%%%%%%%%%%%%%%%%%%%%%%%%%%%%%%%%%%%%%%%%%%%%%%%%%%%
\section{Sources}
\label{C4:Sec:methods_RT}
%%%%%%%%%%%%%%%%%%%%%%%%%%%%%%%%%%%%%%%%%%%%%%%%%%%%%%%
%%%%%%%%%%%%%%%%%%%%%%%%%%%%%%%%%%%%%%%%%%%%%%%%%%%%%%%
In \textsc{Hyperion}, we can set up spherical sources (e.\,g.~stars) which emit photon packets. The parameters for spherical sources are the luminosity $L_*$, radius $R_*$ and spectra $F_{\nu*}$. 
\begin{figure}[t]
\includegraphics[width=0.5\textwidth]{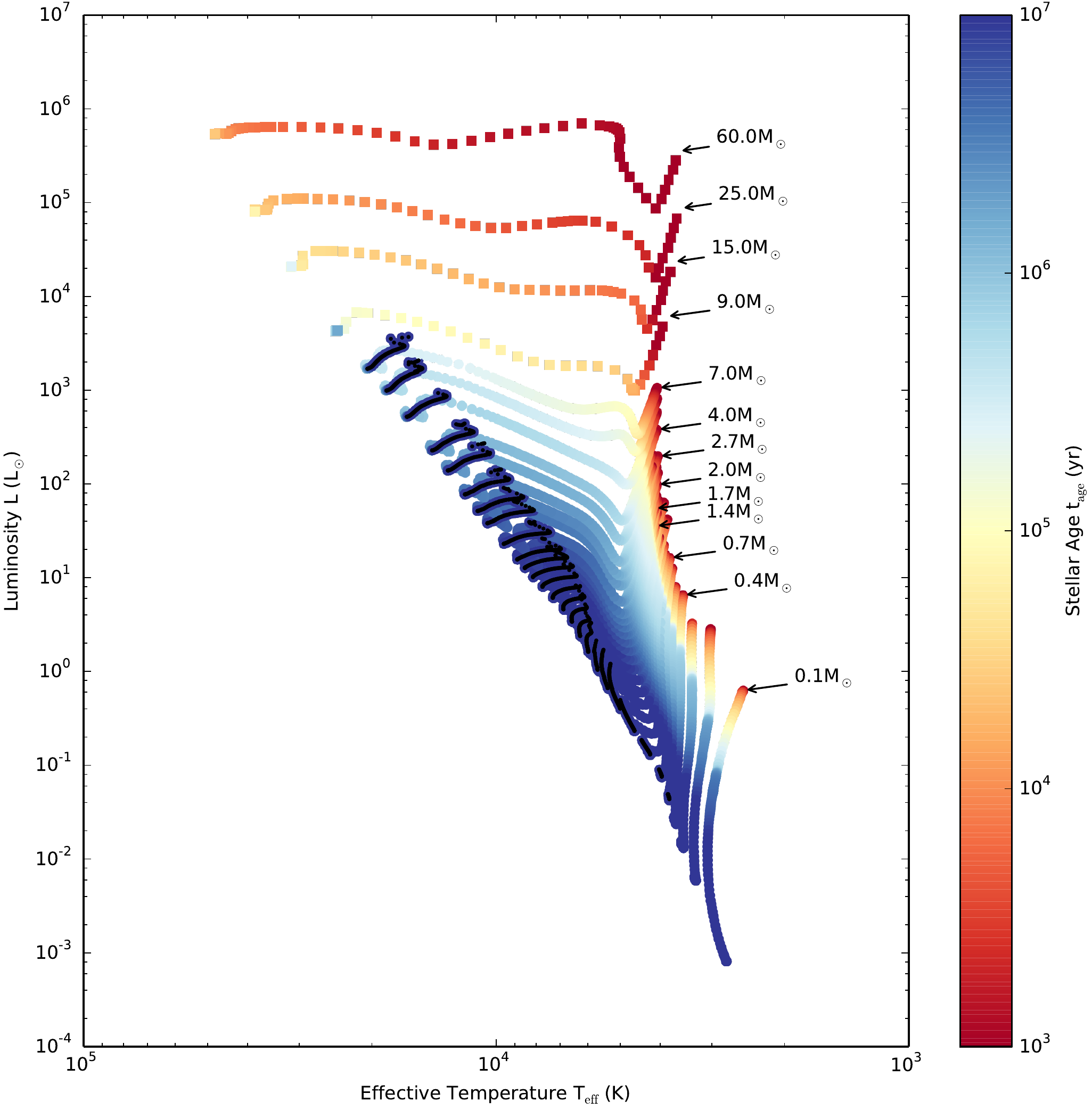}
\caption[Evolutionary pre-main-sequence tracks in luminosity and temperature]{\label{C4:Fig:PMS_tracks} Luminosity versus effective temperature diagram of evolutionary pre-main-sequence tracks. Colors highlight stellar ages of the models. Arrows highlight the beginning of every mass track. Black points highlight stellar evolutionary phases, which have left the pre-main-sequence phase already.}
\end{figure}
\begin{figure*}[t]
\includegraphics[width=1\textwidth]{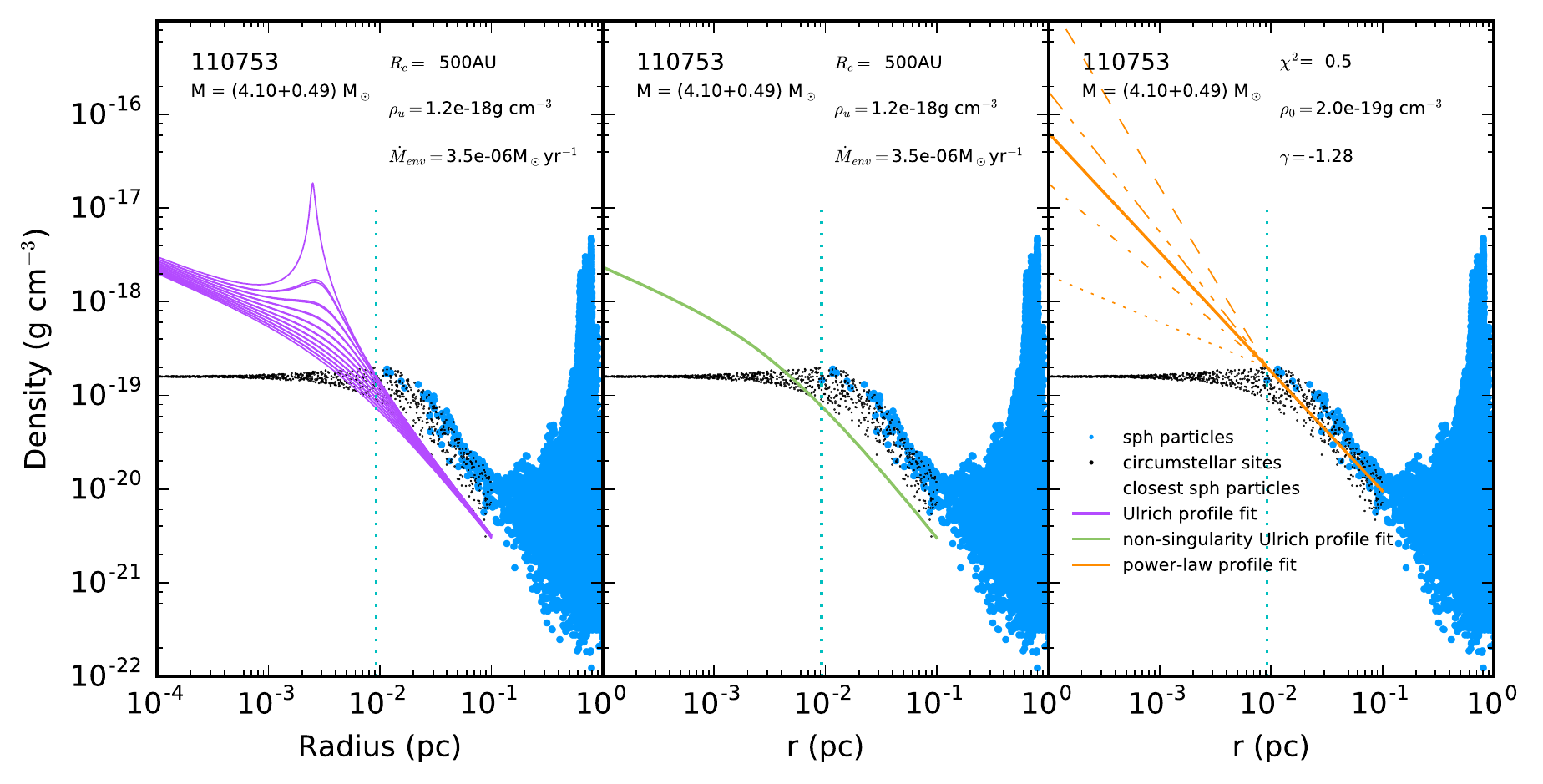}
\caption[Radial density profile with different envelope refinement fits]{\label{C4:Fig:SPH_env_profile} Radial density profile of \ac{SPH} particles (blue) towards the sink particle 110753. The circumstellar sites of the accreting sink particle are over-plotted in black. (Left) rotationally flattened envelope fit (purple) at different angles, (middle) rotationally flattened envelope profile (green) with suppressed singularity at the mid-plane, (right) different power-law envelope profiles (dotted, dashed) and the fitted power-law profile (solid).}
\end{figure*}
%
%%%%%%%%%%%%%%%%%%%%%%%%%%%%%%%%%%%%%%%%%%%%%%%%%%%%%%%
%\newpage
\subsection{Pre-main-sequence Stellar Evolutionary Models}
\label{C4:Sec:methods_PMS}
%%%%%%%%%%%%%%%%%%%%%%%%%%%%%%%%%%%%%%%%%%%%%%%%%%%%%%%
Assuming we want to set up a source resembling a forming star in \acs{D14} \ac{SPH} simulations with mass $M_*$ and age $t_{\textup{age}}$ (see Eq.~\ref{C4:Eq:t_age}), we can calculate the stellar radius $R_*$ and the stellar effective temperature $T_{\textup{eff}}$ by interpolating pre-main-sequence stellar evolutionary tracks. In Figure~\ref{C4:Fig:PMS_tracks}, we plot the pre-main-sequence tracks of \cite{Siess:2000} for stellar masses between $\SI{0.1}{\Msun}\leq M_* \leq \SI{7.0}{\Msun}$ and \cite{Bernasconi:1996} for stellar masses between $\SI{9}{\Msun}\leq M_* \leq \SI{60}{\Msun}$. 

We interpolate $R_*$, $T_{\textup{eff}}$ and $L_*$ from the stellar evolutionary tracks using 3-d natural neighbor interpolation \citep[for more details, see][]{Sibson:1981} of the discrete samples $M_*$ and $t_{\textup{age}}$. In the rare cases, when the stellar masses lie beyond the stellar evolutionary tracks of \cite{Bernasconi:1996}, we interpolate the main-sequence data from Appendix~E of \cite{Carroll:1996} to infer the parameters. 
%%%%%%%%%%%%%%%%%%%%%%%%%%%%%%%%%%%%%%%%%%%%%%%%%%%%%%%
\subsection{Spectra}
\label{C4:Sec:methods_RT_spectra}
%%%%%%%%%%%%%%%%%%%%%%%%%%%%%%%%%%%%%%%%%%%%%%%%%%%%%%%
We use the stellar photosphere models from \cite{Castelli:2004} and \citet[Phoenix models]{Brott:2005} to interpolate the spectra $F_{\nu*}$ based on the parameters $M_*$, $R_*$ and $T_{\textup{eff}}$. The \cite{Castelli:2004} models are defined for stars with temperatures reaching from $\SI{3500}{\kelvin}\leq T_{\textup{eff}} \leq \SI{50000}{\kelvin}$, for logarithmic surface gravities
\begin{eqnarray}
\log_{10}g_* &=& \log_{10}\left(\frac{GM_*}{R_*^2}\right).
\end{eqnarray}
between $\log_{10}g_*\in[\num{0},\num{5}]$ and for solar metallicities. In the rare cases when $T_{\textup{eff}} < \SI{3500}{\kelvin}$, we use the Phoenix models. For a detailled description see Appendix~\ref{C4:Appendix_Spectra}.
%
%%%%%%%%%%%%%%%%%%%%%%%%%%%%%%%%%%%%%%%%%%%%%%%%%%%%%%%
%%%%%%%%%%%%%%%%%%%%%%%%%%%%%%%%%%%%%%%%%%%%%%%%%%%%%%%
\section{Circumstellar Material}
\label{C4:Sec:methods_YSO}
%%%%%%%%%%%%%%%%%%%%%%%%%%%%%%%%%%%%%%%%%%%%%%%%%%%%%%%
%%%%%%%%%%%%%%%%%%%%%%%%%%%%%%%%%%%%%%%%%%%%%%%%%%%%%%%
Now that we have mapped the density and temperature directly onto the Voronoi mesh (see Section~\ref{C4:Sec:methods_sph2voro}) and that we have defined the stellar properties (see Section~\ref{C4:Sec:methods_RT}), it is possible to explore the regions around the accreting objects in more detail. Note that the circumstellar material is the most important place of the radiative transfer calculation, unfortunately unresolved below \SI{1000}{\AU} by the \acs{D14} \ac{SPH} simulations. However, in the following, we describe several methods of how to overcome this and make sure that the density structure is as realistic as possible close to the sources.
%%%%%%%%%%%%%%%%%%%%%%%%%%%%%%%%%%%%%%%%%%%%%%%%%%%%%%%
%\newpage
\subsection{Extrapolating the Envelope Density}
\label{C4:Sec:methods_envelope}
%%%%%%%%%%%%%%%%%%%%%%%%%%%%%%%%%%%%%%%%%%%%%%%%%%%%%%%
As mentioned previously, the \acs{D14} \ac{SPH} simulations are large-scale simulations and do not have as high resolution around the sink particles as other simulations of smaller scale. Even though the \acs{D14} \ac{SPH} simulations lack the resolution to see the inner \SI{0.01}{\pc} close to the sink particles, we can already see a radial trend in the density close to accreting sink particles, which corresponds to the outer regions of an envelope resolved by the \ac{SPH} simulation (c.\,f.~with the accreting sink particles 11640 and 110753 in Figure~\ref{C4:Fig:SPHprofile}); in Figure~\ref{C4:Fig:SPH_env_profile}, we also replotted the sink particle 110753 to illustrate the following envelope refinement.

The blue colored points in Figure~\ref{C4:Fig:SPH_env_profile} represent the \ac{SPH} particles beyond the accretion radius $r_{\textup{acc}}=\SI{0.005}{\pc}$. We can see an accretion stream of \ac{SPH} particles towards the sink particle 110753. But there is a gap of about $\SI{e-2}{\pc}$ (cyan dashed) between the sink particle and the closest
\ac{SPH} particle. The envelope background density between the stream (outer envelope) and the sink particle is about \SI{2e-19}{\gccm} for object 110753. In Section~\ref{C4:Sec:methods_sites}, we improved this situation by putting more circumstellar particles around the sink particle and evaluated their density in Section~\ref{C4:Sec:methods_parameter} (see black points in Figure~\ref{C4:Fig:SPH_env_profile}). The evaluated refined circumstellar density is somewhat only a background density of the outer envelope in the simulation and does not represent a continuous envelope with an increasing density profile at smaller scales. In Figure~\ref{C4:Fig:SPH_env_profile}, we can see the leveling off of the density profile for the new sites at \SI{2e-19}{\gramms\per\centi\meter\cubed} within the closest \ac{SPH} particle of the accreting sink particles (black). In what follows, we extrapolate the envelope profile inwards with different profile descriptions. For the extrapolation, we will use the \acs{s3} circumstellar sites rather than the \ac{SPH} particles close to the sink particles because we have better number statistics with the sites. We now describe three different methods for extrapolating the density structure to the smallest
scales:
%%%%%%%%%%%%%%%%%%%%%%%%%%%%%%%%%%%%%%%%%%%%%%%%%%%%%%%
\subsubsection{Method e1 | Ulrich Envelope Profiles}
\label{C4:Sec:method_e1}
%%%%%%%%%%%%%%%%%%%%%%%%%%%%%%%%%%%%%%%%%%%%%%%%%%%%%%%
The Ulrich envelope profile \citep{Ulrich:1976} is a rotational flattened profile and follows the physical process of accretion. The profile $\rho_{\textup{Ulrich}}(r,\theta)$ is described in spherical coordinates and is dependent on the infall rate of material $\dot{M}_{\textup{env}}$, the mass of the stellar object $M_{*}$ and the centrifugal radius $R_c$:
\begin{eqnarray}
    \label{C4:Eq:rho_Ulrich}
    \rho_{\textup{Ulrich}}(r,\theta)&=&\rho_u^{\textup{env}} \left(\frac{r}{R_c}\right)^{-3/2} \left(1+\frac{\cos\theta}{\cos\theta_0}\right)^{-1/2}\times\nonumber\\ &&\left(\frac{\cos\theta}{\cos\theta_0} + \frac{2R_c\cos^2\theta_0}{r}\right)^{-1}\\
    \rho_u^{\textup{env}}&=&\frac{\dot{M}_{\textup{env}}}{4 \pi (GM_*R_c^3)^{-1/2}}.
\end{eqnarray}
The angle $\theta_0$ is given by the streamline equation:
\begin{eqnarray}
    \label{C4:Eq:streamline}
0&=&\cos^3\theta_0 + \cos\theta_0\left(\frac{r}{R_c}-1\right)-\cos\theta\left(\frac{r}{R_c}\right)
\end{eqnarray}
Hereafter, we call the Ulrich envelope profile\footnote{ The Ulrich profiles are part of models in  \textsc{Hyperion}, which sets up \ac{YSO} models from analytical density descriptions (see \url{http://www.hyperion-rt.org}). We will refer to these models in the remainder of this paper as Analytical \ac{YSO} Models.} method \acs{e1}. We calculate the infall rate $\dot{M}_{\textup{env}}$ from the accretion rate (see Eq. \ref{C4:Eq:infall}). As can be seen in Figure~\ref{C4:Fig:SPH_env_profile}, we found that an Ulrich profile (left panel, purple lines) derived from the parameters of sink particle 110753 seamlessly matches the accretion stream inwards but also underestimates it at larger radii. Note that the singularity at the centrifugal radius $R_c$ (when $\cos\theta=\cos^3\theta_0$; see Eq.~\ref{C4:Eq:streamline}) at the mid-plane (for $\theta=\SI{90}{\degree}$; see Eq.~\ref{C4:Eq:rho_Ulrich}) is caused by the pile-up of material as a result of angular momentum conservation. In reality, the material would be distributed radially to form a circumstellar disk.

%%%%%%%%%%%%%%%%%%%%%%%%%%%%%%%%%%%%%%%%%%%%%%%%%%%%%%%
\subsubsection{Method e2 | Ulrich Envelope Profiles without Singularity}
\label{C4:Sec:method_e2}
%%%%%%%%%%%%%%%%%%%%%%%%%%%%%%%%%%%%%%%%%%%%%%%%%%%%%%%
The Ulrich description of the circumstellar disk (i.\,e.~the singularity) is not physically realistic. Therefore, for the density profile we ignore the singularity at the mid-plane and later add a more advanced description of the circumstellar disk in Section~\ref{C4:Sec:methods_disk}. In this method, hereafter named \acs{e2}, we suppress\footnote{ Developed by Thomas Robitaille.} the Ulrich singularity by adjusting the Ulrich density profile of Eq.~\ref{C4:Eq:rho_Ulrich}:
\begin{eqnarray}
    \label{C4:Eq:rho_Ulrich_no_sing}
    \rho_{\textup{Ulrich}}^{\textup{no\ singularity}}(r,\theta)&=&\rho_u^{\textup{env}} \left(\frac{r}{R_c}\right)^{-3/2} \left(1+\frac{\cos\theta}{\cos\theta_0}\right)^{-1/2} \times\nonumber\\&&\left(1 + 2\frac{R_c}{r}\right)^{-1}.
\end{eqnarray}
In Figure~\ref{C4:Fig:SPH_env_profile} (middle), we present the rotationally flattened Ulrich profile with the singularity removed (green). We can see that the Ulrich distribution matches the sites at the inner rim of the envelope resolved by the \ac{SPH} simulation, but underestimates the resolved \ac{SPH} envelope profile at larger radii for object 110753. 

\begin{figure*}[t]
\includegraphics[width=1\textwidth]{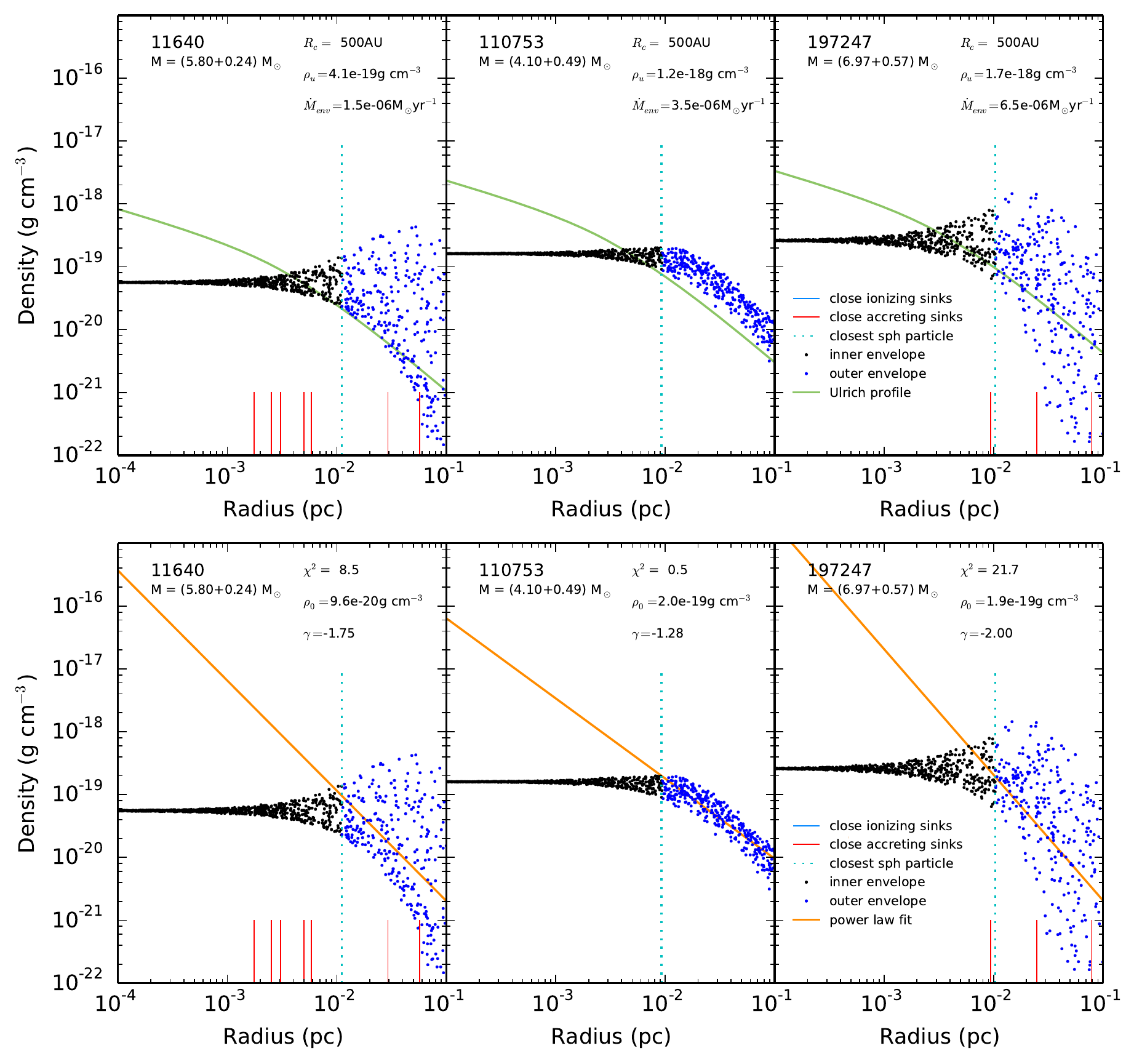}
\caption[Rotationally flattened envelope profile and power-law profile fits]{\label{C4:Fig:site_env_profile} 
Radial density distribution of \acs{s3} sites and their envelope fits in time-step 122 (\SI{6.109}{\Myr}). Red vertical lines highlight the distance of close accreting sink particles; cyan dotted lines show the distance of the closest \ac{SPH} particle.
(top) Green rotational flattened envelope profile: 
    (a) for three relevant sink particles, 
    (b - online material) for all relevant sink particles.
(bottom) Orange lines plot the fit of the blue density sample of \acs{s3} sites: 
    (a) for three relevant sink particles, 
    (b - online material) for all relevant sink particles.
    }
\end{figure*}
%
%%%%%%%%%%%%%%%%%%%%%%%%%%%%%%%%%%%%%%%%%%%%%%%%%%%%%%%
\subsubsection{Method e3 | Power-Law Envelope Profiles}
\label{C4:Sec:method_e3}
%%%%%%%%%%%%%%%%%%%%%%%%%%%%%%%%%%%%%%%%%%%%%%%%%%%%%%%
As an alternative, and in order to understand how important the assumption of the envelope density profile is for the final results, we also consider the case where we use envelopes with spherically-symmetric power-law profiles rather than the Ulrich collapse model (referred to as method \acs{e3}). In the right panel of Figure~\ref{C4:Fig:SPH_env_profile}, we show four different radial power-law profiles (diagonal dashed lines) with exponents $\gamma=[\num{-0.5}, \num{-1.0}, \num{-1.5}, \num{-2.0}]$ for the density function\footnote{ Power-law  envelopes are part of the Analytical \ac{YSO} Models in \textsc{Hyperion} as described on \url{http://www.hyperion-rt.org}.}
\begin{eqnarray}
    \label{C4:Eq:rho_r}
    \rho(r)&=&\rho_0 (r/r_0)^\gamma.
\end{eqnarray}
The orange solid diagonal line is the log-scaled least-squares fit \citep[][]{NumericalRecipes} of the density distribution of the nearby \ac{SPH} particles. The log-scaled least-square fit requires us to fit the following function:
\begin{eqnarray}
    \label{C4:Eq:func_log}
    \log_{10}\rho(r)&=&\log_{10}\rho_0 + \gamma\log_{10}(r/r_0).
\end{eqnarray}
For this accreting sink, the best fit of $\gamma=\num{-1.25}$ is a typical value for a protostar. \\

Note that a fitted power-law profile in density typically produces a better fit in terms of absolute scaling compared to the Ulrich profiles. However, spherical symmetric power-law density profiles with only one slope are not entirely physically motivated, whereas the Ulrich models are self-consistent with respect to the infall rate.
%%%%%%%%%%%%%%%%%%%%%%%%%%%%%%%%%%%%%%%%%%%%%%%%%%%%%%%
\subsubsection{Results | Envelope Profiles}
\label{C4:Sec:results_envelopes}
%%%%%%%%%%%%%%%%%%%%%%%%%%%%%%%%%%%%%%%%%%%%%%%%%%%%%%%
The Ulrich description (method \acs{e1}) is a physically self-consistent envelope-infall model and is therefore consistent with the general physical picture, when improving the resolution on small scales by extrapolating inwards. However, the singularity at the centrifugal radius is unrealistic. Therefore, due to the singularity at the mid-plane, we do not use the standard Ulrich envelope (method \acs{e1}). 

However, we will use the Ulrich description without singularity \acs{e2} and add a circumstellar disk later in the radiative transfer set-up in  Section~\ref{C4:Sec:methods_RT_YSO}. For the Ulrich description, we set the centrifugal radius $R_c$, hence the outer rim of the circumstellar disk, to $R_c=\SI{500}{\AU}$. The choice of the centrifugal radius is dependent on the chosen outer disk radius. \SI{500}{\AU} is a realistic intermediate value of typical values between \SI{100}{\AU} and \SI{1000}{\AU}. Moreover, the disk radius needs to be smaller than the accretion radius from the simulation ($r_{\textup{acc}}=\SI{0.005}{\pc}\approx\SI{1000}{\AU}$), because otherwise the disk would have been resolved by the simulation. We calculate the density $\rho_u^{\textup{env}}$ (see Eq.~\ref{C4:Eq:rho_Ulrich}) for all Ulrich envelope fits. In Figure~\ref{C4:Fig:site_env_profile} (top panel), we show \acs{e2} rotationally flattened envelope profiles for the three accreting sink particles 11640, 110753 and 197247 with the corresponding values for $\rho_u^{\textup{env}}$ and $\dot{M}_{\textup{env}}$. To see all profiles for time-step 122 (\SI{6.109}{\Myr}) see the online extension of Figure~\ref{C4:Fig:site_env_profile} (top). While the majority of sink particle envelopes are reproduced well by Ulrich envelopes without singularities, this model under-predicts the density for some of the accreting objects. However, we cannot arbitrarily scale these particular profiles, because the mass $M_{*}$ of the stellar object needs to drop for increasing $\rho_u^{\textup{env}}$ and constant infall rate $\dot{M}_{\textup{env}}$ (c.\,f.~with Eq.~\ref{C4:Eq:rho_Ulrich}). 

The power-law method \acs{e3}, which fits the envelope profile, follows the slope of the radial profile of the outer envelope resolved by the \ac{SPH} simulation, since it is an extrapolation of the gravitational collapse modeled in the \acs{D14} \ac{SPH} simulations. We fit the slope of the profile $\gamma$ and the density $\rho_0$ at the radius $r_0$, which we set as the distance of the closest \ac{SPH} particle. To extrapolate \citep[see][]{NumericalRecipes} the profile of the density for different cells $j$ within $r_j<r_0$, we are only using \acs{s3} circumstellar sites of the outer envelope ($r_{j}>r_0$), which belong to the sink particles we want to fit the profile for. We only use sites from the outer envelope (see blue sample in Figure~\ref{C4:Fig:site_env_profile}), since the inner density values (black) represent the background density of the outer envelope created by the \ac{SPH} field close to the sink particle. In the bottom panel of Figure~\ref{C4:Fig:site_env_profile}, we show the power-law fits of the accreting sink particles 11640, 110753 and 197247 in time-step 122 (\SI{6.109}{\Myr}), and we show the same for all accreting sink particles of this time-step in the online extension of Figure~\ref{C4:Fig:site_env_profile} (bottom). We limit the slope of the profile to the range $\gamma\in\left[\num{0},\num{-2}\right]$, because for $\gamma=\num{0}$ the density follows the flat background density at the \acs{s3} circumstellar sites calculated in Section~\ref{C4:Sec:methods_parameter}. The limit $\gamma=\num{-2}$ was chosen because it corresponds to the radial profile of the "singular isothermal sphere" in gravitational collapse \citep[see][]{Shu:1977}. From Figure~\ref{C4:Fig:site_env_profile}, we can see that the accretion stream (blue) is not spherically symmetric when the accreting sink particles sit in a close cluster of sink particles (11640, 197247) rather than being isolated (110753). We use a $\chi^2$ value \citep[see][]{NumericalRecipes}, to quantify the quality of the envelope profile fit:
\begin{eqnarray}
    \chi^2&=&\sum\limits_{j=1}^{N_j} \frac{(\log_{10}\rho(r_{j}) - \log_{10}\rho_{\textup{fit}}(r_{j}))^2}{|\log_{10}\rho_{\textup{fit}}(r_{j})|}.
\end{eqnarray}
We need $N_j$, which are all the sites that belong to the refined circumstellar sites of a sink particle with $r_{j}>r_0$, and $\rho_{\textup{fit}}(r_j)$, which is Eq.~\ref{C4:Eq:func_log} evaluated for the fitted parameters $\gamma_{\textup{fit}}$, and $\rho_{0,\textup{fit}}$ to calculate 
\begin{eqnarray}
    \log_{10}\rho_{\textup{fit}}&=&\log_{10}\rho_{0,\textup{fit}} + \gamma_{\textup{fit}}\log_{10}(r_j/r_0).
\end{eqnarray}
Spherically symmetric profiles produce a better fit and have lower $\chi^2$ values, as can be seen in the bottom panel of Figure~\ref{C4:Fig:site_env_profile}. However, when comparing with method \acs{e2}, which uses the rotational flattened envelope profile without singularity, we see that in the central region, the power-law density is several orders of magnitude higher. 

\begin{figure*}[t]
\includegraphics[width=1\textwidth]{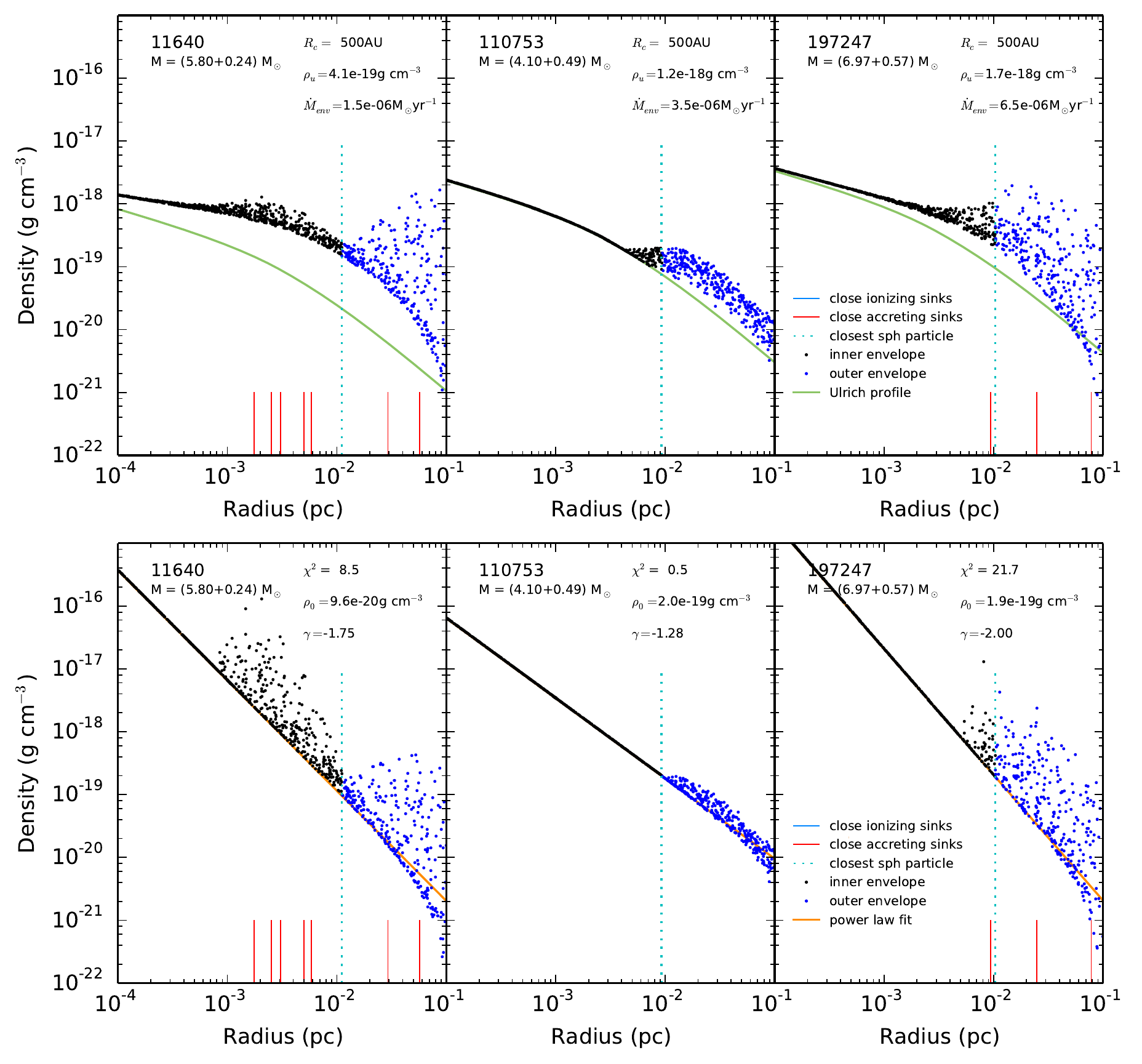}
\caption[Superposition of refined envelope profiles]{\label{C4:Fig:site_env_profile_sup} 
Extrapolated envelope density distribution of \acs{s3} sites including superposition and their envelope profiles in time-step 122 (\SI{6.109}{\Myr}). Red vertical lines highlight the distance of close accreting sink particles; cyan dotted lines show the distance of the closest \ac{SPH} particle.
(top) Green rotational flattened envelope profile:
    (a) for three relevant sink particles, 
    (b - online material) for all relevant sink particles.
(bottom) Orange lines plot the fit of the blue density sample of \acs{s3} sites.
    (a) for three relevant sink particles, 
    (b - online material) for all relevant sink particles.
}
\end{figure*}
While the Ulrich profile is consistent with the physical picture of a gravitational collapse with conservation of angular momentum, the gravitational collapse description in the \acs{D14} \ac{SPH} simulations should reproduce power-law profiles on small scales. The power-law profiles in the \acs{D14} \ac{SPH} simulations are limited by the isothermal accretion flow of $\rho\sim r^{-2}$ at larger radii ($r>\SI{0.1}{\pc}$). Additionally, within the accelerating collapsing shell ($r<\SI{0.1}{\pc}$), the velocity is not constant and the density profile is shallower, with $\rho\sim r^{-3/2}$ \citep{Shu:1977,Ulrich:1976}. This explains the offset of the radial density profile of the outer envelope from the simulations, compared to the rotationally flattened envelope profile and why the power-law profiles produces a better fit. Although the power-law profiles are consistent with the gravitational collapse description used in the \acs{D14} \ac{SPH} simulations, the rotational flattened envelope profiles are consistent with the physical picture of small scale gravitational collapse. 

To explore the effects of the choice of refinement techniques on the synthetic observations, we will provide the results for both methods \acs{e2} and \acs{e3} in this paper.

%%%%%%%%%%%%%%%%%%%%%%%%%%%%%%%%%%%%%%%%%%%%%%%%%%%%%%%
\subsection{Envelope Superposition}
\label{C4:Sec:methods_envelope_super}
%%%%%%%%%%%%%%%%%%%%%%%%%%%%%%%%%%%%%%%%%%%%%%%%%%%%%%%
In some cases, the accreting sink particles lie in compact clusters that include other accreting sink particles. As a result, the different envelope profiles will overlap, increasing the density further. An overlap means different things for the two different envelope descriptions \acs{e2} and \acs{e3}. In the self-consistent picture of an infalling envelope (\acs{e2}: Ulrich profile), the overlap is represented by summation, since the infalling mass is increasing. For the power-law envelope description (\acs{e3}), the maximum of the superimposed envelopes at every position is the resulting envelope profile, since if the envelopes of two sink particles with the same power-law would overlap, the resulting fit would again be the initial power-law envelope, and not a power-law envelope twice as dense. 
\begin{figure*}[t]
\includegraphics[width=1\textwidth]{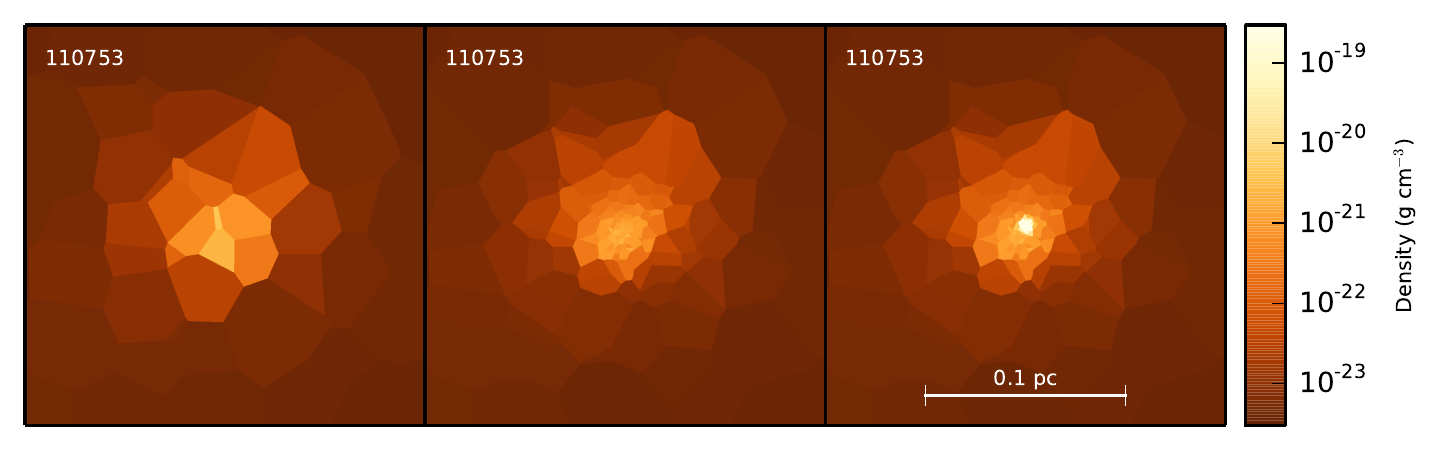}
\caption[Different versions of envelope refinement in a 2-d slice]{\label{C4:Fig:SmoothTransition} Different methods of envelope profiles for the sink particle 110753 in \emph{run I} step 122 (\SI{6.109}{\Myr}). Region around the sink particle with site method \acs{s1} (left); region around the sink particle with site method \acs{s3} (middle); region around the sink particle with site method \acs{s3} and added envelope profile (right).}
\end{figure*}

By using a smooth weighting function, we combine the density of the envelope and the background and recover a smooth transition between the two. Additionally, for sink particles that are close to other sink particles, we then also weight the density of the different envelopes accordingly, before evaluating a final density at each point. We describe these two cases in the next two sections.

%\newpage
\subsubsection{Weighting}
We calculate the distance $r_{js}$, envelope density $\rho^{\textup{env}}_{js}$ of every \acs{s3} circumstellar site $j$ and every accreting sink particle $s$ for which we have already calculated the "background" density of method \acs{p3} $\rho_{\textup{p3},j}$ in Section~\ref{C4:Sec:methods_parameter}. We define that an envelope of a sink particle only contributes when $r_{js}<r_{max}$, where $r_{max}$ is the maximum distribution radius of \acs{s3} circumstellar sites that belong to a sink particle as described in Section~\ref{C4:Sec:methods_sites}, otherwise we set all elements of $\rho^{\textup{env}}_{js}=0$. We weight \citep[][]{NumericalRecipes} the density $\rho^{\textup{env}}_{js}$ by: 

\begin{eqnarray}
    w_{j,s}&=&\Bigg\{\begin{matrix}
  1\ \ \ \ \ \ \ \ \ \ \ \ \forall\ r_{js} < r_{0,s}\\
  1-\frac{r_{js}-r_{0,s}}{r_{max}-r_{0,s}}\ \ \ \ \ \ \ \forall\ r_{0,s} \leq r_{js} \leq r_{max}, \\
  \ 0 \ \ \ \ \ \ \ \ \ \ \ \ \forall\ r_{js} > r_{max}
 \end{matrix}
\end{eqnarray}
where $r_{0,s}$ is the radius of the closest sink particle, as described in Section~\ref{C4:Sec:methods_envelope}. 
The "background" density $\rho_{\textup{p3},j}$ is weighted with the counter weight
\begin{eqnarray}
    \stackrel{\sim}{w}_{j,s}&=& 1 - w_{j,s}.
\end{eqnarray}
We set all weights $w_{j,s}$ and $\stackrel{\sim}{w}_{j,s}$ to zero when the density $\rho^{\textup{env}}_{js}=0$. 

\subsubsection{Weighted Superposition}
For the Ulrich description of method \acs{e2} we sum the weighted density function to get the superposition envelope density $\hat{\rho}_{j,\textup{p3}}$ at every site $j$ with
\begin{eqnarray}
    \label{C4:Eq:rho_final_ul}
    \hat{\rho}_{j,\textup{p3}}^{\textup{e2}}&=& \sum\limits_{s=1}^{N_s} \left[ w_{j,s} \rho^{\textup{env}}_{js}  + \rho_{\textup{p3},j}\stackrel{\sim}{w}_{j,s}\right],
\end{eqnarray}
by using the normalization $N_s$, which is the number of accreting sink particles with envelopes. The summation is necessary, since the Ulrich envelope was constructed from the physical parameters, such as the infall rate $\dot{M}_{\textup{env}}$.

As explained above, for the power-law description \acs{e3}, we take the maximum of the weighted density function when superimposing many envelopes
\begin{eqnarray}
    \label{C4:Eq:rho_final_po}
    \hat{\rho}_{j,\textup{p3}}^{\textup{e3}}&=& \max\limits_{s=1}^{N_s} \left[ w_{j,s} \rho^{\textup{env}}_{js}  + \rho_{\textup{p3},j}\stackrel{\sim}{w}_{j,s}\right].
\end{eqnarray}
When the superposition density is lower than the background density $\rho_{\textup{p3},j}$, for example in some cases the outer regions of the Ulrich envelope fits are too low, we set $\hat{\rho}_{j,\textup{p3}}=\rho_{\textup{p3},j}$.
In Figure~\ref{C4:Fig:site_env_profile_sup}, we show the inward-extrapolated densities of the envelope profiles including superposition. For all objects in time-step 122 (\SI{6.109}{\Myr}), see the online extension of Figure~\ref{C4:Fig:site_env_profile_sup}. For the isolated object 110753, we only see the density extrapolation inwards, while for the object 11640 and 197247 we see a superposition. For instance, in the upper panel in Figure~\ref{C4:Fig:site_env_profile_sup}, where we show the Ulrich description of method \acs{e2}, we can see that for the non-isolated objects 11640 and 197247, the new profile is higher than the initial fit (green) with values approaching the initial fit at small radii. 

For the power-law envelope of method \acs{e3} in the lower panel of Figure~\ref{C4:Fig:site_env_profile_sup}, this effect was suppressed due to the fact that we take the maximum of the envelope densities rather than the sum, as shown in Eq.~\ref{C4:Eq:rho_final_po}. We can see that there is a smooth transition to other envelopes (black particles in object 11640 and 197247) and also a seamless transition\footnote{ Note that superimposed envelopes close to ionization bubbles could now create large high-density cells  in the outer rim of the circumstellar sites for an individual accreting sink particle. To avoid this, we set in those cases (cell volume larger than \SI{5}{\percent} of the envelope volume at the certain radius) the superimposed density to the background density of the cell $\rho_{j,\textup{p3}}$ before superposition and envelope extrapolation.} to the radial profile of other envelopes resolved by the \ac{SPH} simulation (blue particles).

In Figure~\ref{C4:Fig:SmoothTransition}, we present a 2-d slice through the density structure in order to show the improvement in resolution for the accreting sink particle 110753. The left panel shows the density evaluated on the grid where each cell corresponds to an \ac{SPH} particle (\acs{s1}), the
central panel shows the improvement once the circumstellar cells are added (\acs{s3}), and finally the panel on the right shows an inward extension of the envelope including the superposition of envelope densities $\hat{\rho}_{j,\textup{p3}}$ for a rotationally flattened envelope (\acs{e2}).
%%%%%%%%%%%%%%%%%%%%%%%%%%%%%%%%%%%%%%%%%%%%%%%%%%%%%%%
\subsection{Envelope Cavities}
\label{C4:Sec:methods_cavities}
%%%%%%%%%%%%%%%%%%%%%%%%%%%%%%%%%%%%%%%%%%%%%%%%%%%%%%%
\acp{YSO} drive outflows that carve out bipolar cavities in the envelopes, mostly perpendicular to the mid-plane \citep[for more details, see][]{Whitney:2003}. We parameterize the outflow cavity walls\footnote{ The description follows the Analytical \ac{YSO} Models in \textsc{Hyperion} as described on \url{http://www.hyperion-rt.org}.} as:
\begin{eqnarray}
    \label{C4:Eq:cavity}
    z(r)&=& \pm a r^{3/2}.
\end{eqnarray}
We choose the opening constant $a$ in such a way that at the outer radius of the envelope $R_{max}^{\textup{env}}$, the half-opening angle $\theta_{\textup{open}}$ is equal to $\theta_{\textup{open}}=\SI{10}{\degree}$:
\begin{eqnarray}
    \label{C4:Eq:cavity_a}
    a&=& R_{max}^{\textup{env}} \cos(\theta_{\textup{open}}).
\end{eqnarray}
We assume that the material in the cavity is optically thin so that the radiation can escape. Therefore, we set the cavity density
\begin{eqnarray}
    \label{C4:Eq:cavity_dens}
    \rho_{\textup{cavity}}&=&0.
\end{eqnarray}
The cavity removes a certain fraction from the total mass. For the rotational flattened envelopes (\acs{e2}) about \SI{2}{\percent} is removed, while for the power-law envelope (\acs{e3}) up to \SI{4}{\percent} is removed.

%%%%%%%%%%%%%%%%%%%%%%%%%%%%%%%%%%%%%%%%%%%%%%%%%%%%%%%
\subsection{Circumstellar Disks}
\label{C4:Sec:methods_disk}
%%%%%%%%%%%%%%%%%%%%%%%%%%%%%%%%%%%%%%%%%%%%%%%%%%%%%%%
In the last three sections, we focused on the envelope of accreting stellar objects in the \acs{D14} \ac{SPH} simulations. Physically, on smaller scales, the accreting material of the envelope is channeled through the disk to the central object. From the \acs{D14} \ac{SPH} simulations, we extracted the total angular momentum vector $\mathcal{L}_{tot}$ of the accreting sink particles to determine the inclination $\Delta\theta$ of all disks in the simulation at every time-step:
\begin{eqnarray}
    \label{C4:Eq:L_tot}
    \mathcal{L}_{tot} &=& \sqrt{\mathcal{L}_x^2 + \mathcal{L}_y^2 + \mathcal{L}_z^2}\\
    \label{C4:Eq:incl}
    \Delta\theta &=& \arccos(\mathcal{L}_z/\mathcal{L}_{tot})
\end{eqnarray}
We can evaluate the density contributed by the disk $\rho_{\textup{disk}}(r,\theta)$ for the \acs{s3} circumstellar sites of a sink particle at a certain radius $r$ and at an angle $\theta$ (see  Eq.~\ref{C4:Eq:angles}) through the flared disk\footnote{ The description follows the Analytical \ac{YSO} Models in \textsc{Hyperion} as described on \url{http://www.hyperion-rt.org}.} \citep[][]{ArmitageBook} density profile \citep[e.\,g.][]{Whitney:2003,Shakura:1973} below. By replacing $\theta$ by $\theta - \Delta\theta$, we incline the disk, as set by the angular momentum vector (see Eq.~\ref{C4:Eq:L_tot} and Eq.~\ref{C4:Eq:incl}):
\begin{eqnarray}
    \label{C4:Eq:rho_disk}
    \rho_{\textup{disk}}(r,\theta)&=& \Bigg\{\begin{matrix}
  \rho_{disk}& \forall&\ r^d_{min} < r \leq r^d_{max},\\
  0 & \forall&\ r > r^d_{max}
 \end{matrix}
\end{eqnarray}
\begin{eqnarray}
 \rho_{\textup{disk}} &=& \rho_{\textup{norm}} \left(\frac{r_0}{r}\right)^{\beta-p} \exp\left[-\left(\frac{r\cos(\theta-\Delta\theta)}{4h}\right)^2\right]\ \ \ 
\end{eqnarray}
with the disk flaring parameter $\beta$, the surface density exponent $p$, the maximum disk radius $r^d_{max}$, the minimum disk radius $r^d_{min}$ and disk scale height $h$, which is defined as a relation of the height $h_0$ at radius $r_0$:
\begin{eqnarray}
   h&=& h_0 \left(\frac{r}{r_0}\right)^{\beta}.
\end{eqnarray}
The integration normalization constant $\rho_{\textup{norm}}$ is defined to satisfy that the volume integral of the disk density function of Eq.~\ref{C4:Eq:rho_disk} is equal to the disk mass:
\begin{eqnarray}
    \label{C4:Eq:M_disk}
   M_{\textup{disk}} &=&\int\limits_{0}^{2\pi} d\phi \int\limits_{-\pi}^{\pi}d\theta\int\limits_{0}^{r^d_{max}} dr\ r^2 \sin\theta \rho_{\textup{disk}}(r,\theta) 
\end{eqnarray}
which can be solved numerically. Since we assume the mass of the disk $M_{\textup{disk}}\approx\num{e-2}M_*$, we can solve, with the help of Eq.~\ref{C4:Eq:rho_disk} for $\rho_{\textup{norm}}$ . We calculate the disk density distribution $\rho_{\textup{disk}}(r,\theta, j)$ for all the \acs{s3} circumstellar sites $j$ with the following values:
\begin{table*}[t]
\caption[Summary of Set-ups for different Circumstellar Configurations.]{Summary of Set-ups for different Circumstellar Configurations.}
\label{C4:Tab:CM}
\begin{center}
\begin{tabular*}{\textwidth}{lccc}
\hline\\[-13pt]
\hline\\[-5pt]
&&&\\
\multicolumn{1}{l}{method} &
\multicolumn{1}{c}{\acs{CM1}} &
\multicolumn{1}{c}{\acs{CM2}} &
\multicolumn{1}{c}{\acs{CM3}}\\[6pt]
\hline\\[-5pt]
envelope fitting            &-                                      &\acs{e2}                                                                                                           &\acs{e3}\\
sites                       &\acs{s3}                               &\acs{s3}                                                                                                           &\acs{s3}\\
density                     &$\rho_{\textup{p3}}$                   &$\hat{\rho}_{j,\textup{p3}}^{\textup{e2}}(r_{j,s}>r^d_{max})$                                                      &$\hat{\rho}^{\textup{e3}}_{j,\textup{p3}}(r_{j,s}>r^d_{max})$\\
density reference           &Eq.~\ref{C4:Eq:rho_v3}                 &Eq.~\ref{C4:Eq:rho_final_po}                                                                                       &Eq.~\ref{C4:Eq:rho_final_po}\\
Analytical \ac{YSO} Model   &no                                     &yes                                                                                                                &yes\\
source mass                 &$M_*=M_{\textup{sink}}$                &$M_*=M_{\textup{sink}} - (M_{\textup{env}}^{\textup{e2}} - M_{\textup{cavity}}^{\textup{e2}})- M_{\textup{disk}}$  &$M_*=M_{\textup{sink}} - (M_{\textup{env}}^{\textup{e3}} - M_{\textup{cavity}}^{\textup{e3}})- M_{\textup{disk}}$\\
mass references             &\acs{D14} \ac{SPH} simulations    &Eq.~\ref{C4:Eq:M_disk}&Eq.~\ref{C4:Eq:M_env}, \ref{C4:Eq:M_disk}\\[6pt]
\hline\\[-5pt]
 \end{tabular*}
% \vspace{-0.5cm}
\end{center}
\end{table*}
\begin{eqnarray}
    M_{\textup{disk}}&=&\num{e-2}M_*\\
    r_0&=&\SI{1000}{\AU}\\
    h_0&=&\SI{50}{\AU}\\
    r^d_{min}&=&r_{\textup{OptThin}}(T_{\textup{dust}}=\SI{1600}{\kelvin})\\
    \label{C4:Eq:r_disk_max}
    r^d_{max}&=&R_c=\SI{500}{\AU}\\
    \alpha&=&\num{-1.0}\\
    \beta&=&\num{1.25}
\end{eqnarray}
where $r_{\textup{OptThin}}(T_{\textup{dust}}=\SI{1600}{\kelvin})$ is the sublimation radius\footnote{ Description from \url{http://www.hyperion-rt.org} and dereived by Thomas Robitaille.}, given by the radius at which the temperature would drop to 1600K in an optically-thin medium: 
\begin{eqnarray}
r_{\textup{OptThin}}(T_{\textup{dust}})&=&R_*\left[1-\left(1-2\frac{T^4_{\textup{dust}}}{T^4_{\textup{eff}}}\frac{\kappa_B(T_{\textup{dust}})}{\kappa_*}\right)^2\right]^{-1/2}
\end{eqnarray}
where $\kappa_B$ is the Planck mean opacity, $\kappa_*$ is the mean opacity of the dust weighted by the stellar photosphere and $T_{\textup{eff}}$ is the effective temperature of the star. Note that when dust properties, including \ac{PAH} molecules, are used, we set the inner radius to $r_{\textup{OptThin}}(T_{\textup{dust}})$ for the largest dust grain.

%%%%%%%%%%%%%%%%%%%%%%%%%%%%%%%%%%%%%%%%%%%%%%%%%%%%%%%
\subsection{Circumstellar Set-up}
\label{C4:Sec:methods_RT_YSO}
%%%%%%%%%%%%%%%%%%%%%%%%%%%%%%%%%%%%%%%%%%%%%%%%%%%%%%%
In Section~\ref{C4:Sec:methods_sph2voro}, we discussed different methods for how to distribute sites and how to evaluate the density and temperature when transferring from a particle-based \ac{SPH} simulation to a Voronoi tessellation. With a description of the temperature $T_{\textup{p3}}$ and the "background" density $\rho_{\textup{p3}}$, we described how to evaluate the properties of the stellar sources in Section~\ref{C4:Sec:methods_RT}. Further, the circumstellar density can be computed by using the envelope profiles \acs{e2} and \acs{e3} (see Section~\ref{C4:Sec:methods_envelope}) in order to evaluate the overlapping envelope density $\hat{\rho}_{\textup{p3}}$ (see Section~\ref{C4:Sec:methods_envelope_super}). We described the bipolar outflow cavity set-up in Section~\ref{C4:Sec:methods_cavities}. For smaller scales the disk density $\rho_{\textup{disk}}$ (Section~\ref{C4:Sec:methods_disk}) can contribute at \acs{s3} circumstellar sites close to the accreting sink particle.

\begin{figure}[t]
\includegraphics[trim=0cm 0cm 28.6cm 0cm, width=\textwidth]{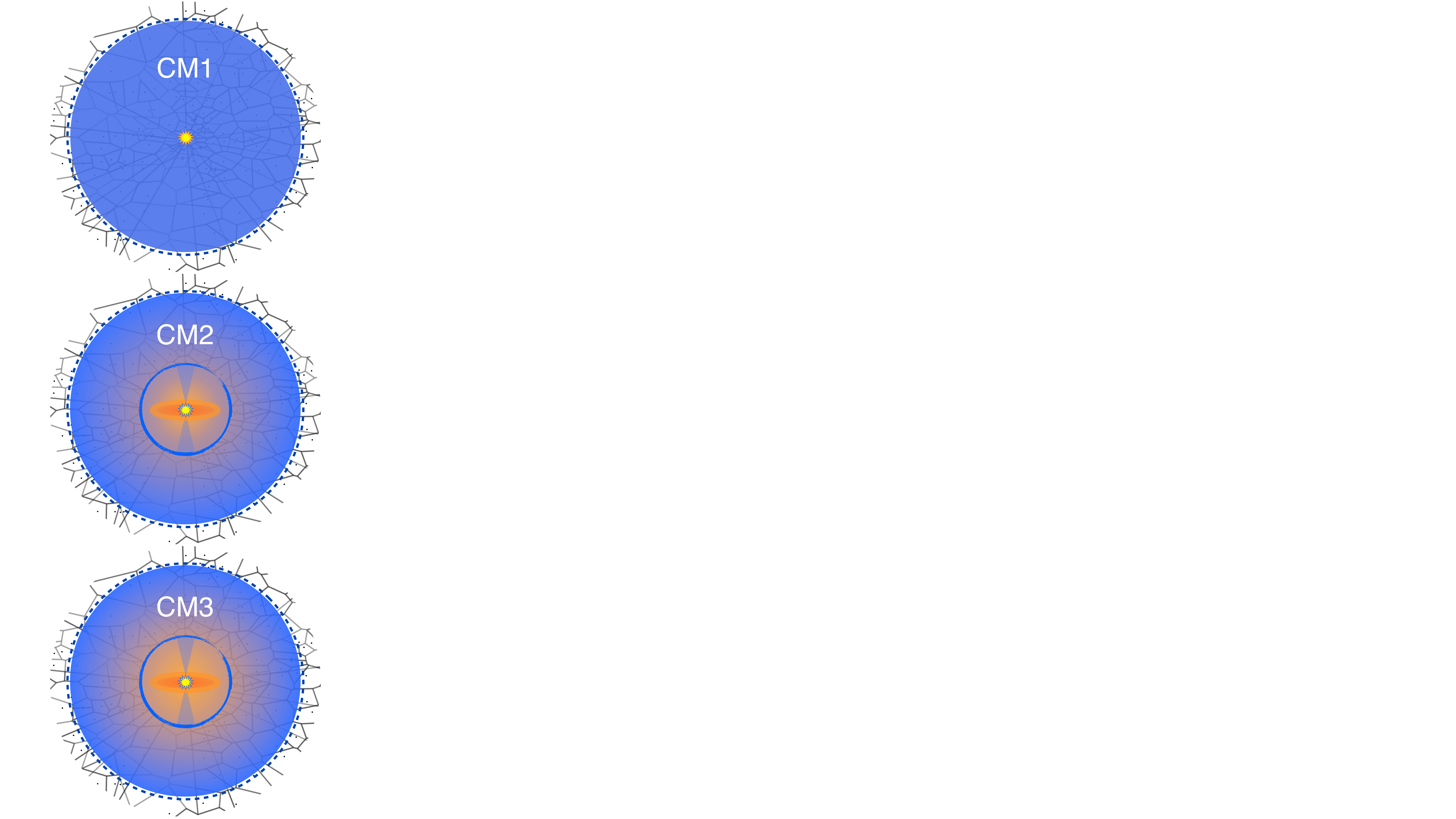}
\caption[Three methods of circumstellar material set-up in the radiative transfer calculations]{\label{C4:Fig:RTversion} Three methods of circumstellar material set-up in the radiative transfer calculations.}
\end{figure}

In this section, we will combine the above prescriptions and explore three different approaches of setting up the circumstellar material from the radiative transfer calculation. In Figure~\ref{C4:Fig:RTversion}, we plot sketches of the different methods of the circumstellar material set-up and summarize the different configuration in Table~\ref{C4:Tab:CM}.
%%%%%%%%%%%%%%%%%%%%%%%%%%%%%%%%%%%%%%%%%%%%%%%%%%%%%%%
\subsubsection{Method CM1 | No Circumstellar Matter}
\label{C4:Sec:method_CM1}
%%%%%%%%%%%%%%%%%%%%%%%%%%%%%%%%%%%%%%%%%%%%%%%%%%%%%%%
The simplest approach (hereafter, \acs{CM1}), is not to add any circumstellar material on top of the density given by the \ac{SPH} simulation, but still include circumstellar sites to ensure that the temperature profile around the sources is resolved. The radiative transfer calculation is then set up as listed in Table~\ref{C4:Tab:CM}.

%%%%%%%%%%%%%%%%%%%%%%%%%%%%%%%%%%%%%%%%%%%%%%%%%%%%%%%
\subsubsection{Method CM2 | e2 Envelope, Cavity \& Disk from "Analytical YSO Models"}
\label{C4:Sec:method_CM2}
%%%%%%%%%%%%%%%%%%%%%%%%%%%%%%%%%%%%%%%%%%%%%%%%%%%%%%%
We extend the first approach by using the densities from the rotationally flattened envelopes (\acs{e2}), we evaluated in Section~\ref{C4:Sec:methods_envelope}, and also account for their superposition (see Section~\ref{C4:Sec:methods_envelope_super}) and hereafter call this circumstellar material method, method \acs{CM2}. We used the superimposed density $\hat{\rho}_{j,\textup{p3}}$ for sites of accreting sink particles, however we set the superimposed envelope density $\hat{\rho}_{j,\textup{p3}}$ of sites $j$ within the disk radius $r_{j}<r^d_{max}=\SI{500}{\AU}$ (as defined in Eq.~\ref{C4:Eq:r_disk_max}) to zero and rather include an Analytical \ac{YSO} Model in the center. For illustration, see Figure~\ref{C4:Fig:RT_temp_test} (middle).

For the inner $\SI{500}{\AU}$, we set up the Analytical \ac{YSO} Model using the radiative transfer code \textsc{Hyperion}, which is adding the \ac{SED} of the \ac{YSO} (star, envelope, cavity \& disk). 

With the higher density material present, we need to adjust the mass of the source in the radiative transfer by the mass of the added circumstellar material (envelope, cavity, disk) between $r_0$ and $r_{min}\approx 0$. Therefore, we calculate the envelope mass $M_{\textup{env}}^{\textup{e2}}$ by numerically integrating over the envelope density profile. We use the bi-polar cavity set-up described in Section~\ref{C4:Sec:methods_cavities} and set the outer radius of the envelope in the cavity description (Eq.~\ref{C4:Eq:cavity_a}) to $R_{max}^{\textup{env}}=\SI{500}{\AU}$. The cavity removes about \SI{2}{\percent} from the mass of the envelope within $R_{max}^{\textup{env}}$. We scale the disk density $\rho_{\textup{disk}}$ thus that the disk mass $M_{\textup{disk}}$ (from Eq.~\ref{C4:Eq:M_disk}) is \SI{1}{\percent} of the stellar mass $M_*$. The other parameters of the Analytical \ac{YSO} Models are set up with the parameters given in Section~\ref{C4:Sec:methods_disk}. The new mass of the stellar particle is then calculated as follows:
\begin{eqnarray}
M_*&=&M_{\textup{sink}} - (M_{\textup{env}}^{\textup{e2}} - M_{\textup{cavity}}^{\textup{e2}}) - M_{\textup{disk}}.
\end{eqnarray}
The method \acs{CM2} sets up with the parameters listed in Table~\ref{C4:Tab:CM}. We use this set-up to precompute the \ac{SED} of the Analytical \ac{YSO} Model. Since the Analytical \ac{YSO} Model is not spherically symmetric, we compute the models for different viewing angles and determine an angle-averaged \ac{SED} that we then use in the radiative transfer calculation. Therefore, we average the computed \ac{SED} over \num{12} isotropically spaced viewing angles. 
\begin{figure*}[t]
\includegraphics[width=\textwidth]{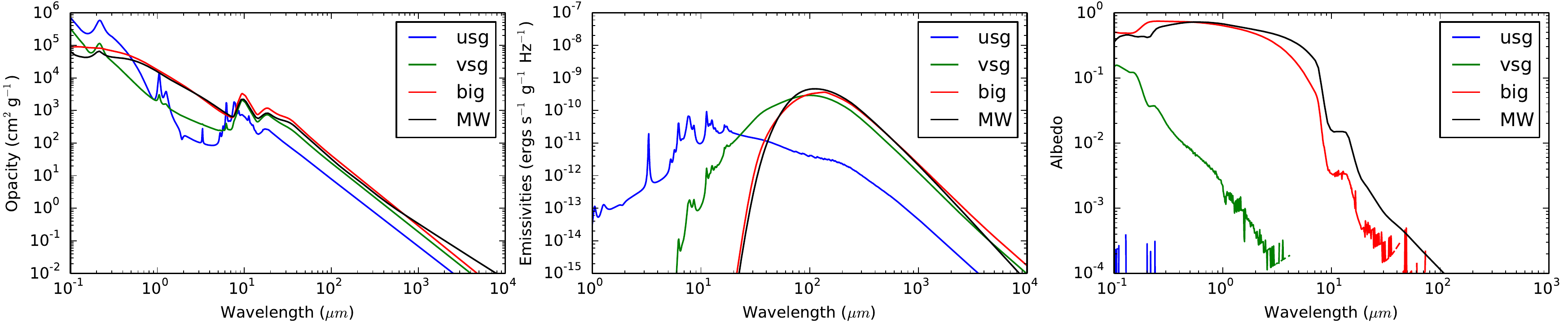}
\caption[Dust opacity, emissivity and albedo for different dust types]{\label{C4:Fig:Albedo}Dust opacity, emissivity and albedo of the ultra-small (usg), very small (vsg) and big grains (big) of the \ac{PAH} dust models and the non-\ac{PAH} Milky Way dust (MW). Note that the panels have a different scale in wavelength.}
\end{figure*}
%
%%%%%%%%%%%%%%%%%%%%%%%%%%%%%%%%%%%%%%%%%%%%%%%%%%%%%%%
\subsubsection{Method CM3 | e3 Envelope, Cavity \& Disk from Analytical YSO Models}
\label{C4:Sec:method_CM3}
%%%%%%%%%%%%%%%%%%%%%%%%%%%%%%%%%%%%%%%%%%%%%%%%%%%%%%%
Method \acs{CM3} is similar to \acs{CM2} but with power-law envelopes instead of rotationally flattened Ulrich envelopes. The envelope mass $M_{\textup{env}}^{\textup{e3}}$ can be, therefore, calculated analytically as the spherical volume integral of the density function from Eq.~\ref{C4:Eq:rho_r} between $r_0$ and $r_{min}\approx 0$.
\begin{eqnarray}
    \label{C4:Eq:M_env}
   M_{\textup{env}}^{\textup{e3}} &=&\int\limits_{0}^{2\pi} d\phi \int\limits_{-\pi}^{\pi}d\theta\int\limits_{r_{min}}^{r_0} dr\ r^2 \sin\theta \rho(r)\\
   &=&4\pi\rho_0\int\limits_{r_{min}\ \approx\ 0}^{r_0} dr\ \left(\frac{r}{r_0}\right)^{\gamma}r^2 \\
   &\approx& \frac{4\pi}{\gamma+3}\ \rho_0\ r_0^3
\end{eqnarray}

In Table~\ref{C4:Tab:CM}, we summarize the set-up parameters of method \acs{CM3}. Note that the cavity removes about \SI{4}{\percent} from the mass of the envelope within $R_{max}^{\textup{env}}$.
%%%%%%%%%%%%%%%%%%%%%%%%%%%%%%%%%%%%%%%%%%%%%%%%%%%%%%%
\subsubsection{Results | Circumstellar Material Set-up}
\label{C4:Sec:results_YSO}
%%%%%%%%%%%%%%%%%%%%%%%%%%%%%%%%%%%%%%%%%%%%%%%%%%%%%%%
Above we presented the circumstellar material set-up versions \acs{CM1} (star alone), \acs{CM2} (star \& \acs{e2} Analytical \ac{YSO} Models) and \acs{CM3} (star \& \acs{e3} Analytical \ac{YSO} Models). 

Method \acs{CM1} is the simplest approach when setting up a radiative transfer model from simulations, because it does not require an inward envelope extrapolation. For instance, \cite{Offner:2012} used this approach without circumstellar refinement because the circumstellar material was too optically thick to contribute to the flux that they were interested in. 

However, in this work we will present a parameter study: From now on we will use method \acs{CM1}, \acs{CM2} and \acs{CM3} to set up the circumstellar material in the radiative transfer set-up and therefore explore the effects of circumstellar material on the \ac{MIR} flux.

%%%%%%%%%%%%%%%%%%%%%%%%%%%%%%%%%%%%%%%%%%%%%%%%%%%%%%%
%%%%%%%%%%%%%%%%%%%%%%%%%%%%%%%%%%%%%%%%%%%%%%%%%%%%%%%
\section{Dust \& Temperatures}
\label{C4:Sec:methods_dust_temp}
%%%%%%%%%%%%%%%%%%%%%%%%%%%%%%%%%%%%%%%%%%%%%%%%%%%%%%%
%%%%%%%%%%%%%%%%%%%%%%%%%%%%%%%%%%%%%%%%%%%%%%%%%%%%%%%
Now that the density has been successfully mapped onto the radiative transfer grid and stellar sources including the circumstellar material have been set up, we will focus in this section on the dust and the temperature set-up.
%%%%%%%%%%%%%%%%%%%%%%%%%%%%%%%%%%%%%%%%%%%%%%%%%%%%%%%
\subsection{Dust Properties}
\label{C4:Sec:methods_RT_dust}
%%%%%%%%%%%%%%%%%%%%%%%%%%%%%%%%%%%%%%%%%%%%%%%%%%%%%%%
We set the dust-to-gas ratio to a typical value of $\frac{\textup{dust}}{\textup{gas}} = \num{0.01}$ \citep{DraineBook}. When computing the radiative transfer, we explore two different sets of dust properties: 
\begin{itemize}
     \item one that assumes thermal emission for all grain sizes \citep{Draine:2003,Weingartner:2001} and\\[-0.5cm] 
     \item one that also takes into account the transient heating of very small grains and \ac{PAH} molecules \citep{Draine:2007}.
\end{itemize}
We will refer to these as non-\ac{PAH} dust and \ac{PAH} dust respectively. In Figure~\ref{C4:Fig:Albedo}, we show the opacity, emissivity and albedo of the different dust components.
%%%%%%%%%%%%%%%%%%%%%%%%%%%%%%%%%%%%%%%%%%%%%%%%%%%%%%%
\subsubsection{Non-PAH Dust}
\label{C4:Sec:dust_nonPAH}
%%%%%%%%%%%%%%%%%%%%%%%%%%%%%%%%%%%%%%%%%%%%%%%%%%%%%%%
In Figure~\ref{C4:Fig:Albedo}, we show the properties of the Milky Way non-\ac{PAH} dust from \cite{Weingartner:2001} with the re-normalization of abundances from \cite{Draine:2003} with $R_V=\num{5.5}$ and $b_c=\num{3.0}$, where $R_V$ is the ratio of the visual extinction to reddening magnitude, and $b_c$ is the concentration of carbon atoms in the medium. These dust properties reproduce the fluxes of the \ac{FIR} well but lacks emission in the \ac{NIR} and \ac{MIR} up to \SI{20}{\microns}. 
%%%%%%%%%%%%%%%%%%%%%%%%%%%%%%%%%%%%%%%%%%%%%%%%%%%%%%%
\subsubsection{PAH Dust}
\label{C4:Sec:dust_PAH}
%%%%%%%%%%%%%%%%%%%%%%%%%%%%%%%%%%%%%%%%%%%%%%%%%%%%%%%
We use an approximate method for the treatment of the emission from dust grains and PAH molecules,
following \citet[and private communication with Bruce Draine therein]{Robitaille:2012}. We separate the dust size distribution into three grain species: \SI{80.63}{\percent} big grains ($>$\SI{200}{\angstrom}); \SI{13.51}{\percent} of a smaller dust species within \SI{200}{\angstrom} and \SI{20}{\angstrom}, called very small grains (vsg); and \SI{5.86}{\percent} of the \ac{PAH} molecules, called ultra-small grains ($<$\SI{20}{\angstrom},usg). 

The very small and ultra-small grains are transiently heated as mentioned above, while the big grains are in \ac{LTE}, similarly to the non-\ac{PAH} dust. When comparing the non-\ac{PAH} dust (black) with the big grains (red) in Figure~\ref{C4:Fig:Albedo}, we can see that the opacities are not that different, while the ultra-small grains (blue) and very small grains (green) have pronounced features in the \ac{MIR}. The strong emissivity values of the ultra-small grains below \SI{20}{\microns} causes the \ac{PAH} emission typically observed towards high-mass star-forming regions.
\begin{figure*}[t]
\includegraphics[width=\textwidth]{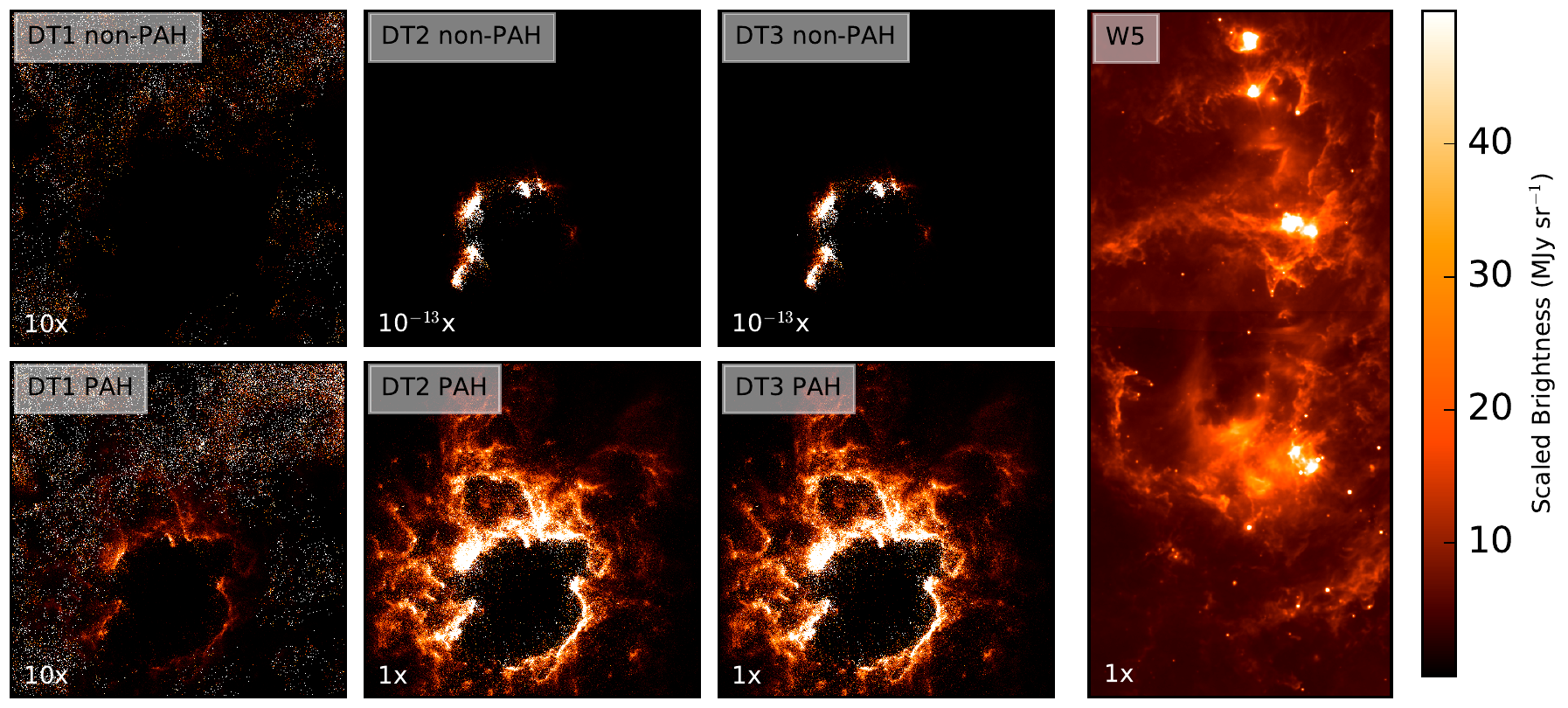}
\caption[Ideal synthetic observations for different temperature combinations and dust types]{\label{C4:Fig:RT_temp_test} Ideal synthetic observations for different temperature combinations and dust types. The right panel shows the background subtracted image at \ac{WISE} \SI{22}{\microns} of the Soul Nebula (c.\,f.~with Figure~\ref{C4:Fig:apples_light}), a real star-forming region. Note that the images do not all have the same color scaling. The white numbers in the individual panels represent the scaling factors of the images in respect to the colorbar. A low scaling factor stands for low brightnesses.}
\end{figure*}
%
%%%%%%%%%%%%%%%%%%%%%%%%%%%%%%%%%%%%%%%%%%%%%%%%%%%%%%%
\subsection{Temperature Combinations}
\label{C4:Sec:methods_RT_temp}
%%%%%%%%%%%%%%%%%%%%%%%%%%%%%%%%%%%%%%%%%%%%%%%%%%%%%%%
In \textsc{Hyperion}, it is possible to couple the calculated radiative transfer temperature of the dust $T^{\textup{dust}}_{\textup{RT}}$ to the hydrodynamical temperature mapped onto the Voronoi mesh $T_{\textup{p3}}$. We will now explore the effects for different dust types and temperature combinations:
%%%%%%%%%%%%%%%%%%%%%%%%%%%%%%%%%%%%%%%%%%%%%%%%%%%%%%%
\subsubsection{Method DT1}
\label{C4:Sec:method_DT1}
%%%%%%%%%%%%%%%%%%%%%%%%%%%%%%%%%%%%%%%%%%%%%%%%%%%%%%%
The method \acs{DT1} combines the radiative transfer temperature $T^{\textup{dust}}_{\textup{RT}}$, calculated from the non-\ac{PAH} and \ac{PAH} dust with the hydrodynamical temperature $T_{\textup{p3}}$.
\begin{eqnarray}
    \label{C4:Eq:DT1}
T^{tot}_{\textup{RT}} &:& T^{\textup{dust}}_{\textup{RT}}\ \ \&\ \ T_{\textup{p3}}
\end{eqnarray}
Note that for the \ac{PAH} dust only the big grains can heat thermally, and therefore, we couple only big grains with the hydrodynamical temperature. The temperatures are not directly summed, since this would not be physical. In \textsc{Hyperion}, the temperature is actually represented by the specific energy rate $\dot{\epsilon}$ (changes in absorption and emission process), which under the assumption of \ac{LTE} can be transferred into a corresponding temperature. Therefore, the temperatures can be combined by converting them to $\dot{\epsilon}(T)$, and adding those energies before they are converted back. 
%%%%%%%%%%%%%%%%%%%%%%%%%%%%%%%%%%%%%%%%%%%%%%%%%%%%%%%
\subsubsection{Method DT2}
\label{C4:Sec:method_DT2}
%%%%%%%%%%%%%%%%%%%%%%%%%%%%%%%%%%%%%%%%%%%%%%%%%%%%%%%
We will further explore the effects when coupling $T^{\textup{dust}}_{\textup{RT}}$ with an isothermal temperature $T_{\textup{iso}}=\SI{18}{\kelvin}$ for the two different dust types in method \acs{DT2}. The constant temperature of \SI{18}{\kelvin} on average matches the temperature in relatively empty patches in the Galactic plane (see \citetalias{KDR2a:inprep} for more details), and is therefore a good choice for an ambient temperature.
\begin{eqnarray}
    \label{C4:Eq:DT2}
T^{tot}_{\textup{RT}} &:& T^{\textup{dust}}_{\textup{RT}}\ \ \&\ \ T_{\textup{iso}}
\end{eqnarray}
Note that for the \ac{PAH} dust, the temperature is only significant for the big grains, which satisfies \ac{LTE}. Therefore, we couple only big grains with the isothermal temperature. 

%%%%%%%%%%%%%%%%%%%%%%%%%%%%%%%%%%%%%%%%%%%%%%%%%%%%%%%
\subsubsection{Method DT3}
\label{C4:Sec:method_DT3}
%%%%%%%%%%%%%%%%%%%%%%%%%%%%%%%%%%%%%%%%%%%%%%%%%%%%%%%
In the method \acs{DT3}, as a sanity check, we only run the radiative transfer without combining with an extra temperature.
\begin{eqnarray}
    \label{C4:Eq:DT3}
T^{tot}_{\textup{RT}} &:& T^{\textup{dust}}_{\textup{RT}}
\end{eqnarray}
%
%%%%%%%%%%%%%%%%%%%%%%%%%%%%%%%%%%%%%%%%%%%%%%%%%%%%%%%
\subsubsection{Results | Dust \& Temperature Combinations}
\label{C4:Sec:result_DT}
%%%%%%%%%%%%%%%%%%%%%%%%%%%%%%%%%%%%%%%%%%%%%%%%%%%%%%%
We produce ideal synthetic observations of the methods \acs{DT1}, \acs{DT2} and \acs{DT3} using non-\ac{PAH} and \ac{PAH} dust at \SI{20}{\microns} and present them in Figure~\ref{C4:Fig:RT_temp_test}, where we also show a background subtracted image at \ac{WISE} \SI{22}{\microns} of the Soul Nebula \citep{Koenig:2012,Wright:2010}, a real star-forming region, in the right panel.

We now focus on the non-\ac{PAH} dust in the upper panels of Figure~\ref{C4:Fig:RT_temp_test}. We note that the non-\ac{PAH} dust produces much too bright emission\footnote{ Note that the white pixels at the rim of the set-up \acs{DT1} are more than $\num{10}\times$ brighter than the maximum value in the W5 panel. Therefore, also the total flux of the \acs{DT1} set-up is much higher than other comparable values.} in comparison to the real region, when coupling with the hydrodynamical temperature (see method \acs{DT1}). The coupling of isothermal and radiative transfer temperature (\acs{DT2}) produces drastically too low brightnesses (note the colorbar scaling factor \num{e-13}) as does the radiative transfer temperature alone (\acs{DT3}).

We found that the \ac{PAH} dust in the lower panels of Figure~\ref{C4:Fig:RT_temp_test} can reproduce the typical surface brightness of the Galactic star-forming region better. While the coupling with the hydrodynamical temperatures (see method \acs{DT1}) causes even higher brightnesses than its non-\ac{PAH} dust counterpart, neglecting the hydrodynamical temperatures produces better results: When combining the radiative transfer temperature with an ambient isothermal temperature (\acs{DT2}), the brightness of the object is consistent with real observed regions. Also method \acs{DT3} produces a comparable result to the real star-forming region but lacks emission in the outer regions for longer wavelengths.

From Figure~\ref{C4:Fig:RT_temp_test}, we can see that when combining the radiative transfer temperature with the hydrodynamical temperature (see method \acs{DT1}), the brightness in the \ac{MIR} is overestimated for non-\ac{PAH} and \ac{PAH} dust. The high brightness components are produced by the high hydrodynamical temperatures $T_{\textup{p3}}$ in the outer regions, which was coupled with the dust radiative transfer temperature in Eq.~\ref{C4:Eq:DT1}. The gas temperatures are highest in the low density regions, due to the \cite{Larson:2005} \ac{EOS} (c.\,f.~Figure~\ref{C4:Fig:EOS} and Section~\ref{C4:Sec:clipping}), but this is not observed for real star-forming regions. Although the temperatures of neutral \ac{SPH} particles lie below the dust sublimation temperature of \SI{1600}{\kelvin} (see Section~\ref{C4:Sec:methods_disk}), the brightness due to the hydrodynamical temperature is several orders too high, due to the erroneous assumption that the gas and dust are well-thermally coupled everywhere. In fact, the heating of the gas at low densities due to the \cite{Larson:2005} \ac{EOS} used in the \acs{D14} \ac{SPH} simulations, is physically due to the decoupling of the dust from the gas and therefore it does not make sense to use these high temperatures for the dust. Therefore, having explored the different ways of setting up the radiative transfer temperatures, we recommend using method \acs{DT2} by coupling the radiative transfer dust temperature $T^{\textup{dust}}_{\textup{RT}}$ with the ambient isothermal temperature $T_{\textup{iso}}$ for big grains of the \ac{PAH} dust, which recovers the brightness features in the \ac{MIR} much better.
%%%%%%%%%%%%%%%%%%%%%%%%%%%%%%%%%%%%%%%%%%%%%%%%%%%%%%%
%%%%%%%%%%%%%%%%%%%%%%%%%%%%%%%%%%%%%%%%%%%%%%%%%%%%%%%
%\newpage
\section{Realistic Synthetic Observations}
\label{C4:Sec:methods_synobs}
%%%%%%%%%%%%%%%%%%%%%%%%%%%%%%%%%%%%%%%%%%%%%%%%%%%%%%%
%%%%%%%%%%%%%%%%%%%%%%%%%%%%%%%%%%%%%%%%%%%%%%%%%%%%%%%
In \citetalias{KDR2a:inprep} and \citetalias{KDR2b:inprep}, we will test and calibrate techniques commonly used by observers to infer star-formation properties. For this task, it is critical that we have synthetic observations which are as realistic as possible. We set up \textsc{Hyperion} with the preferred methods described from Section~\ref{C4:Sec:methods_HD} to Section~\ref{C4:Sec:methods_dust_temp} and produce ideal synthetic observations. We then simulate all the effects introduced by telescopes and the Galactic environment with the \textsc{FluxCompensator} (\citeauthor[in preparation]{KoepferlRobitaille:inprep}, hereafter referred to \citetalias{KoepferlRobitaille:inprep}). In the following sections, we will give a brief description of these simulated effects by the \textsc{FluxCompensator}.
%%%%%%%%%%%%%%%%%%%%%%%%%%%%%%%%%%%%%%%%%%%%%%%%%%%%%%%
\subsection{Imaging}
\label{C4:Sec:methods_RT_imaging}
%%%%%%%%%%%%%%%%%%%%%%%%%%%%%%%%%%%%%%%%%%%%%%%%%%%%%%%
After the total temperature $T^{tot}_{\textup{RT}}$ calculation/combination in \textsc{Hyperion}, the images are produced by the raytracing algorithm \citep{Robitaille:2011}. Note that for computational reasons we are not computing the scattering: For all the dust properties that we use, the albedo has values smaller than \SI{1}{\percent} above \SI{8}{\microns} (compare with Figure~\ref{C4:Fig:Albedo}) and therefore, scattering is not important in the wavelength regimes we are interested in.

\begin{figure*}[t]
\includegraphics[width=\textwidth]{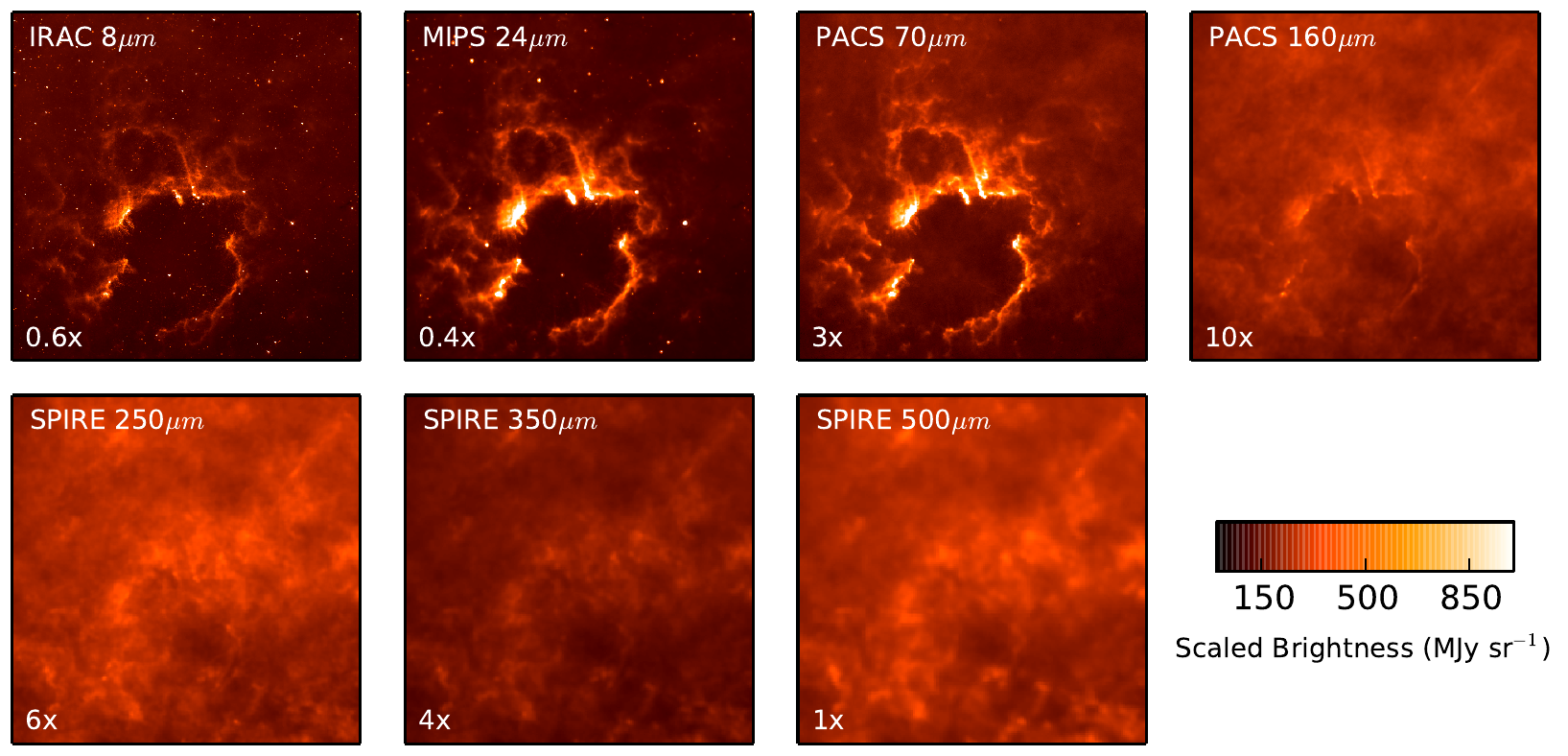}
\caption[Realistic synthetic observations for different bands of \emph{Spitzer} and \emph{Herschel}]{\label{C4:Fig:all_bands} Realistic synthetic observations of a star-forming region of the \acs{D14} \ac{SPH} simulations in \ac{IRAC} \SI{8}{\microns}, \ac{MIPS} \SI{24}{\microns}, \ac{PACS}  \SI{70}{\microns}, \ac{PACS} \SI{160}{\microns}, \ac{SPIRE} \SI{250}{\microns}, \ac{SPIRE}  \SI{350}{\microns} and \ac{SPIRE} \SI{500}{\microns} combined with a real background of time-step 122 (\SI{6.109}{\mega\yr}). Note that not all images have the same scale. The white numbers in the individual panels represent the scaling factors with respect to the colorbar.}
\end{figure*}
We produce ideal synthetic observations for 3 viewing angles looking down the x, y, and z axes respectively. We compute the ideal synthetic observations in a spectral cube of 30 steps in wavelength from \SI{1}{\microns} to \SI{750}{\microns}. The 2-d image slices of the cube span 1200$\times$1200 pixels.
%%%%%%%%%%%%%%%%%%%%%%%%%%%%%%%%%%%%%%%%%%%%%%%%%%%%%%%
\subsection{Distance}
%%%%%%%%%%%%%%%%%%%%%%%%%%%%%%%%%%%%%%%%%%%%%%%%%%%%%%%
We push the ideal synthetic observations from \textsc{Hyperion} to a distance of $d=\SI{3}{\kpc}$ to be roughly comparable to close high-mass star-forming regions, such as the Carina Nebula, the Eagle Nebula or the Heart and Soul Nebulae (Westerhout 4 \& 5), and $d=\SI{10}{\kpc}$ to represent more distant star-forming regions across the Galactic plane. The resulting intrinsic pixel resolution $\vartheta_{0}$ is roughly \ang{;;1.7} and \ang{;;0.5}. For more details of the mentioned clusters in the Milky Way, see \cite{Smith:2008, Oliveira:2008, Megeath:2008} respectively.
%%%%%%%%%%%%%%%%%%%%%%%%%%%%%%%%%%%%%%%%%%%%%%%%%%%%%%%
%\newpage
\subsection{Filter Bands}
%%%%%%%%%%%%%%%%%%%%%%%%%%%%%%%%%%%%%%%%%%%%%%%%%%%%%%%
%
For the analysis of images and photometry in \citetalias{KDR2a:inprep} and \citetalias{KDR2b:inprep}, we require observations in \ac{IRAC} \SI{8}{\microns}, \ac{MIPS} \SI{24}{\microns}, \ac{PACS}  \SI{70}{\microns}, \ac{PACS} \SI{160}{\microns}, \ac{SPIRE} \SI{250}{\microns}, \ac{SPIRE}  \SI{350}{\microns} and \ac{SPIRE} \SI{500}{\microns}. For a detailed description of spectral transmission, see the \citetalias{KoepferlRobitaille:inprep}. We convolve the spectral cube with the transmission curves for these bands and extract 2-d images of the respective bands. We use the \textsc{FluxCompensator}'s built-in database filters. We do not take into account the effects of saturation\footnote{ Note that saturation is currently not implemented in the \textsc{FluxCompensator} which might create synthetic point sources with a large dynamic range in flux. This is enhanced by the effect that in the \acs{D14} \ac{SPH} simulations stars below \SI{20}{\Msun} do not ionize their surrounding or blow winds. A more realistic feedback implementation in future will improve our realistic synthetic observations.}.
%%%%%%%%%%%%%%%%%%%%%%%%%%%%%%%%%%%%%%%%%%%%%%%%%%%%%%%
\newpage
\subsection{Extinction \& Background}
%%%%%%%%%%%%%%%%%%%%%%%%%%%%%%%%%%%%%%%%%%%%%%%%%%%%%%%
The \textsc{FluxCompensator} can also account for effects introduced by the extinction between the simulation object and the observer. We redden the synthetic observations with the extinction law from \cite{Kim:1994}. We set the optical extinction to $A_V=\SI{10}{\mag}$ for the distance $d=\SI{3}{\kpc}$ and to $A_V=\SI{20}{\mag}$ for the distance $d=\SI{10}{\kpc}$. See the \citetalias{KoepferlRobitaille:inprep} for the description of the built-in extinction function of the \textsc{FluxCompensator}.

However, the observations are not only affected by interstellar extinction. One of the difficulties when analyzing real observations is the presence of non-uniform background, and therefore, disentangling the background from the astronomical objects is challenging. Therefore, to make sure we do not compare apples with oranges in \citetalias{KDR2a:inprep} and \citetalias{KDR2b:inprep}, namely to ensure that the simulated observations are as realistic as possible, in this paper \citepalias{KDR1:inprep} we attempt to combine the synthetic observations with real observations. We use relatively empty patches of the Galactic plane 
from the \acl{GLIMPSE} \citep[\acs{GLIMPSE};][]{Benjamin:2003, Churchwell:2009} for \ac{IRAC}, 
from the \acl{MIPSGAL} \citep[\acs{MIPSGAL};][]{Carey:2009} for \ac{MIPS}, 
from the \acl{Hi-GAL} \citep[\acs{Hi-GAL};][]{Molinari:2010} for \ac{PACS} and \ac{SPIRE}.
\begin{figure*}[t]
\includegraphics[width=\textwidth]{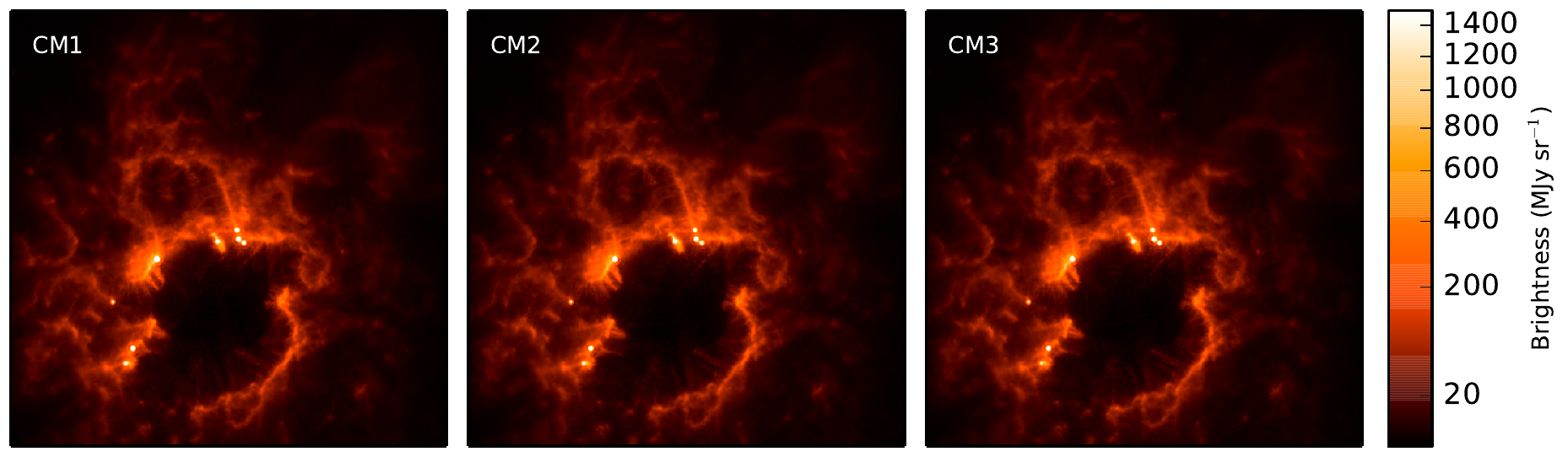}
\caption[Realistic synthetic observations for different circumstellar set-ups]{\label{C4:Fig:all_CM} \ac{MIPS} \SI{24}{\microns} realistic synthetic observations of different circumstellar set-ups. No additional circumstellar material (\acs{CM1}), rotational flattened envelope refinement (\acs{CM2}), symmetrical power-law envelope density distribution (\acs{CM3}).}
\end{figure*}
The observations have an observed pixel resolution $\vartheta_{obs}$ of \ang{;;1.2} (\ac{IRAC} \SI{8}{\microns}), \ang{;;2.4} (\ac{MIPS} \SI{24}{\microns}), \ang{;;3.2} (\ac{PACS} \SI{70}{\microns}), \ang{;;4.5} (\ac{PACS} \SI{160}{\microns}), \ang{;;6.0} (\ac{SPIRE} \SI{250}{\microns}), \ang{;;8.0} (\ac{SPIRE} \SI{350}{\microns}) and \ang{;;11.5} (\ac{SPIRE} \SI{500}{\microns}). We combine relatively empty patches of real observations with our computed realistic synthetic observations for all seven bands. We place our synthetic observations roughly half a degree away ($\ell\simeq\ang{17.1915}$, $b\simeq-\ang{0.61870445}$) from the Eagle Nebula. See the \citetalias{KoepferlRobitaille:inprep} for the description of the built-in observation-combination functionality in the \textsc{FluxCompensator}.
%%%%%%%%%%%%%%%%%%%%%%%%%%%%%%%%%%%%%%%%%%%%%%%%%%%%%%%
\subsection{Resolution \& PSF}
%%%%%%%%%%%%%%%%%%%%%%%%%%%%%%%%%%%%%%%%%%%%%%%%%%%%%%%
Since the intrinsic pixel resolution $\vartheta_{0}$ is different from the observed pixel resolutions $\vartheta_{obs}$, therefore, before we can combine the synthetic observations with real observations, we need to adjust the resolution of synthetic observations to the resolution of the real observations. We change the intrinsic pixel resolution $\vartheta_{0}$ to the pixel resolution of the observations $\vartheta_{obs}$. See the \citetalias{KoepferlRobitaille:inprep} for the description of the built-in re-gridding function of the \textsc{FluxCompensator}. 

To further account for diffraction, we convolve the re-gridded images in the respective bands with the \ac{PSF} of the respective telescopes. See the \citetalias{KoepferlRobitaille:inprep} for the description of the built-in convolution function of the \textsc{FluxCompensator}. We use the \ac{PSF} files provided by the \textsc{FluxCompensator}'s built-in database. 
%%%%%%%%%%%%%%%%%%%%%%%%%%%%%%%%%%%%%%%%%%%%%%%%%%%%%%%
\subsection{Set of Realistic Synthetic Observations}
\label{C4:Sec:result_synobs}
%%%%%%%%%%%%%%%%%%%%%%%%%%%%%%%%%%%%%%%%%%%%%%%%%%%%%%%
For every selected time-step in the \acs{D14} \ac{SPH} simulations, we produced realistic synthetic observations as described in Section~\ref{C4:Sec:methods_synobs}, using our favored techniques presented and discussed in Section~\ref{C4:Sec:methods_HD} to Section~\ref{C4:Sec:methods_dust_temp}. 

For the time-step 122 (\SI{6.109}{\Myr}), we show all seven bands of the realistic synthetic observations, combined with a real observed background in Figure~\ref{C4:Fig:all_bands} for a distance of \SI{3}{\kpc} and one viewing angle. While for the \ac{IRAC}, \ac{MIPS} and \ac{PACS} bands, the features of the simulated region dominate the image, in \ac{SPIRE} the features are overpowered by the background. Clearly, this star-forming region does not contain enough mass for the \ac{SPIRE} emission to dominate, when compared to the other material in the Galactic plane. Since the Galactic plane is full of clouds, other background patches within the \ac{Hi-GAL} survey could not likely improve this. Therefore, intermediate-mass clouds, such as the simulated region here, are "lost" in \ac{SPIRE} bands. This has to be kept in mind when analyzing the realistic synthetic observations in \citetalias{KDR2a:inprep} and \citetalias{KDR2b:inprep}. However, we also provide the realistic synthetic observations without background. Naturally, a desired background observation in \ac{SPIRE} would be slightly off the Galactic plane of a relatively empty region. However, off the Galactic plane, the \emph{Herschel} Space Observatory only observed star-forming clouds directly. We could have selected a region off the Galactic plane with lower background, but this would not have been typical of most observed star-forming regions, which at that distance are all located in the Galactic plane.

In Figure~\ref{C4:Fig:all_CM}, we show one example for the three circumstellar set-ups described in Section~\ref{C4:Sec:methods_RT_YSO}. We can see that we can "observe" point sources for all three set-ups. In \citetalias{KDR2b:inprep} and \citetalias{KDR3:inprep}, we will explore the differences when counting the young stellar population in our synthetic observations.

In Figure~\ref{C4:Fig:evolution_PACS1}, we show an evolutionary sequence of the \emph{run I} star-forming region of the \acs{D14} \ac{SPH} simulations in \ac{PACS} \SI{70}{\microns} at a distance of \SI{3}{\kpc} and for one viewing angle. We provide all (5796) realistic synthetic observations for the 3 circumstellar material set-ups, 23 time-steps, 3 viewing angles, 2 distances, 7 spectral bands and both with and without combined realistic background as \acs{FITS} files in Appendix~\ref{C4:Appendix_data}.

%%%%%%%%%%%%%%%%%%%%%%%%%%%%%%%%%%%%%%%%%%%%%%%%%%%%%%%
%%%%%%%%%%%%%%%%%%%%%%%%%%%%%%%%%%%%%%%%%%%%%%%%%%%%%%%
\section{Summary}
\label{C4:Sec:discuss}
%%%%%%%%%%%%%%%%%%%%%%%%%%%%%%%%%%%%%%%%%%%%%%%%%%%%%%%
%%%%%%%%%%%%%%%%%%%%%%%%%%%%%%%%%%%%%%%%%%%%%%%%%%%%%%%

In this paper \citepalias{KDR1:inprep}, we produced realistic synthetic observations of the coupled high-mass stellar feedback \emph{run I} of winds and ionization of the \acs{D14} \ac{SPH} simulations. We showed that mapping of \ac{SPH} particles onto a Voronoi mesh has to be done with caution. We have tested several techniques which can perform this task and found that the following methods produce the best results:
\begin{figure*}
    \centering
\includegraphics[width=0.9\textwidth]{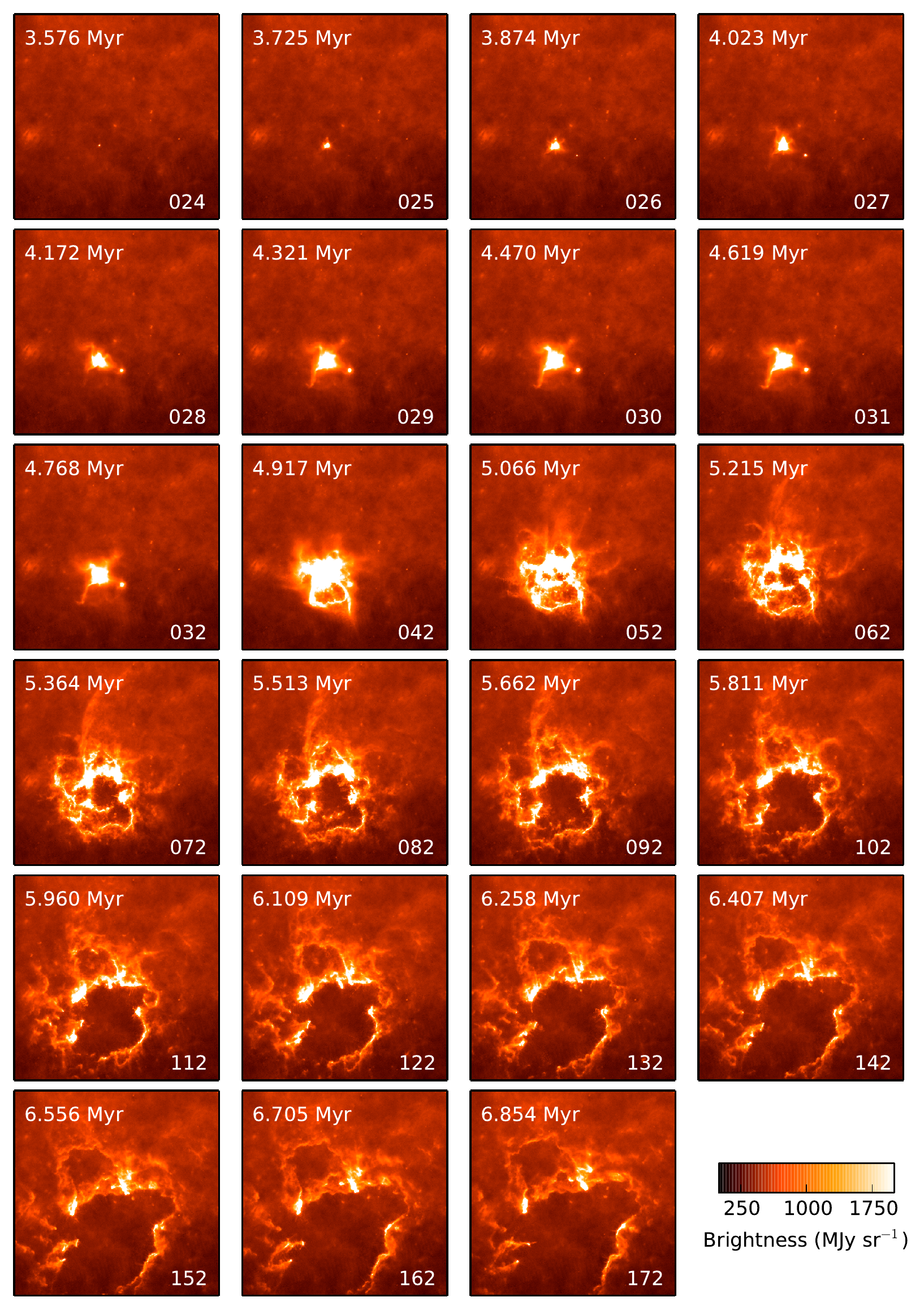}
\caption[Realistic synthetic observations for different evolutionary steps]{\label{C4:Fig:evolution_PACS1} \ac{PACS} \SI{70}{\microns} realistic synthetic observations (including background) over all time-steps of the \acs{D14} \ac{SPH} simulations.}
\end{figure*}
\begin{itemize}
    \item {\bf Pre-processing}\\
    Different radiative transfer codes treat different physics, such as dust continuum or molecular line transfer. Therefore, it is 
    important to filter away those \ac{SPH} particles from the \ac{SPH} distribution (e.\,g.~ionized particles), which are meaningless for the radiative transfer code in question, before mapping onto the radiative transfer mesh.
    \item {\bf Grid}\\
    Within a Voronoi mesh, resolution can be increased by gradually putting in more cells at regions of interest, for instance around accreting stars. The absence of a preferred direction of the mesh creates smoother images than Cartesian grids when "observing" along a grid axis. 
    \item {\bf Sites}\\
    When a Voronoi mesh is set up to represent an \ac{SPH} simulation, including large low-density regions can cause artifacts when mapping only the positions of the \ac{SPH} particles onto the mesh. Artifacts are reduced by placing Voronoi sites also at the positions of the sink particles in addition to the sites at the positions of the \ac{SPH} particles. Additionally, adding circumstellar sites around the accreting sink particles reduces artifacts and increases the resolution in the circumstellar material.
    \item {\bf Mapping}\\
    The mapping of the properties, such as density and temperature, from the \ac{SPH} simulation onto the radiative transfer mesh has to preserve the gradients of the original distributions. We presented a random sampling technique which ensures a smooth density and temperature mapping from the \ac{SPH} simulations onto the mesh, while conserving mass.
    \item {\bf Envelope Fitting}\\
    We present three different ways of refining the circumstellar material beyond the resolution limit of the \ac{SPH} simulation. We presented methods to extrapolate the inner circumstellar material with rotational flattened and spherically symmetric power-law profiles. However, inward extrapolations are only an approximation. In order to have a consistent physical picture, in future, simulations with higher resolution would be ideal. 
   \item {\bf Circumstellar Material}\\
   We produced a set of realistic synthetic observations with and without circumstellar material to evaluate the impact of circumstellar material on observed fluxes. We will explore the effects on the total flux in \citetalias{KDR2b:inprep} and the point-source counting in \citetalias{KDR2b:inprep} and \citetalias{KDR3:inprep}. 
    \item {\bf Dust}\\
    To recover the \ac{NIR} and \ac{MIR} flux (below \SI{70}{\microns}), dust distributions which include \ac{PAH} molecules (e.\,g.~\citealt{Draine:2007}) should be used when calculating the radiative transfer.
    \item {\bf Temperatures}\\
    Coupling the \ac{SPH} temperature, following \cite{Larson:2005}'s \ac{EOS}, with the iterated radiative transfer dust temperature produces unphysical emission within a dust radiative transfer code, although the hydrodynamical temperature is below the dust sublimation temperature. The line cooling part of \cite{Larson:2005}'s \ac{EOS} causes a gas temperature rise in low density regions at the edges of the simulation, which produces unphysical/unobserved fluxes when comparing to real star-forming regions, when blindly coupling the hydrodynamical temperature with the dust radiative transfer temperature. However, the heating term in \cite{Larson:2005}'s \ac{EOS} at low densities implies a decoupling from gas and dust. Therefore, we conclude that, at least for dust radiative transfer codes, the hydrodynamical temperature (hence gas temperature) should be replaced by a constant background temperature.
\end{itemize}

In this paper \citepalias{KDR1:inprep}, we will provide online about 5800 realistic synthetic observations of the 3 circumstellar material set-ups, 23 time-steps, 3 viewing angles, 2 distances, 7 spectral bands and with combined background and without of the synthetic star-forming region. The \acs{FITS} files of the "observations" are stored in Appendix~\ref{C4:Appendix_data}. In \citetalias{KDR2a:inprep}, \citetalias{KDR2b:inprep} and \citetalias{KDR3:inprep}, we will analyze these "observations".

\section{Acknowledgements}
We thank the referee for a constructive report that helped us improve the clarity and the strength of the results presented in our paper.
This work was carried out in the Max Planck Research Group \textit{Star formation throughout the Milky Way Galaxy} at the Max Planck Institute for Astronomy. C. K. is a fellow of the International Max Planck Research School for Astronomy and Cosmic Physics (IMPRS) at the University of Heidelberg, Germany and acknowledges support. C. K. acknowledges support from STFC grant ST/M001296/1.
J. E. D. was supported by the DFG cluster of excellence \textit{Origin and Structure of the Universe}. 
This research made use of Astropy, a community-developed core Python package for Astronomy \citep{Astropy:2013}, matplotlib, a Python plotting library \citep{Hunter:2007}, Scipy, an open source scientific computing tool \citep{Scipy}, the NumPy package \citep{NumPy} and IPython, an interactive Python application \citep{IPython}.

\bibliographystyle{latex_apj}
\bibliography{latex_ref}

\appendix
\input{latex_header}

\section{Online Observations of Synthetic Star-forming Regions}
\label{C4:Appendix_data}

We provide $\sim$5800 realistic synthetic observations (\acs{FITS} files) of the synthetic star-forming region for the combinations of the 3 circumstellar material set-ups, 23 time-steps, 3 viewing angles, 2 distances, 7 spectral bands, with combined background and without. We analyze these "observations" and we will publish the results in \citetalias{KDR2a:inprep}, \citetalias{KDR2b:inprep} and \citetalias{KDR3:inprep}. The realistic synthetic observations can be accessed here: 
\begin{center}
\url{http://dx.doi.org/10.5281/zenodo.31293}
\end{center}
The filename of the "observations" is constructed of the combinations of abbreviations which refer to methods/algorithms from \citetalias{KDR1:inprep} and are separated by underscores:
\begin{lstlisting}[numbers=none,basicstyle=\small]
both_I122_s3_p3_voronoi_DraineLiPAH_c1_DT2_CM1_O1_D1_MIPS1_R1_B2.fits
\end{lstlisting}
The filenames are set up through the following system:
\begin{lstlisting}[numbers=none,basicstyle=\small]
both        : <feedback type>              # both (winds & ionization)
I122        : <SPH ID>                     # SPH simulation set-up type & time-step
s3          : <site distribution>          # sites at SPH particles (s1), 
                                           # additional sites at sink particles (s2), 
                                           # additional circumstellar sites (s3)
p3          : <parameter distribution>     # SPH function (p1), 
                                           # SPH splitting (p2), 
                                           # random sampling method (p3)
voronoi     : <grid type>                  # Voronoi tessellation
DraineLiPAH : <dust type>                  # DraineLiPAH, DraineMW
c1          : <clipping>                   # for equation of state (c1), 
                                           # for temperature (c2)
DT2         : <RT temperature & dust>      # combined temperatures (DT1), 
                                           # isothermal temperature (DT2), 
                                           # radiative transfer temperature alone (DT3)
CM1         : <circumstellar material>     # none (CM1), 
                                           # Ulrich analytical model (CM2), 
                                           # power-law analytical model (CM3)
O1          : <orientation plane>          # O1 (xy), O2 (xz), O3(yz)
D1          : <distance>                   # D1 (3kpc), D2 (10kpc)
MIPS1       : <detector>                   # Detector of realistic synthetic observation
R1          : <resolution type>            # detector resolution (R1), 
                                           # technique resolution (R2)
B2          : <background type>            # without background (B1), 
                                           # with background (B2)
\end{lstlisting}
The different methods and their abbreviations are described in \citetalias{KDR1:inprep}. In the header of the respective \acs{FITS} file, the information from the file name is stored additionally. The \acs{FITS} files include world coordinate information similarly to real observations from the respective detectors. As an example, we show the \acs{FITS} header of a \ac{MIPS} \SI{24}{\microns} observation below: 

\begin{lstlisting}[numbers=none,basicstyle=\small]
SIMPLE  =                    T / conforms to FITS standard                      
BITPIX  =                  -64 / array data type                                
NAXIS   =                    2 / number of array dimensions                     
NAXIS1  =                  860                                                  
NAXIS2  =                  860                                                  
WCSAXES =                    2 / Number of coordinate axes                      
CRPIX1  =              -781.75 / Pixel coordinate of reference point            
CRPIX2  =              1358.25 / Pixel coordinate of reference point            
CDELT1  =      -0.000666666666 / [deg] Coordinate increment at reference point  
CDELT2  =       0.000666666666 / [deg] Coordinate increment at reference point  
CUNIT1  = 'deg'                / Units of coordinate increment and value        
CUNIT2  = 'deg'                / Units of coordinate increment and value        
CTYPE1  = 'GLON-CAR'           / galactic longitude, plate caree projection     
CTYPE2  = 'GLAT-CAR'           / galactic latitude, plate caree projection      
CRVAL1  =                 18.0 / [deg] Coordinate value at reference point      
CRVAL2  =               0.0001 / [deg] Coordinate value at reference point      
LONPOLE =                  0.0 / [deg] Native longitude of celestial pole       
LATPOLE =              89.9999 / [deg] Native latitude of celestial pole        
BUNIT   = 'MJy/sr  '           / Units of array                                 
DISTANCE=                  3.0 / Distance to object in units of kpc             
FEEDBACK= 'both    '           / High-mass stellar feedback                     
RUNTYPE = 'I       '           / SPH simulation set-up type                     
TIMESTEP= '122     '           / SPH simulation time-step                       
SITETYPE= 's3      '           / Site distribution: s1, s2, s3                  
PARATYPE= 'p3      '           / Parameter mapping: p1, p2, p3                  
GRIDTYPE= 'voronoi '           / Grid types: Voronoi tessellation               
DUSTTYPE= 'DraineLiPAH'        / Dust types: DraineLiPAH, DraineMW              
CLIPTYPE= 'c1      '           / Clipping SPH sample: c1, c2                    
DTTYPE  = 'DT2     '           / Radiative transfer set-up: DT1, DT2, DT3       
CMTYPE  = 'CM1     '           / Circumstellar material: CM1, CM2, CM3          
ORIETYPE= 'O1      '           / Orientation plane: O1 (xy), O2 (xz), O3(yz)    
DISTTYPE= 'D1      '           / Distance: D1 (3kpc), D2 (10kpc)                
RTCODE  = 'HYPERION'           / 3-d dust continuum radiative transfer code     
SYNOBS  = 'FluxCompensator'    / Tool to create realistic synthetic observations
DETECTOR= 'MIPS1   '           / Detector of realistic synthetic observations   
RESTYPE = 'R1      '           / Resolution type: R1 (detector), R2 (other)     
BGTYPE  = 'B2      '           / Background type: B1 (without), B2 (with)       
COMMENT For more description see Appendix.                                      
END                                                                             
\end{lstlisting}

\section{Random Sampling with Probability Distribution Functions}
\label{C4:Appendix_PDF}

A \ac{PDF} is dependent on a variable $x$ that we want to distribute. A random number $\xi$ is sampled between $[0,1]$ and the distributed variable $x(\xi)$ is equated through analytical or numerical integration of the \ac{PDF} through inverse transform sampling \citep[for more information on \acp{PDF}, see][]{NumericalRecipes}:
    \begin{eqnarray}
        \label{C1:Eq:PDF} \xi&=&\frac{\int\limits_{x_{min}}^{x}P(x')dx'}{\int\limits_{x_{min}}^{x_{max}}P(x')dx'}  \rightarrow x(\xi)
    \end{eqnarray}
Following Section~\ref{C4:Sec:method_s3}, we can sample the radial log-spaced \ac{PDF} through a step-by-step evaluation of the integral over the normalized \ac{PDF}. The log-spaced probability distribution produces more sites at small radii. Hence, the size of the Voronoi cells will decrease and we recover a smoother gradient of density across the dynamic range (this will be discussed further in Section~\ref{C4:Sec:methods_envelope}). With a constant $\textup{PDF}=const=c$, we can solve the integral for the radius $r$ analytically using inverse transform sampling:
\begin{eqnarray}
    \label{C4:Eq:PDF}
    \xi&=&\int\limits_{r_{min}}^{r} d\log_{10}(r)\textup{PDF}\stackrel{\textup{PDF}=c}{=} c\int\limits_{r_{min}}^{r} d\log_{10}(r) = c\left[{\log_{10}{r} - \log_{10}{r_{min}}}\right]\\
    r &=& 10^{\xi/c+ \log_{10}{r_{min}}}.\nonumber
\end{eqnarray}
Eq.~\ref{C4:Eq:PDF} uses the random numbers $\xi$ from 1 to zero and the normalization constant $c$
\begin{eqnarray}
    1&=&\int\limits_{r_{min}}^{r_{max}} d\log_{10}(r)\textup{PDF} \stackrel{\textup{PDF}=c}{=} c\int\limits_{r_{min}}^{r_{max}} d\log_{10}(r) = c\left[\log_{10}{r_{max}} - \log_{10}{r_{min}}\right]\\
    c&=&1/\left[\log_{10}{r_{max}} - \log_{10}{r_{min}}\right],
\end{eqnarray}
to calculate the distribution function:
\begin{eqnarray}
r &=& 10^{\xi\left[\log_{10}{r_{max}} - \log_{10}{r_{min}}\right]+ \log_{10}{r_{min}}}.
\end{eqnarray}

Following Section~\ref{C4:Sec:method_p2}, to space the split \ac{SPH} particle $k$ with respect to the \ac{SPH} particle $i$ again, we use the \ac{PDF} (as in Eq.~\ref{C4:Eq:PDF}). However, this time the 
\begin{eqnarray}
    \textup{PDF}&=&\stackrel{\sim}{c}W_{ik}(r_{ik},h_i)\ \neq\ const
\end{eqnarray}
with the spacing $r_{ik}$ and the \ac{SPH} particle smoothing length $h_i$. We determine the spacing $r_{ik}$, by setting up the integral over the \ac{PDF} (i.\,e.~Eq.~\ref{C4:Eq:PDF}) but this time with linear spacing:
\begin{eqnarray}
    \xi&=&\int\limits_{r_{min}}^{r_{ik}} dr\ r^2 \ \textup{PDF} = \stackrel{\sim}{c}\int\limits_{r_{min}}^{r_{ik}} dr\ r^2 W_{ik}(r_{ik},h_i=1)
\end{eqnarray}
for a constant smoothing length $h_i=\num{1}$, with the random numbers $\xi$ and the normalizing constant $\stackrel{\sim}{c}$. Since the \acs{PDF} is not constant, we have to solve for the values of distance $r_{ik}$ numerically \citep[see][]{NumericalRecipes}.

\section{Spectra Interpolation}
\label{C4:Appendix_Spectra}
Following Section~\ref{C4:Sec:methods_RT_spectra}, we will describe here in more detail the spectra interpolation. We pick the four closest model spectra $F_{\textup{mod}}(T_1,T_2,g_1,g_2)$, which have values between 
\begin{eqnarray}
\log_{10}{T_{2}}&\geq&\log_{10}{T_{\textup{eff}}}\geq\log_{10}{T_{1}}\\
\log_{10}{g_{2}}&\geq&\log_{10}{g_{*}}\geq\log_{10}{g_{1}}
\end{eqnarray} 
and define the model differences as follows:
\begin{eqnarray}
    \label{C4:Eq:delta_T}
        \Delta T &=& \log_{10}{T_{2}}-\log_{10}{T_{1}}\\
        \Delta g &=& \log_{10}{g_{2}}-\log_{10}{g_{1}}\\
        \Delta T_{e} &=& \log_{10}{T_{2}}-\log_{10}{T_{\textup{eff}}}\\
    \label{C4:Eq:delta_g}
        \Delta g_* &=& \log_{10}{g_{2}}-\log_{10}{g_{*}}.
\end{eqnarray}
Through bilinear interpolation \citep[for more details, see][]{NumericalRecipes}, we weight the four spectra 
\begin{eqnarray}
F_{\textup{mod}} &=&\begin{pmatrix}
        F_{\textup{mod}}(T_{1}, g_{1}) &  F_{\textup{mod}}(T_{1}, g_{2})\\
        F_{\textup{mod}}(T_{2}, g_{1}) &  F_{\textup{mod}}(T_{2}, g_{2})\\
\end{pmatrix}
\end{eqnarray}
with the differences from Eq.~\ref{C4:Eq:delta_T} to Eq.~\ref{C4:Eq:delta_g} and extract the spectra $F_{\nu*}$ of a source with mass $M_*$:
\begin{eqnarray}
F_{\nu*} &=& 
\begin{pmatrix}
        \Delta T_{e} && \Delta T - \Delta T_{e}
\end{pmatrix}
\frac{F_{\textup{mod}}}{\Delta T\Delta g}
\begin{pmatrix}
        \Delta g_*\\
        \Delta g -\Delta g_{*}
\end{pmatrix}.
\end{eqnarray}

\end{document}

%% file: latex_header.tex
\begin{acronym}[ATLASGAL]
\acro{2MASS}{Two Micron All-Sky Survey}
%A
\acro{AGB}{Asymptotic Giant Branch}
\acro{ALMA}{Atacama Large Millimeter/Submillimeter Array}
\acro{AMR}{Adaptive Mesh Refinement}
\acro{ATLASGAL}{APEX Telescope Large Area Survey of the Galaxy}
%B
\acro{BGPS}{Bolocam Galactic Plane Survey}
%C
\acro{c1}{sample clipping of only neutral particles within a box of \SI{30}{\pc}}
\acro{c2}{sample clipping of \acs{c1} particles and for a certain threshold temperature}
\acro{c2d}{Cores to Disks Legacy}
\acro{CASA}{Common Astronomy Software Applications package}
\acro{cm}{centimeter}
\acro{CMF}{core mass function}
\acro{CMZ}{central molecular zone}
 %D
\acro{D1}{distance at \SI{3}{\kpc}}
\acro{D14}{\acs{SPH} simulations performed by Jim Dale and collaborators \citep{DaleI:2011,DaleIoni:2012,DaleIoni:2013,DaleWind:2013,DaleBoth:2014}}
\acro{D2}{distance at \SI{10}{\kpc}}
\acro{DT1}{temperature coupling of radiative transfer \& hydrodynamical temperature}
\acro{DT2}{temperature coupling of the radiative transfer \& isothermal temperature}
\acro{DT3}{no temperature coupling of the radiative transfer temperature}
 %E
\acro{e1}{using the \cite{Ulrich:1976} envelope profile to extrapolate the envelope inwards}
\acro{e2}{using the \cite{Ulrich:1976} envelope profile with suppressed singularity to extrapolate the envelope inwards}
\acro{e3}{using a power-law envelope profile to extrapolate the envelope inwards}
\acro{EOS}{equation of state}
 %F
\acro{FIR}{far-infrared}
\acro{FITS}{Flexible Image Transport System}
\acro{FWHM}{full-width at half-maximum}
 %G
\acro{GLIMPSE}{Galactic Legacy Infrared Mid-Plane Survey Extraordinaire}
\acro{GMC}{Giant Molecular Clouds}
 %H
\acro{Hi-GAL}{\emph{Herschel} Infrared Galactic Plane Survey}
\acro{HST}{\emph{Hubble} Space Telescope}
\acro{HWHM}{half-width at half-maximum}
 %I
\acro{IMF}{initial mass function}
\acro{IR}{infrared}
\acro{IRAC}{Infrared Array Camera}
\acro{IRAS}{Infrared Astronomical Satellite}
\acro{ISM}{interstellar medium}
 %J
 %K
\acro{$K$}{K band}
 %L
\acro{LTE}{local thermodynamical equilibrium}
 %M
\acro{MIPS}{Multiband Imaging Photometer for \emph{Spitzer}}
\acro{MIPSGAL}{\acs{MIPS} Galactic Plane Survey}
\acro{MIR}{mid-infrared}
\acro{MS}{main-sequence}
\acro{mm}{millimeter}
 %N
\acro{NASA}{National Aeronautics and Space Administration}
\acro{NIR}{near-infrared}
\acro{N-PDF}{column density \acs{PDF}}
 %O
 \acro{O1}{xy plane}
 \acro{O2}{xz plane}
 \acro{O3}{yz plane}
  
 %P
\acro{p1}{parameter evaluation version from \acs{SPH} kernel function}
\acro{p2}{parameter evaluation version from \acs{SPH} splitted kernel distribution}
\acro{p3}{parameter evaluation version from \acs{SPH} random distribution}
\acro{PACS}{Photoconductor Array Camera and Spectrometer}
\acro{PAH}{polycyclic aromatic hydrocarbon}
\acro{PDF}{probability distribution function}
\acro{PDR}{Photon Dominated Region}
\acro{PSF}{point-spread-function}
\acro{px}{one of the parameter evaluation version \acs{p1}, \acs{p2}, \acs{p3}}
 %Q
 %R
\acro{RGB}{red, green and blue}
\acro{CM1}{circumstellar setup with background density and sink mass as stellar mass}
\acro{CM2}{circumstellar setup by a toy model with \acs{e2} envelope superposition density and corrected stellar mass, protoplanetary disk and envelope cavity}
\acro{CM3}{circumstellar setup by a toy model with \acs{e3} envelope superposition density and corrected stellar mass, protoplanetary disk and envelope cavity}
 %S
\acro{s1}{Voronoi site placement version at \acs{SPH} particle position}
\acro{s2}{Voronoi site placement version as \acs{s1} including sites at sink particles}
\acro{s3}{Voronoi site placement version as \acs{s2} including circumstellar sites}
\acro{SAO}{Smithsonian Astrophysical Observatory}
\acro{SED}{spectral energy distribution}
\acro{SFE}{star-formation efficiency}
\acro{SFR}{star-formation rate}
\acro{SFR24}{technique to measure the \acs{SFR} using the \SI{24}{\microns} tracer}
\acro{SFR70}{technique to measure the \acs{SFR} using the \SI{70}{\microns} tracer}
\acro{SFRIR}{technique to measure the \acs{SFR} using the total infrared tracer}
\acro{Sgr}{Sagittarius}
\acro{SMA}{Submillimeter Array}
\acro{SPH}{smoothed particle hydrodynamics}
\acro{SPIRE}{Spectral and Photometric Imaging Receiver}
\acro{sub-mm}{sub-millimeter}
 %T
 %U
\acro{UKIDSS}{UKIRT Infrared Deep-Sky Survey}
\acro{UKIRT}{UK Infrared Telescope}
\acro{UV}{ultra-violet}
 %V
 %W
\acro{WFCAM}{\acs{UKIRT} Wide Field Camera}
\acro{WISE}{Wide-field Infrared Survey Explorer}
 %X
 %Y
\acro{YSO}{young stellar object}
\end{acronym}